\documentclass[secnumarabic,amssymb, nobibnotes, aps, prd]{revtex4-2}
\usepackage{graphicx}
\usepackage{amsmath}
\usepackage{subcaption}

\usepackage{multirow}

\begin{document}
	
	\title{Thermodynamic topology of black holes in f(R) gravity }%

	\author{Bidyut Hazarika$^1$}
	
	\email{$rs_bidyuthazarika@dibru.ac.in$}
	
	\author{Prabwal Phukon$^{1,2}$}
	\email{prabwal@dibru.ac.in}
	
	\affiliation{$1.$Department of Physics, Dibrugarh University, Dibrugarh, Assam,786004.\\$2.$Theoretical Physics Division, Centre for Atmospheric Studies, Dibrugarh University, Dibrugarh, Assam,786004.}

	\maketitle
	\tableofcontents
	\clearpage
	\section*{Abstract}
	In this work, we study the thermodynamic topology of a static, a charged static and a charged, rotating black hole in $f(R)$ gravity. For charged static black holes, we work in two different ensembles: fixed charge$(q)$ ensemble and fixed potential$(\phi)$ ensemble. For charged, rotating black hole, four different types of ensembles are considered: fixed $(q, J)$, fixed $(\phi, J)$, fixed $(q,\Omega)$ and fixed $(\phi,\Omega)$ ensemble, where $J$ and $\Omega$ denotes the angular momentum and the angular frequency respectively. Using the generalized off-shell free energy method, where the black holes are treated as topological defects in their thermodynamic spaces, we investigate the local and global topology of these black holes via the computation of winding numbers at these defects. For static black hole we work in three model. We find that the topological charge for a static black hole is always $-1$ regardless of the values of the thermodynamic parameters and the choice of $f(R)$ model. For a charged static black hole, in the fixed charge ensemble, the topological charge is found to be zero. Contrastingly, in the fixed $\phi$ ensemble, the topological charge is found to be $-1.$ For charged static black holes, in both the ensembles, the topological charge is observed to be independent of the thermodynamic parameters. For charged, rotating black hole, in fixed $(q, J)$ ensemble, the topological charge is found to be $1.$  In $(\phi, J)$ ensemble, we find the topological charge to be $1.$ In case of fixed $(q,\Omega)$ ensemble, the topological charge is $1$ or $0$ depending on the value of the scalar curvature($R$). In fixed $(\Omega,\phi)$ ensemble, the topological charge is $-1,0$ or $1$ depending on the values of $R,\Omega$ and $\phi.$ Therefore, we conclude that the thermodynamic topologies of the charged static black hole and charged rotating black hole are influenced by the choice of ensemble. In addition, the thermodynamic topology of the charged rotating black hole also depends on the thermodynamic parameters. 
	\section{Introduction:}
	Black hole thermodynamics has advanced significantly over the last fifty years since its initiation in the 1970s \cite{1,2,3,4,5,6,7,8,9,10,11,12,13}, culminating in new frameworks of study such as extended black hole thermodynamics \cite{14,15,16,17,18,19,20,21,22,23,24,25,new}, restricted phase space thermodynamics and holographic thermodynamics \cite{rp1,rp2,rp3,rp4,rp5,rp6,rp7,rp8,rp9,rp10,rp11}. A relatively recent development in the context of understanding critical phenomena in black hole thermodynamics is the introduction of topology in black hole thermodynamics \cite{28,29}. In the approach shown in \cite{29} and known as the off-shell free energy method, the topology of black hole thermodynamics can be studied by treating the black hole solutions as topological defects in their thermodynamic space. The local and global topology of the black hole are then analyzed by computing the winding numbers at these defects. Based on the total winding number or the topological charge, all black hole solutions are conjectured to be classified into three topological classes. Such studies of thermodynamic topology have been generalized to a variety of black hole solutions in different theories of gravity \cite{30,31,32,33,34,35,36,37,38,39,40,41,42,43,44,45,46,47,48,49,50,51,52,53,54,55,56,57,58,59,60,61,62,63,64,65,66,67,68,69,70,71,72,73}. \\
	
	In the off-shell free energy method, one begins with an expression for the off-shell free energy of a black hole with arbitrary mass given by:
	\begin{eqnarray}
		\mathcal{F}=E-\frac{S}{\tau} \label{8}  
	\end{eqnarray}
	Here, $E$ and $S$ are the energy and entropy of the black hole respectively. $\tau$ is the time scale parameter which can be considered as the reciprocal of the cavity's  temperature  that encloses the black hole :
	\begin{equation}
		\tau=\frac{1}{T}
	\end{equation}
	Here,  $T$ is the equilibrium temperature at the surface of the cavity and the time parameter $\tau$ is set freely to vary. Utilizing the generalized free energy, a vector field is constructed as follows \cite{29} :
	\begin{eqnarray}
		\phi=\left(\phi^r,\phi^\Theta \right)=\left(\frac{\partial\mathcal{F}}{\partial r_{+}},-\cot\Theta ~\csc\Theta  \right)
	\end{eqnarray}
	Where $\theta=\pi/2$ and $\tau=1/T$ represent a zero point of the vector field $\phi$. The topological property associated with the zero point of a field is its winding number or topological charge. The topological charge can be calculated by constructing a contour $C$ around each zero point which is parametrized as:
	\begin{equation}
		\begin{cases}
			r_+=a cos\nu+r_0\\
			\theta=b sin \nu +\frac{\pi}{2}\\
		\end{cases}
	\end{equation}
	where $\nu \in (0,2\pi)$
	followed by calculating the deflection of vector field $n$ along the contour $C$ as:
	\begin{equation}
		\Omega(\nu)=\int_{0}^{\nu} \epsilon_{12} n^1\partial_\nu n^2 d\nu
	\end{equation}
	The unit vectors $\left(n^1, n^2\right)$ are given by: 
	$$n^1=\frac{\phi^r}{\sqrt{(\phi^r)^2+(\phi^\Theta)^2}}  \hspace{0.5cm}\text{and} \hspace{0.5cm} n^2=\frac{\phi^\theta}{\sqrt{(\phi^r)^2+(\phi^\Theta)^2}}$$
	Finally, the winding numbers $w$ and topological charge $W$ can be calculated as follows:
	\begin{equation}
		\begin{cases}
			w=\frac{1}{2\pi} \Omega(2\pi)\\
			W=\sum_{i} w_i\\
		\end{cases}
	\end{equation}
	In cases where the parameter region does not encompass any zero points, the overall topological number or charge is determined to be $0$. This approach for computing the topological number or charge is referred to as Duan's $\phi$ mapping technique \cite{g1,g2}.\\
	An alternative method used to calculate the winding number has been proposed in \cite{64}. In this approach, the winding number $(w_{i})$ for each solution can be calculated by using the residue theorem. First, a solution for $\tau$ is obtained for the following equation.
	\begin{equation}
		\frac{\partial\mathcal{F}}{\partial r_{+}}=0
		\label{sol}
	\end{equation}
	$$\frac{\partial\mathcal{F}}{\partial r_{+}}=0$$ 
	the solution for $\tau$, thus obtained, is a function of the horizon radius $r_+.$ This is followed by replacing $r_+$ with a complex variable $z.$ and renaming the solution to equation \ref{sol} as $ \mathcal{G}(z)$. Then a rational complex function $\mathcal{R}(z)$ is constructed as follows :
	\begin{equation}
		\mathcal{R}(z)=\frac{1}{\tau - \mathcal{G}(z)}  \label{residue2}
	\end{equation}
	In the final step, the residues at the poles of $\mathcal{R}(z)$ are computed to find the winding number at the defects.
	\begin{eqnarray}
		w_{i}=\frac{Res\mathcal{R}(z_{i})}{|Res\mathcal{R}(z_{i})|}=Sign [Res\mathcal{R}(z_{i}]\label{residue1}
	\end{eqnarray}
	The total winding number or the topological charge, $W$ is given by $W=\sum_{i} w_{i}$.\\
	
	In this study, we investigate the thermodynamic topology of black holes within the framework of $f(R)$ gravity. Applications of modified theories of gravity have recently garnered significant attention across several fields of theoretical physics. Among these alternatives to Einstein's gravity, $f(R)$ theories which incorporate features of both cosmological and astrophysical significance, are particularly noteworthy \cite{f1,f2,f3,f4,f5}. In $f(R)$ gravity, the gravitational action is expressed as a general function of the scalar curvature $R.$Various facets associated with modified theories of gravity, including black hole solutions,cosmic inflation, cosmic acceleration, cosmic rays, dark matter, correction of solar system anomalies etc have been studied within the realm of $f(R)$ gravity \cite{f6,f7,f8,f9a,f9b,f10,f11a,f11b,f12,f13,f14,f15,f16,f17,f18,f19,f20,f21,f22,f23,f24,f25,f26,f27,f28,f29,f30,f31,f32,f33,f34,f35,f36,f37,f38,f39,f40,rc1,fr1,fr2,new1,new2,new3,new4,new5,new6,new7,new8,new9,new10,new11,new12}.\\
	
	For our analysis, We have considered three black hole solutions in $f(R)$ gravity. These are:  static black holes,  static charged black holes, and  rotating charged black holes. For static black hole, we consider three different $f(R)$ model. For the charged static black hole, we work in two ensembles. In one ensemble, the charge $q$ is kept fixed, while in the other, its conjugate potential, $\phi$ is kept fixed. The rotating charged black hole is analyzed in four ensembles: fixed $(q, J)$, fixed $(\phi, J)$, fixed $(q,\Omega)$, and fixed $(\phi,\Omega)$. In this paper, we address the issues related to the dependence of thermodynamic topology on the $f(R)$ model and the choice of ensemble. In particular, the choice of ensemble is often found to be a determining factor in the nature of thermodynamic properties and phase transitions of black holes \cite{ec1,ec2,ec3,ec4,ec5,ec6}. For all these black holes we study their thermodynamic topologies by computing the topological charge in different ensembles with different values of thermodynamic parameters.\\
	
	This paper is organized into the following sections: in Section \textbf{II}, we have studied the thermodynamic topology of static black holes in $f(R)$ gravity where we have considered three $f(R)$ models in the subsections  Model I, Model II and Model III respectively. In Section \textbf{III}, subsection \textbf{I}, we have analyzed the thermodynamic topology of charged static black hole in fixed charge ensemble followed by extending the same in fixed $\phi$ ensemble in subsection \textbf{II}. In Section \textbf{IV}, we examine the rotating charged black hole in fixed $(q,J)$(Subsection \textbf{I}), fixed $(\phi,J)$(Subsection \textbf{II}), fixed $(q,\Omega)$(Subsection \textbf{III}), and fixed $(\phi,\Omega)$(subsection \textbf{IV}). Finally, the conclusions are presented in Section \textbf{V}.
	\section{Static Black hole in $f(R)$ gravity}
	We consider a static black hole solution in $f(R)$ gravity which appears as a solution to the following action \cite{fr1}
	$$ S=\frac{1}{2k} \int d^4x \sqrt{-g} f(R) +S_{m}$$
	Varying the action with respect to the metric gives :
	\begin{equation}
		F(R) R_{\mu \nu} - \frac{1}{2} f(R) g_{\mu \nu}-\left(\nabla _\mu \nabla_\nu -g_{\mu \nu}\Box\right) F(R)=kT_{\mu \nu}
	\end{equation}
	for vacuum space
	\begin{equation}
		F(R) R_{\alpha \beta} - \frac{1}{2} f(R) g_{\alpha \beta}-\left(\nabla _\alpha \nabla_\beta -g_{\alpha \beta}\Box\right) F(R)=0
		\label{contracted}
	\end{equation}
	where $F(R)=\frac{df(R)}{dR}$ and $\Box=\nabla _\alpha \nabla^\alpha.$\\	
	The generic form of the metric for spherically symmetric space-time is :
	\begin{equation}
		ds^2=-N(r)dt^2+M(r) dr^2+r^2(d\theta^2+sin^2\theta d\phi^2)
		\label{le1}
	\end{equation}
	where $M(r)=\frac{1}{N(r)}$.
	\section*{Model I  }
	The first model considered in this work is given by \cite{f8} :
	\begin{equation}
		f(R)=  \frac{\alpha  R^{2 m}-\beta  R^m}{\gamma  R^m+1}
	\end{equation}
	To find the static black hole solution using this model we will be using a technique showcased in \cite{m0,m1}.\\
	Contracting the equation \ref{contracted}, we obtain :
	\begin{equation}
		F(R) R - 2 f(R)+3 \Box F(R)=0
		\label{mocontracted}
	\end{equation}
	Differentiating the above equation we get the consistency relation as :
	\begin{equation}
		R F'-R'F+3(\Box F)'=0
		\label{condition}
	\end{equation}
	Using equation \ref{mocontracted}, modified Einstein's field equation becomes :
	\begin{equation}
		F R_{\alpha \beta}-\nabla_\alpha \nabla_\beta F =\frac{1}{4} g_{\alpha \beta} (F R-\Box F)
		\label{modifiedeinstein}
	\end{equation}
	So, any solution of equation \ref{modifiedeinstein} and \ref{contracted} must satisfy the relation \ref{condition}. Equation \ref{modifiedeinstein} can be seen as a set of differential equations for $F(r)$, $M(r)$, and $N(r)$ since the metric only depends on $r$. As only diagonal elements are non-zero for the metric, we get four equations.\\ Considering $$I_\alpha=\frac{F R_\alpha-\nabla_\alpha \nabla_\alpha F}{g_{\alpha \alpha}}$$ we get two equation as follows :
	\begin{equation}
		I_{[t]}-I_{[r]}= 2F \frac{Y'}{Y} + r F' \frac{Y'}{Y}-2 r F''=0
		\label{e1}
	\end{equation}
	where $Y=M(r)N(r)$
	\begin{equation}
		I_{[t]}-I_{[\theta]}= N'' + \left(\frac{F'}{F}-\frac{Y'}{2 Y}\right) N'-\frac{2}{r}\left(\frac{F'}{F}-\frac{Y'}{2 Y}\right) N -\frac{2}{r^2} N+\frac{2}{r^2} Y=0
		\label{e2}
	\end{equation}
	For the model we have considered here, we will have to take solution with a constant curvature. Hence taking $F'=0,F''=0$, equation \ref{e1} and \ref{e2} takes the form :
	\begin{equation}
		\begin{cases}
			M N'+N M'=0 \\
			1-N+\frac{r}{2}\left(\frac{N'}{N}+\frac{M'}{M}\right)\left(\frac{r}{2} \frac{M'}{M}-1\right)-\frac{r^3}{2}\frac{M''}{M}=0
		\end{cases}
		\label{diff}
	\end{equation}
	solving equation \ref{diff} we obtain
	\begin{equation}
		\begin{cases}
			N(r)=c_0+\frac{c_1}{r}+c_2 r^2\\
			M(r)=\frac{c_0}{N(r)}
		\end{cases}
		\label{solution}
	\end{equation}
	
	The Schwarzschild-de Sitter-spacetime(SdS), which is the Schwarzschild solution in the presence of a cosmological constant has the form $$ds^2=-B(r) dt^2+\frac{1}{B(r)}dr^2+r^2(d\theta^2+sin^2\theta d\phi^2)$$ Here
	\begin{equation}
		B(r)=1-\frac{2M}{r}-\frac{\Lambda r^2}{3}
		\label{schw}
	\end{equation} 
	The relation between constant scalar curvature $R$ and cosmological constant in this case is given by  
	\begin{equation}
		R=-4\Lambda
	\end{equation}
	comparing equation \ref{solution} with equation \ref{schw}, we get 
	$$c_0=1,c_1=-2M,c_2=-\frac{\Lambda}{3}=\frac{R}{12}$$
	constant curvature $R=R_0$ can be obtain from equation \ref{mocontracted} as : 
	$$R_0= \frac{2 f(R_0)}{ F(R_0)}$$
	For the model we have considered in this section 
	$$R_0^2=144 c_2^2=4^{-1/m} \left(-\frac{\left(2 \beta  \gamma \pm 2 \sqrt{(\alpha +\beta  \gamma ) \left(\beta  \gamma +\alpha  (m-1)^2\right)}\right)+2 \alpha  (m-1)}{\alpha  \gamma  (m-2)}\right)^{2/m}$$
	and $N(r)$ of the solution \ref{le1} will take the form :
	\begin{equation}
		N(r)=1-\frac{2M}{r}+\kappa r^2
		\label{nr}
	\end{equation}
	where $$\kappa=\frac{1}{12} \sqrt{4^{-1/m} \left(-\frac{\left(2 \beta  \gamma \pm 2 \sqrt{(\alpha +\beta  \gamma ) \left(\beta  \gamma +\alpha  (m-1)^2\right)}\right)+2 \alpha  (m-1)}{\alpha  \gamma  (m-2)}\right)^{2/m}}$$
	Solving the equation $N(r_+)=0$ ($r_+$ is the event horizon radius), we calculate the mass as :
	\begin{equation}
		M=\frac{1}{2} r_+ \left(1-\kappa  r_+^2\right)
		\label{m1}
	\end{equation}
	The temperature is calculated as :
	$$T=\frac{N'(r_+)}{4 \pi}=\frac{1-3 \kappa  r_+^2}{4 \pi  r_+}$$
	and the entropy is given by :
	\begin{equation}
		S=\int \frac{dM}{T}=\pi r_+^2
		\label{s1}
	\end{equation}
	Using equation \ref{m1} and equation \ref{s1}, the free energy $\mathcal{F}=M-S/\tau$  is found to be:
	\begin{equation}
		\mathcal{F}=\frac{r_+ \left(-\kappa  r_+^2 \tau -2 \pi  r_++\tau \right)}{2 \tau }
	\end{equation}
	
	\noindent  The components of the vector $\phi$ are found to be 
	\begin{equation}
		\phi^{r}=-\frac{1}{2} 3 \kappa  r_+^2-\frac{2 \pi  r_+}{\tau }+\frac{1}{2}
	\end{equation}
	\begin{equation}
		\phi^\Theta=-\cot\Theta ~\csc\Theta 
	\end{equation}
	\noindent The unit vectors $\left(n^1, n^2\right)$ are computed using the following prescription : 
	
	$$n^1=\frac{\phi^r}{\sqrt{(\phi^r)^2+(\phi^\Theta)^2}}  \hspace{0.5cm}\text{and} \hspace{0.5cm} n^2=\frac{\phi^\theta}{\sqrt{(\phi^r)^2+(\phi^\Theta)^2}}$$

\begin{figure}[h]
		\centering
		\begin{subfigure}{0.32\textwidth}
			\includegraphics[width=\linewidth]{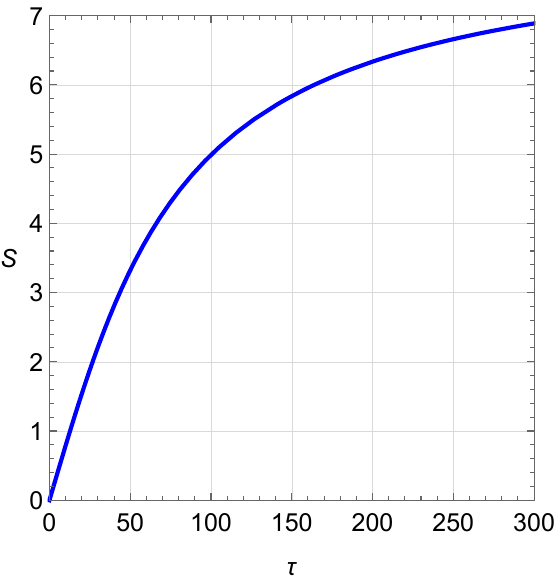}
			\caption{}
			\label{c1a}
		\end{subfigure}
		\begin{subfigure}{0.32\textwidth}
			\includegraphics[height=5cm,width=6cm]{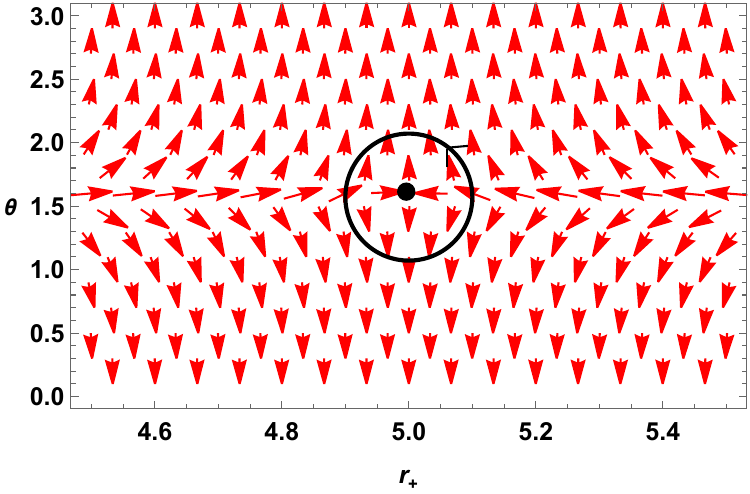}
			\caption{}
			\label{c1b}
		\end{subfigure}
		\begin{subfigure}{0.32\textwidth}
			\includegraphics[width=\linewidth]{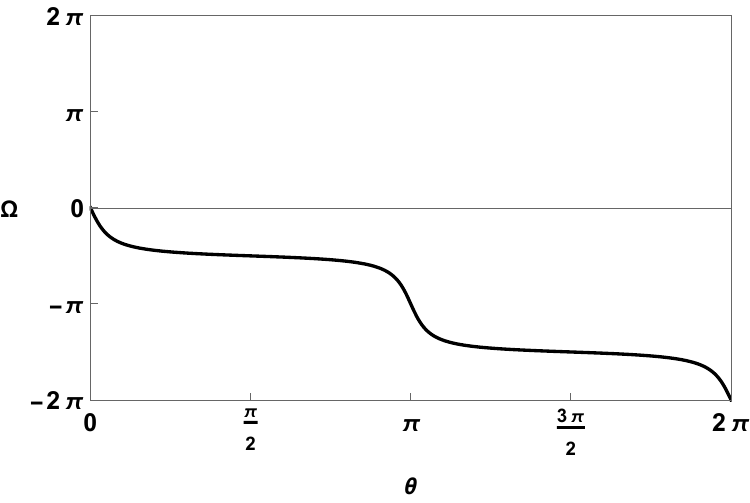}
			\caption{}
			\label{c1c}
		\end{subfigure}
		
		\caption{Plots for static black hole considering model I, where $\kappa=0.005$. Figure $\left(a\right)$ shows $\tau$ vs $r_+$ plot. Figure $\left(b\right)$ is the plot of vector field $n$ on a portion of $r_+-\theta$ plane for $\tau=100$. The zero point is located at $r_+=4.9879$ and $\left(c\right)$ shows the computation of the winding number for the contour around the zero point,  $r_+=4.9879$.}
		\label{c1}
	\end{figure}

	\noindent The expression for $\tau$ corresponding to zero points is obtained by setting $\phi^r=0$.
	\begin{equation}
		\tau=\frac{4 \pi  r_+}{1-3 \kappa  r_+^2}
	\end{equation}
	We plot $\tau$ vs $r_+$ plot for $\kappa=0.005$ in figure.\ref {c1a}, where we observe one single black hole branch. In figure \ref{c1b}, vector plot is shown for $\phi^r$ and $\phi^\theta$ component taking $\tau=100$, where we observe the zero point of the vector field at $r_+=4.9879$. From figure.\ref {c1c} it is observed, that the winding number or the topological charge corresponding to $r_+=4.9879$ is found to be $-1$ which is represented by the black-colored solid line. Further analysis shows, that for all values of $\kappa$, the topological charge is always $-1$. \\
	\section*{Model II }
	The second model we have considered is \cite{f7} :
	\begin{equation}
		f(R)=-\frac{\left(R-R_0\right){}^{2 n+1}+R_0^{2 n+1}}{f_1 \left\{\left(R-R_0\right){}^{2 n+1}+R_0^{2 n+1}\right\}+f_0}
	\end{equation}
	or
	\begin{equation}
		f(R)=-\frac{1}{f_1}+\frac{f_0}{f_1 \left(f_1 \left\{\left(R-R_0\right){}^{2 n+1}+R_0^{2 n+1}\right\}+f_0\right)}
	\end{equation}
	where $$\frac{1}{f_1}=\Lambda_i \hspace{0.5cm}\text{and}\hspace{0.5cm} f_0=\frac{R_0^{2n}}{2}$$
	Here, $n=1,2,3....$ is a positive integer. $R_0$ is the current curvature ($R_0\sim (10^{-33}eV)^2$) and $\Lambda_i$ is the effective cosmological constant $(\Lambda_i\sim 10^{20\sim38})$. The details about this model are available in \cite{f7}.
	For simplicity, we take $n=1$ and following the above-mentioned procedure, we find out the second order approximated metric solution of equation \ref{le1}  :
	$$ds^2=-N(r)dt^2+M(r) dr^2+r^2(d\theta^2+sin^2\theta d\phi^2)$$
	here, 
	\begin{equation}
		N(r)=1-\frac{2M}{r}+\frac{1}{12} \left(-\sqrt{3} R_0+\frac{6 \sqrt{3} \left(7 \sqrt{3}+12\right) R_0^2}{\Lambda _i}\right) r^2
	\end{equation}
	The mass is calculated as:
	\begin{equation}
		M=\frac{r_+ \left(12 \Lambda _i-\sqrt{3} r_+^2 R_0 \Lambda _i+72 \sqrt{3} r_+^2 R_0^2+126 r_+^2 R_0^2\right)}{24 \Lambda _i}
		\label{massstatic2}
	\end{equation}
	The temperature is given by:
	\begin{equation}
		T=\frac{4 \Lambda _i+r_+^2 R_0 \left(18 \left(4 \sqrt{3}+7\right) R_0-\sqrt{3} \Lambda _i\right)}{16 \pi  r_+ \Lambda _i}
	\end{equation}
	And the entropy is calculated as :
	\begin{equation}
		S=\pi r_{+}^2
		\label{entropystatic2}
	\end{equation} 
	from equation \ref{massstatic2} and \ref{entropystatic2}, free energy is calculated as :  
	\begin{equation}
		\mathcal{F}=\mathcal{F}=M-S/\tau=\frac{1}{24} \left(\frac{r_+^3 R_0 \left(18 \left(4 \sqrt{3}+7\right) R_0-\sqrt{3} \Lambda _i\right)}{\Lambda _i}-\frac{24 \pi  r_+^2}{\tau }+12 r_+\right)
	\end{equation}
	
\begin{figure}[h]
		\centering
		\begin{subfigure}{0.32\textwidth}
			\includegraphics[width=\linewidth]{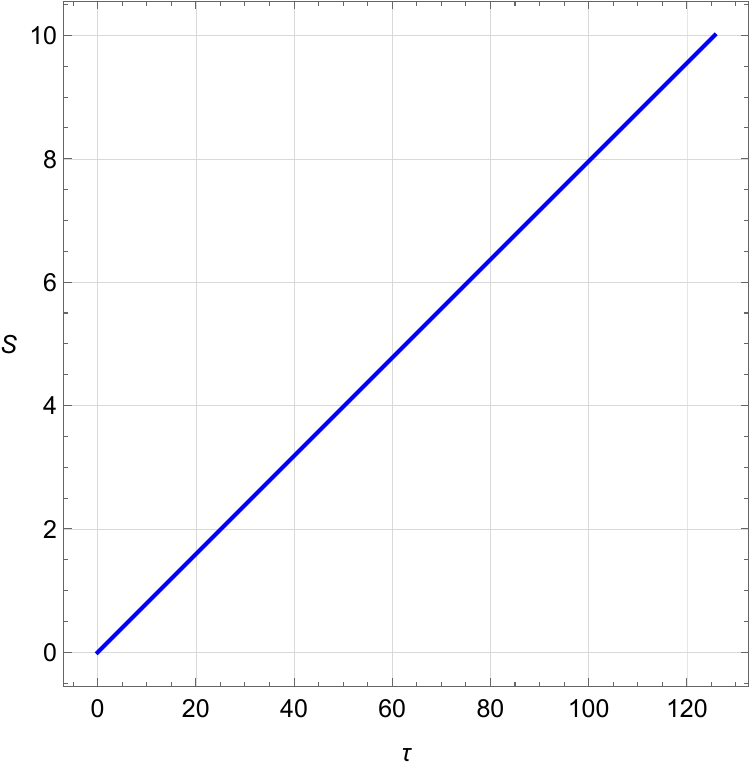}
			\caption{}
			\label{c2a}
		\end{subfigure}
		\begin{subfigure}{0.32\textwidth}
			\includegraphics[height=5cm,width=6cm]{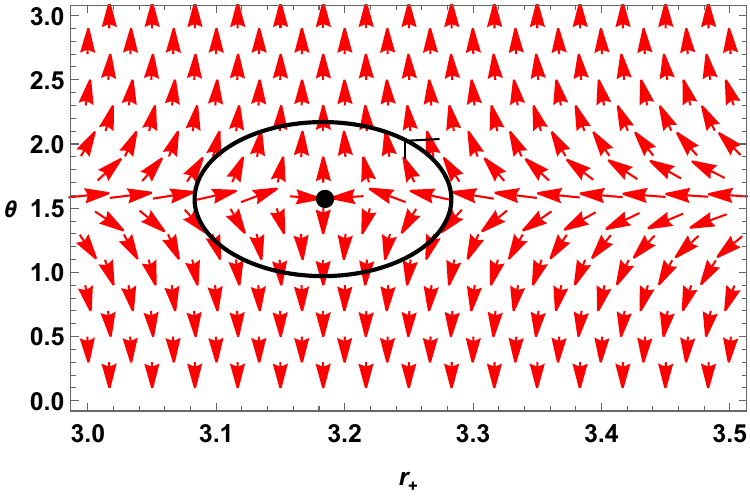}
			\caption{}
			\label{c2b}
		\end{subfigure}
		\begin{subfigure}{0.32\textwidth}
			\includegraphics[width=\linewidth]{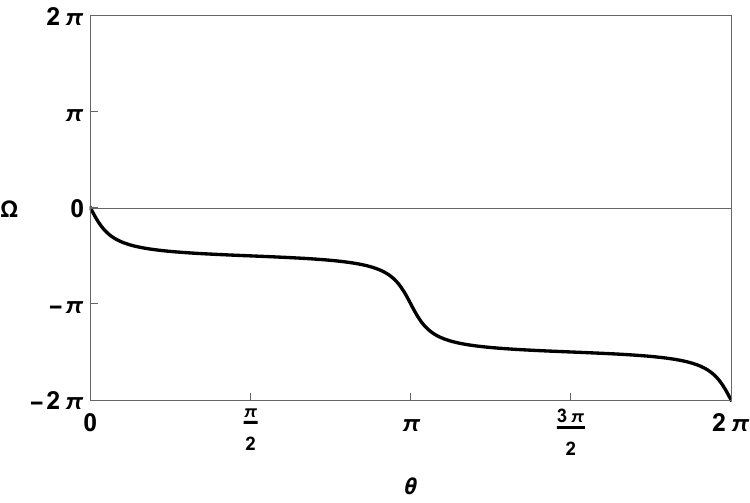}
			\caption{}
			\label{c2c}
		\end{subfigure}
		
		\caption{Plots for static black hole considering model II, where $R_0=(10^{-33})^2 eV$ and $\Lambda_i=10^{26}$. Figure $\left(a\right)$ shows $\tau$ vs $r_+$ plot. Figure $\left(b\right)$ is the plot of vector field $n$ on a portion of $r_+-\theta$ plane for $\tau=40$. The zero point is located at $r_+=3.18309$ and $\left(c\right)$ shows the computation of the winding number for the contour around the zero point,  $r_+=3.18309$.}
		\label{c1}
	\end{figure}

	\noindent  The components of the vector $\phi$ are found to be 
	\begin{equation}
		\phi^{r}=\frac{r_+^2 R_0 \left(18 \left(4 \sqrt{3}+7\right) R_0-\sqrt{3} \Lambda _i\right)}{8 \Lambda _i}-\frac{2 \pi  r_+}{\tau }+\frac{1}{2}
	\end{equation}
	\begin{equation}
		\phi^\Theta=-\cot\Theta ~\csc\Theta 
	\end{equation}
	\noindent The expression for $\tau$ corresponding to zero points is obtained by setting $\phi^r=0$.
	\begin{equation}
		\tau=-\frac{16 \pi  r_+ \Lambda _i}{r_+^2 R_0 \left(\sqrt{3} \Lambda _i-18 \left(4 \sqrt{3}+7\right) R_0\right)-4 \Lambda _i}
	\end{equation}
	Next, we plot $\tau$ vs $r_+$ plot taking $R_0=(10^{-33})^2 eV$ and $\Lambda_i=10^{26}$ in figure.\ref {c2a}, where we again observe one single black hole branch whose topological charge is $-1.$  The Vector plot in Figure \ref{c2b} and contourplot \ref{2c} shows the same. In fact the topological charge is always $-1$ for all values of $R_0$ and $\Lambda_i$. \\
	\section*{Model III}
	The next model that will be used in this work is \cite{m1,m2}:
	\begin{equation}
		f(R)=R+\Lambda+\frac{R+\Lambda}{R/R_0+2/\alpha}ln\left(\frac{R+\Lambda}{R_c}\right)
	\end{equation}
	
	where $R_c$ is the integration constant and $R_0=6\alpha^2/d^2$;$d$ and $\alpha$ are the free parameters of the action.$\Lambda$ is the cosmological constant.\\
	
	The metric solution of equation \ref{le1} is obtained as \cite{m1,m2}:
	\begin{equation}
		N(r)=1-\frac{2M}{r}+\beta r -\frac{\Lambda r^2}{3}
		\label{staticsolution}
	\end{equation}
	where $\beta=\alpha/d \geq 0$ is a real constant.\\
	From equation \ref{staticsolution} ,mass $M$ can be obtained by setting $N(r=r_+)=0$,which gives:
	\begin{equation}
		M=\frac{r_+ \left(\beta  l^2 r_++l^2-r_+^2\right)}{2 l^2}
		\label{massstatic}
	\end{equation}
	We have used $\Lambda=\frac{3}{l^2}$ \cite{m4} in which $l$ is the radius of curvature of the de sitter space.\\
	The entropy can be obtained by using the radius of the event horizon as:
	\begin{equation}
		S=\pi r_+^2
		\label{entropystatic}
	\end{equation}
	Using equation \ref{massstatic} and equation \ref{entropystatic}, the free energy $\mathcal{F}=M-S/\tau$ ,for static black hole in this $f(R)$ model is found to be:
	\begin{equation}
		\mathcal{F}=\frac{1}{2} r_+ \left(-\frac{r_+^2}{l^2}+r_+ \left(\beta -\frac{2 \pi }{\tau }\right)+1\right)
	\end{equation}
	
	\noindent  The components of the vector $\phi$ are found to be 
	\begin{equation}
		\phi^{r}=\frac{\partial\mathcal{F}}{\partial r_{+}}=-\frac{3 r_+^2}{2 l^2}+r_+ \left(\beta -\frac{2 \pi }{\tau }\right)+\frac{1}{2}
	\end{equation}
	\begin{equation}
		\phi^\Theta=-\cot\Theta ~\csc\Theta 
	\end{equation}
	
	\noindent The expression for $\tau$ corresponding to zero points is obtained by setting $\phi^r=0$.
	\begin{equation}
		\tau=\frac{4 \pi  l^2 r_+}{2 \beta  l^2 r_++l^2-3 r_+^2}
	\end{equation}
	
\begin{figure}[h]
		\centering
		\includegraphics[width=7cm,height=5cm]{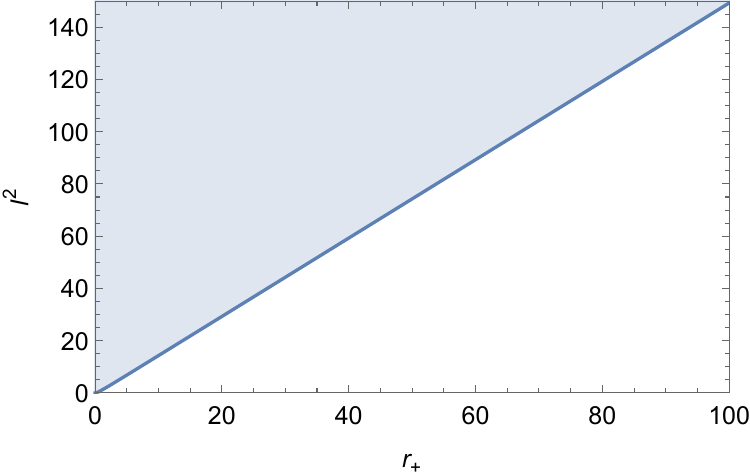}
		
		\caption{The relation between $l^2$
			and $r$ for the positive temperature. Temperature is positive only on the shaded portions. We have taken $\beta=1$.}
		\label{1}
	\end{figure}

\begin{figure}[h]
		\centering
		\begin{subfigure}{0.32\textwidth}
			\includegraphics[width=\linewidth]{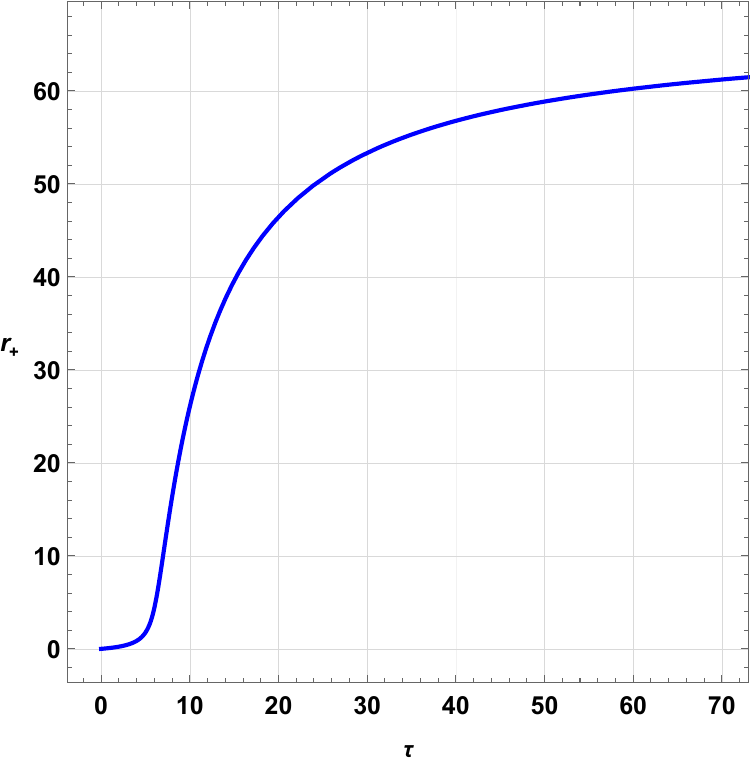}
			\caption{}
			\label{2a}
		\end{subfigure}
		\begin{subfigure}{0.32\textwidth}
			\includegraphics[height=5cm,width=6cm]{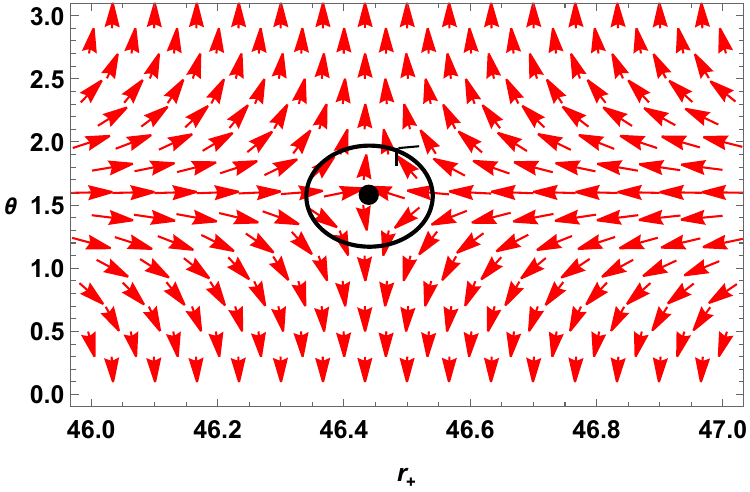}
			\caption{}
			\label{2b}
		\end{subfigure}
		\begin{subfigure}{0.32\textwidth}
			\includegraphics[width=\linewidth]{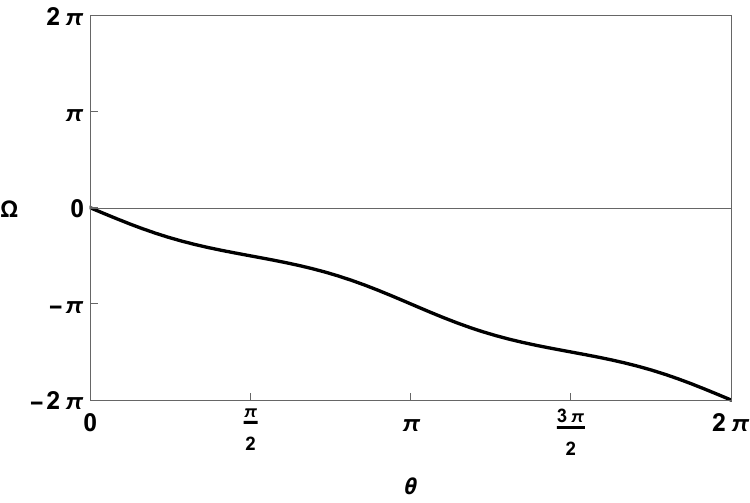}
			\caption{}
			\label{2c}
		\end{subfigure}
		
		\caption{Plots for static black hole in model III, at $\l^2=100 $ with $ \beta=1$. Figure $\left(a\right)$ shows $\tau$ vs $r_+$ plot. Figure $\left(b\right)$ is the plot of vector field $n$ on a portion of $r_+-\theta$ plane for $\tau=20$. The zero point is located at $r_+=46.44048$ and $\left(c\right)$ shows the computation of the winding number for the contour around the zero point,  $r_+=46.44048$.}
		\label{2}
	\end{figure}

	It is to be noted that, not all values of $r_+$ and $l$ are allowed for specific values of $\beta$. The allowed combination of values of  $r_+$ and $l^2$ for $\beta=1$ are shown in Figure \ref{1} as shaded portions with semi-positive temperature.
	From the $\tau$ vs $r_+$ plot in Figure.\ref {2a}, we observe one single black hole branch. The winding number is calculated by keeping $\beta=1$, $l^2=100$ and $\tau=20$. For this combination of values, we observe the zero point of the vector field $n$ at $r_+=46.44048$ (Figure \ref{2b}). From figure.\ref {2c} it is observed, that the winding number or the topological charge corresponding to $r_+=46.44048$ (represented by the black colored solid line) is found to be $-1$. Further analysis shows, that for all sets of values of $\beta$, $l^2$, the topological charge remains constant, which is $-1$. \\
	
	\section{Charged Static Black hole in $f(R)$ gravity}
	The second black hole solution in $f(R)$ gravity that we have chosen to study is a charged static black hole solution originating from the following action \cite{fr1}
	$$S=\frac{1}{16 \pi G} \int d^4x \sqrt{-g} \left(R+f(R)-F_{\mu \nu} F^{\mu \nu}\right)$$
	Varying the action with respect to the metric gives :
	$$ R_{\mu \nu}(1+f'(R)) - \frac{1}{2} \left(R+f(R)\right) g_{\mu \nu}+\left(g_{\mu \nu} \nabla^2-\nabla _\mu \nabla_\nu \right) f'(R)=2T_{\mu \nu}$$
	where $f'(R)=\frac{df(R)}{dR}$ and $T_{\mu \nu}$ is the stress-energy tensor of the electromagnetic field.\\
	
	The trace of the above equation at $R=R_0$ results in :
	$$R_0(1+f'(R_0))-2((R_0+f(R_0))=0$$
	which eventually gives the constant curvature scalar as:
	$$R_0=\frac{2 f(R_0)}{f'(R_0)-1}$$
	Finally, the metric of the spherically symmetric spacetime is obtained as follows:
	\begin{equation}
		ds^2=P(r)dt^2-\frac{1}{P(r)} dr^2-r^2(d\theta^2+sin^2\theta d\phi^2)
		\label{le2}
	\end{equation}
	Where,
	\begin{equation}
		P(r)=1-\frac{2 G M}{r}+\frac{Q^2}{\left(1+f'(R_0)\right) r^2}-\frac{R_0 r^2}{12}
	\end{equation}
	for the details about the metric see \cite{fr2}.\\
	
	By putting $G=1$ and $q^2=\frac{Q^2}{\left(1+f'(R_0)\right)}$ in the above equation:
	\begin{equation}
		P(r)=1-\frac{2  M}{r}+\frac{q^2}{r^2}-\frac{R_0 r^2}{12}
		\label{staticchargedsolution}	
	\end{equation}
	where $R_0=4 \Lambda$=$\frac{12}{l^2}$ is the constant curvature
	\subsection{Fixed charge Ensemble}
	For charged static black hole, from the equation \ref{staticchargedsolution},mass $M$ in canonical ensemble is obtained as \cite{fr1} :
	\begin{equation}
		M=\frac{\pi ^2 l^2 q^2+\pi  l^2 S-S^2}{2 \pi ^{3/2} l^2 \sqrt{S}}
		\label{masscharged}
	\end{equation}
	
	In the context of the charged static black hole and rotating charged black hole, formulating the vector field component in terms of the horizon radius $r_+$ is difficult, particularly when working in different ensembles. Consequently, we have performed all calculations in terms of the entropy $S$. Using equation \ref{masscharged}, free energy is computed as :
	\begin{equation}
		\mathcal{F}=M-S/\tau=\frac{\pi ^2 l^2 q^2 \tau -2 \pi ^{3/2} l^2 S^{3/2}+\pi  l^2 S \tau -S^2 \tau }{2 \pi ^{3/2} l^2 \sqrt{S} \tau }
	\end{equation}
	The  components of the vector field $\phi$ is obtained as :
	$$\left(\phi^S,\phi^\Theta\right)=\left(\frac{\partial\mathcal{F}}{\partial S},-\cot\Theta ~\csc\Theta\right)$$
	Where,
	\begin{equation}
		\phi^S=\frac{\pi  l^2 \left(-\pi  q^2 \tau -4 \sqrt{\pi } S^{3/2}+S \tau \right)-3 S^2 \tau }{4 \pi ^{3/2} l^2 S^{3/2} \tau }
	\end{equation}
	and
	\begin{equation}
		\phi^\Theta=-\cot\Theta ~\csc\Theta 
	\end{equation}
	The zero points of $\phi^r$ is also obtained as:
	\begin{equation}
		\tau=\frac{4 \pi ^{3/2} l^2 S^{3/2}}{-\pi ^2 l^2 q^2+\pi  l^2 S-3 S^2}
	\end{equation} 
	
\begin{figure}[h]
		\centering
		\begin{subfigure}{0.32\textwidth}
			\includegraphics[width=\linewidth]{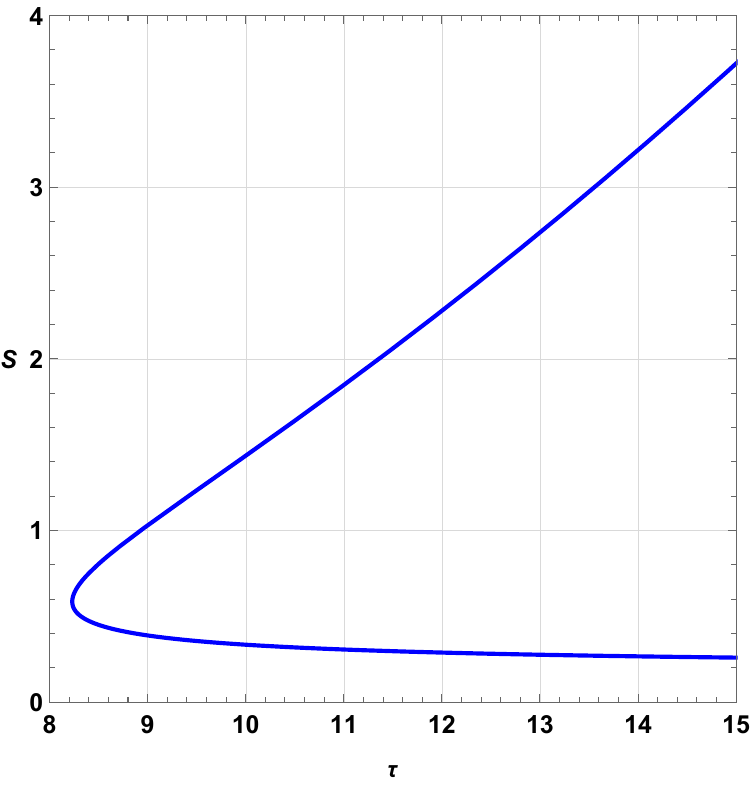}
			\caption{}
			\label{3a}
		\end{subfigure}
		\begin{subfigure}{0.32\textwidth}
			\includegraphics[height=5cm,width=6cm]{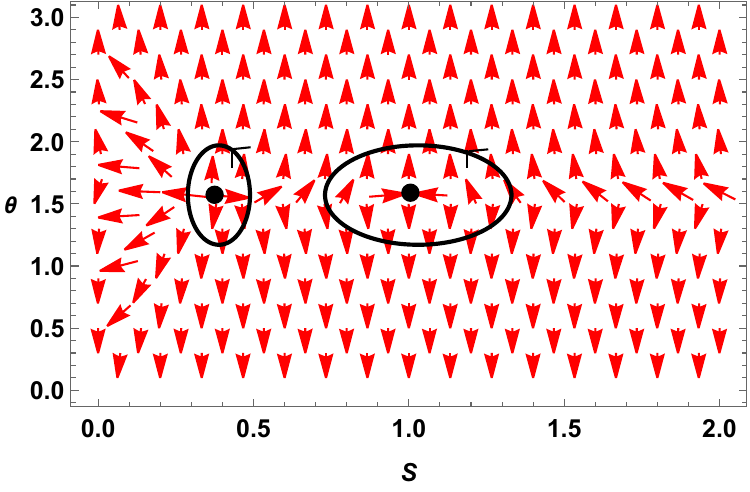}
			\caption{}
			\label{3b}
		\end{subfigure}
		\begin{subfigure}{0.32\textwidth}
			\includegraphics[width=\linewidth]{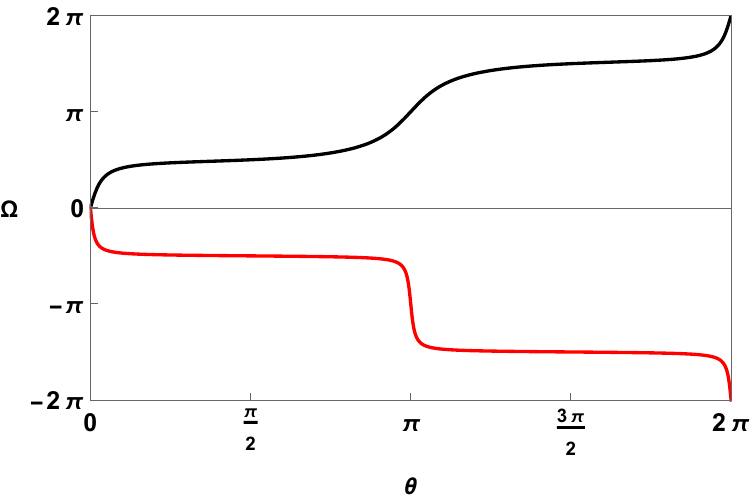}
			\caption{}
			\label{3c}
		\end{subfigure}
		
		\caption{Plots for static charged black hole in $f(R)$ gravity in fixed charge ensemble. Here $q=0.25 $ with $ l=10$. Figure $\left(a\right)$ shows $\tau$ vs $S$ plot. Figure $\left(b\right)$ is the plot of vector field $n$ on a portion of $S-\theta$ plane for $\tau=40$. The zero points are located at  $S=0.38868$ and $S=1.033022$. In figure $\left(c\right)$, the computation of the winding numbers for the contours around the zero points $S=0.38868$ and $S=1.033022$ are shown in black and red colored solid lines respectively.}
		\label{3}
	\end{figure}

	For all values of $q$ and $l$, the static charged black hole has topological charge equal to $0$. At $q=0.25$ and $l=10$, for example, we encounter two black hole branches with total topological charge equaling $1-1=0$ as shown in figure.\ref{3}. In figure.\ref{3a},  $\tau$ vs $S$ is plotted in the allowed range of $S$. For $\tau=9$, there are two zero points located at  $S=0.38868$ and $S=1.033022$ as shown in figure \ref{3b}. From figure.\ref {3c}, the winding numbers corresponding to   $S=0.38868$ and $S=1.033022$ (represented by the black and the red colored solid line respectively) are found to be $-1$, and $+1$ respectively. The total topological charge of the black hole, therefore, is $1-1=0$.\\
	
\begin{figure}[h]
		\centering
		\begin{subfigure}{0.32\textwidth}
			\includegraphics[width=\linewidth]{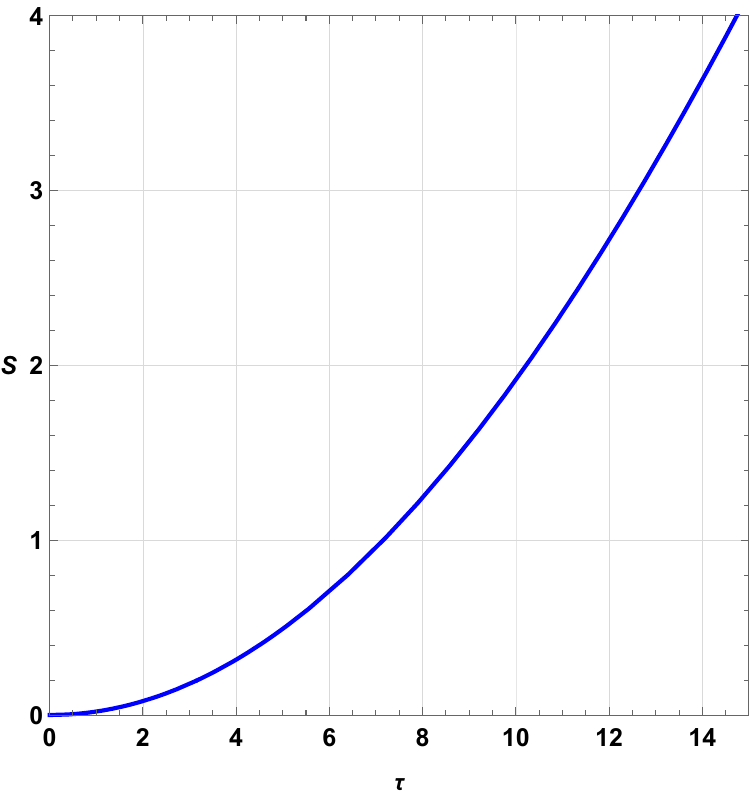}
			\caption{}
			\label{4a}
		\end{subfigure}
		\begin{subfigure}{0.32\textwidth}
			\includegraphics[height=5cm,width=6cm]{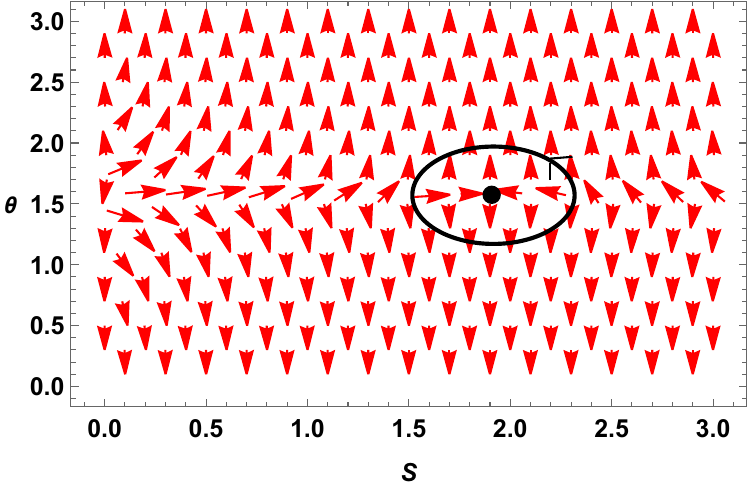}
			\caption{}
			\label{4b}
		\end{subfigure}
		\begin{subfigure}{0.32\textwidth}
			\includegraphics[width=\linewidth]{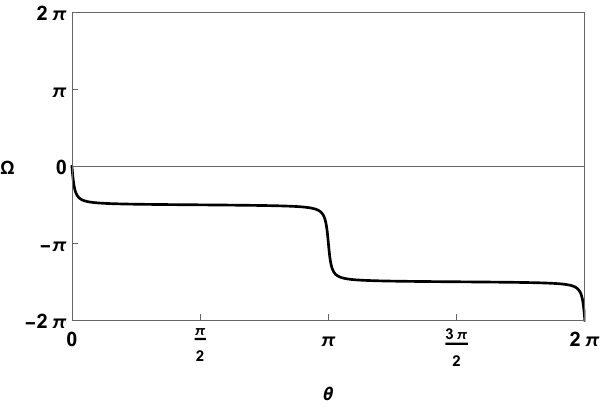}
			\caption{}
			\label{4c}
		\end{subfigure}
		
		\caption{Plots for static charged black hole in $f(R)$ gravity in fixed charge ensemble. Here, $q=0 $ with $ l=10$. Figure $\left(a\right)$ shows $\tau$ vs $S$ plot. Figure $\left(b\right)$ is the plot of vector field $n$ on a portion of $S-\theta$ plane for $\tau=10$. The zero point is located at $S=1.91726$. In figure $\left(c\right)$, the computation of the winding number for the contour around the zero point  $S=2.5581$ is shown in black-colored solid lines.}
		\label{4}
	\end{figure}

	If we set $q=0$, then we encounter one single black hole branch with topological charge $-1$ as shown in figure.\ref{4}. In Figure.\ref{4a} $\tau$ vs $S$ is plotted in the allowed range of $S$. For $\tau=10$, zero point is located at $S=1.91726$ as shown in Figure \ref{4b}. From figure.\ref {4c}, the winding number corresponding to  $S=1.91726$ (represented by the black-colored solid line) is found to be $-1$. So, when the charge $q$ equals zero, the thermodynamic topology of static charged black holes becomes equivalent to that of static black holes as expected.\\
	\subsection{Fixed potential($\phi$) Ensemble}
	In fixed $\phi$ ensemble, we define a potential  $\phi$ conjugate to $q$ and keep it fixed. These two parameters are related by $$\phi=\frac{\partial M}{\partial q}=\frac{\sqrt{\pi } q}{\sqrt{S}}$$.\\
	The mass in the fixed $\phi$ canonical ensemble is given by : 
	$$\Tilde{M}=M-q\phi$$
	We get,
	\begin{equation}
		\tilde{M}=\frac{\sqrt{S} \left(\pi  l^2 \left(1-\phi ^2\right)-S\right)}{2 \pi ^{3/2} l^2}
	\end{equation}
	Accordingly, the free energy is calculated using: 
	\begin{equation}
		\tilde{\mathcal{F}}=\tilde{M}-\frac{S}{\tau}
	\end{equation}
	or
	\begin{equation}
		\tilde{\mathcal{F}}=-\frac{\sqrt{S} \left(2 \pi ^{3/2} l^2 \sqrt{S}+\pi  l^2 \tau  \phi ^2-\pi  l^2 \tau +S \tau \right)}{2 \pi ^{3/2} l^2 \tau }
	\end{equation}
	The components of the vector field are obtained as :
	$$\left(\phi^S,\phi^\Theta\right)=\left(\frac{\partial\mathcal{F}}{\partial S},-\cot\Theta ~\csc\Theta\right)$$
	where
	\begin{equation}
		\phi^S=\frac{\pi  l^2 \left(-4 \sqrt{\pi } \sqrt{S}-\tau  \phi ^2+\tau \right)-3 S \tau }{4 \pi ^{3/2} l^2 \sqrt{S} \tau }
	\end{equation}
	and
	\begin{equation}
		\phi^\theta=-\cot\Theta\csc\Theta
	\end{equation}
	Finally, we obtain the expression for $\tau$ as :
	\begin{equation}
		\tau=\frac{4 \pi ^{3/2} l^2 \sqrt{S}}{-\pi  l^2 \phi ^2+\pi  l^2-3 S}
	\end{equation} 
	
	\begin{figure}[h]	
		\centering
		\begin{subfigure}{0.32\textwidth}
			\includegraphics[width=\linewidth]{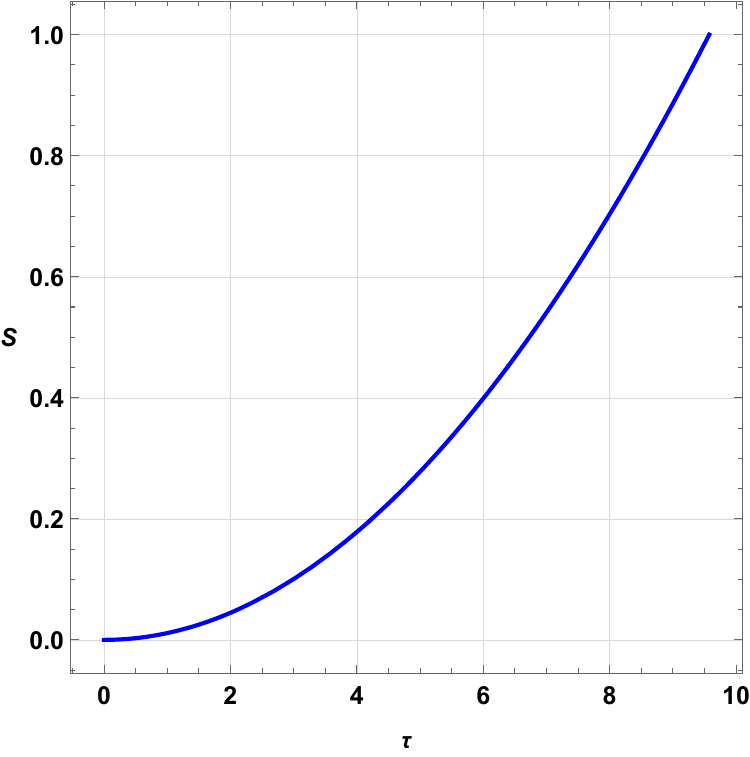}
			\caption{}
			\label{5a}
		\end{subfigure}
		\begin{subfigure}{0.32\textwidth}
			\includegraphics[height=5cm,width=6cm]{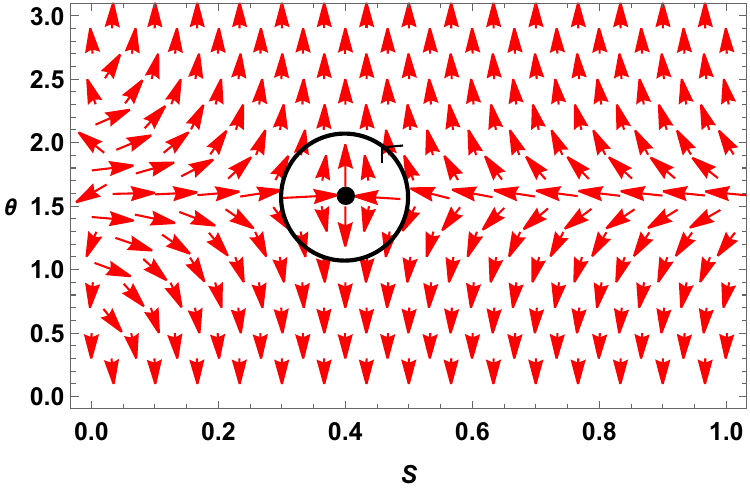}
			\caption{}
			\label{5b}
		\end{subfigure}
		\begin{subfigure}{0.32\textwidth}
			\includegraphics[width=\linewidth]{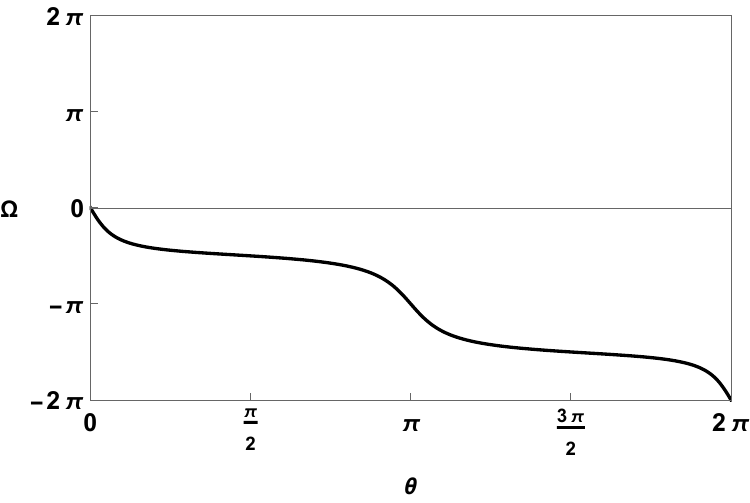}
			\caption{}
			\label{5c}
		\end{subfigure}
		\caption{ Plots for the static charged black hole in the fixed potential ensemble at $\phi=0.5$ with $l=10$. Figure $\left(a\right)$ shows $\tau$ vs $S$ plot, figure $\left(b\right)$  is the plot of vector field $n$ on a portion of $S-\theta$ plane for $\tau=6$. The red arrows represent the vector field $n$. The zero point is located at $S=0.39878.$ In figure $\left(c\right)$, computation of the contour around the zero point $\tau=100$ and  $S=0.39878.$ is  shown.}
		\label{5}
	\end{figure}
	
	It is seen that for all the values of $\phi$ and $l$, static charged black hole in the fixed potential ensemble has a topological charge equal to $-1$. In Figure.\ref {5} $\tau$ vs $S$ plot is shown where $\phi=0.5$ and $l=10$. Here, we observe a single black hole branch. For $\tau=6$, the zero point is located at  $S=0.39878.$. This is also confirmed from the vector plot of $n$ in the $S-\theta$ plane as shown in  Figure.\ref {5b}. To find out the winding number/topological number associated with this zero point, we perform a contour integration around  $S=0.39878.$ which is shown in Figure.\ref {5c}. The topological charge, in this case, is equal to $-1$. We have explicitly verified that the topological charge of any zero point on the black hole branch remains the same and is equal to $-1$. Strikingly, in the fixed potential ensemble, the topological charge is different from that in the fixed charge ensemble.
	\section{Rotating Charged Black hole}
	In this section, we study the thermodynamic topology of a rotating charged black hole solution in $f(R)$ gravity for four ensemble: fixed $(q, J)$, fixed $(\phi, J)$, fixed $(q,\Omega)$ and fixed $(\phi,\Omega)$ ensemble, where $q,\phi$,$J$ and $\Omega$ denotes the charge, potential, angular momentum and the angular frequency respectively.
	The black hole of our interest originates from the following action :
	\begin{equation}
		S=\frac{1}{16 \pi}\left(\int d^D x \sqrt{|g|} (R+f(R))-\int d^4 x  \sqrt{|g|} [F_{\mu \nu}F^{\mu \nu}]\right)
	\end{equation}
	where the first part represents the gravitational action and the second part represents the four dimensional Maxwell term. $R$ is the scalar curvature and $R+f(R)$ is a function of scalar curvature.The field equations in the metric formalism are \cite{rc1}
	\begin{equation}
		R_{\mu \nu}(1+f'(R))-\frac{1}{2}(R+f(R)) g_{\mu \nu} +(g_{\mu \nu} \nabla^2 -\nabla_\mu \nabla_\nu)f'(R)=2 T_{\mu \nu}
		\label{rceq1}
	\end{equation}
	where, $\nabla$ is the covariant derivative,$R_{\mu \nu}$ is the Ricci tensor, and $T_{\mu \nu}$  is the stress-energy tensor of the electromagnetic field given by:
	$$T_{\mu \nu}=F_{\mu \rho} F^\rho_\nu-\frac{g_{\mu \nu}}{4} F_{\rho \sigma}F^{\rho \sigma}$$
	The trace of equation \ref{rceq1} gives the expression for  $R=R_0$ as:
	\begin{equation}
		R_{0}=\frac{2 f(R_0)}{f'(R_0)-1}
	\end{equation}
	The axisymmetric ansatz, utilizing Boyer-Lindquist type coordinates $(t, r, \theta, \varphi)$, derived from the Kerr-Newman-Ads black hole solution, as shown in \cite{rc1} is:
	\begin{equation}
		ds^2=-\frac{\Delta_r}{\rho^2}\left [ dt-\frac{a sin^2\theta d\varphi}{\Xi} \right ]^2+\frac{\rho^2}{\Delta_r} dr^2 +\frac{\rho^2}{\Delta_\theta} d\theta^2+\frac{\Delta_\theta sin^2 \theta} {\rho^2}\left[a dt-\frac{r^2+a^2}{\Xi} d\varphi\right]^2
		\label{rcmetric}
	\end{equation}
	where 
	$$\Delta_{r}=(r^2+a^2)\left(1+\frac{R_0}{12} r^2 \right)-2mr+\frac{Q^2}{(1+f'(R_0))},$$
	$$\Xi=1-\frac{R_0}{12}a^2 \hspace{0.5cm},\hspace{0.5cm} \rho^2=r^2+a^2cos^2\theta,$$
	$$\Delta_\theta=1-\frac{R_0}{12}a^2 cos^2\theta$$
	in which $R_0=-4 \Lambda$, Q is the electric charge and $a$ is the angular momentum per mass of the black hole. By setting $dr=dt=0$ in equation \ref{rcmetric}, we can calculate the area of the two-dimensional horizon, which eventually gives the expression for area as\cite{fr1}:
	\begin{equation}
		S=\frac{\pi(r_+^2+a^2)}{1-\frac{R_0}{12}a^2}
	\end{equation}
	where $r_+$ is the radius of the horizon.\\
	The expressions for total mass and the angular momentum are \cite{rc1} :
	\begin{equation}
		M=\frac{m}{\Xi^2}
	\end{equation}
	and
	\begin{equation}
		J=\frac{a m}{\Xi^2}
	\end{equation}
	From which the generalized Smarr formula of the rotating charged black hole is obtained as\cite{rc1}:
	\begin{equation}
		M^2=\frac{S}{4 \pi }-\frac{J^2 R_0}{12}+\frac{\pi  \left(4 J^2+q^4\right)}{4 S}-\frac{R_0 S \left(q^2-\frac{R_0 S^2}{24 \pi ^2}+\frac{S}{\pi }\right)}{24 \pi }+\frac{q^2}{2}
		\label{smr}
	\end{equation}
	\subsection{Fixed (q, J) ensemble}
	In the fixed $(q, J)$ ensemble, we keep $q$ and $J$ as fixed parameters.The mass expression, given by equation \ref{smr}, remains unchanged within this ensemble,which is :
	\begin{equation}
		M=\sqrt{\frac{S}{4 \pi }-\frac{J^2 R_0}{12}+\frac{\pi  \left(4 J^2+q^4\right)}{4 S}-\frac{R_0 S \left(q^2-\frac{R_0 S^2}{24 \pi ^2}+\frac{S}{\pi }\right)}{24 \pi }+\frac{q^2}{2}}
	\end{equation}
	The off-shell free energy is calculated to be:
	$$\mathcal{F}=-\frac{24 \pi ^{3/2} S-\tau  \sqrt{\frac{-48 \pi ^3 J^2 R S+576 \pi ^4 J^2-24 \pi ^2 q^2 R S^2+288 \pi ^3 q^2 S+144 \pi ^4 q^4+R^2 S^4-24 \pi  R S^3+144 \pi ^2 S^2}{S}}}{24 \pi ^{3/2} \tau }$$
	The  components of the vector field $\phi$ are obtained as :
	$$\left(\phi^S,\phi^\Theta\right)=\left(\frac{\partial\mathcal{F}}{\partial S},-\cot\Theta ~\csc\Theta\right)$$
	or
	$$\phi^S=\frac{\partial\mathcal{F}}{\partial S}=\frac{-16 \pi ^{3/2} S^2 \sqrt{\frac{48 \pi ^3 J^2 (12 \pi -R S)+\left(12 \pi ^2 q^2-R S^2+12 \pi  S\right)^2}{S}}-48 \pi ^4 \tau  \left(4 J^2+q^4\right)-8 \pi ^2 S^2 \tau  \left(q^2 R-6\right)+R^2 S^4 \tau -16 \pi  R S^3 \tau }{16 \pi ^{3/2} S^2 \tau  \sqrt{\frac{48 \pi ^3 J^2 (12 \pi -R S)+\left(12 \pi ^2 q^2-R S^2+12 \pi  S\right)^2}{S}}}$$
	and
	$$\phi^\Theta=-\cot\Theta ~\csc\Theta $$
	The expression for $\tau$ corresponds to the zero point of $\phi^S$ can be obtained by setting $\phi^S=0$.
	\begin{equation}
		\tau=\frac{4 \pi ^{3/2} l^4 S^2 \sqrt{\frac{4 \pi ^3 J^2 l^2 \left(\pi  l^2+S\right)+\left(\pi  l^2 \left(\pi  q^2+S\right)+S^2\right)^2}{l^4 S}}}{-4 \pi ^4 J^2 l^4+l^4 \left(\pi ^2 S^2-\pi ^4 q^4\right)+2 \pi  l^2 S^2 \left(\pi  q^2+2 S\right)+3 S^4}
	\end{equation}
	where we have substituted $R_0=-\frac{12}{l^2}.$
	\begin{figure}[h]	
		\centering
		\includegraphics[width=6cm,height=6cm]{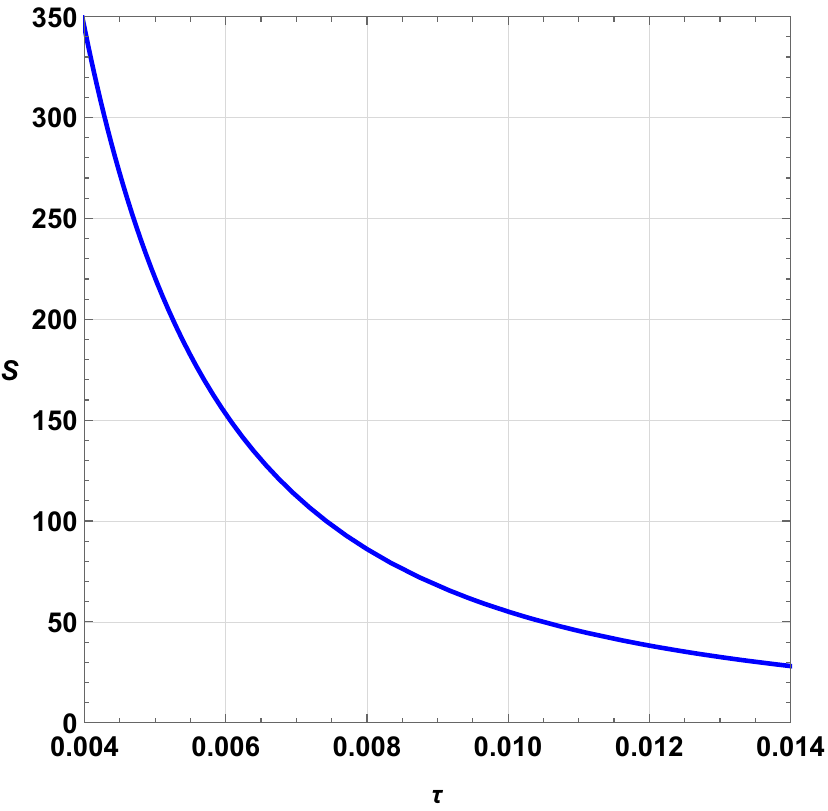}
		\caption{$\tau$ vs $S$ plots for a rotating charged black hole in the fixed $(q,J)$ ensemble at $q=0.05$, $J=1.5$, $l=0.1$.}
		\label{6}
	\end{figure}
	We plot the entropy $S$ against $\tau$ for a fixed length scale $l=0.1$ while keeping $J$ and $q$ fixed at $J=1.5$ and $q=0.05$ which is shown in figure \ref{6}. Here we observed a single black hole branch.\\
	In this section, we have adopted another method to calculate the topological charge \cite{64}.The complex function $\mathcal{R}_{rc}(z)$ defined in section \textbf{I} is given by : 	
	\begin{equation}
		\mathcal{R}_{rc}(z)=-\frac{-4 \pi ^4 J^2 l^4+2 \pi ^2 l^2 q^2 z^2-\pi ^4 l^4 q^4+\pi ^2 l^4 z^2+4 \pi  l^2 z^3+3 z^4}{4 \pi ^{3/2} l^4 z^2 \sqrt{\frac{4 \pi ^3 J^2 l^2 \left(\pi  l^2+z\right)+\left(\pi  l^2 \left(\pi  q^2+z\right)+z^2\right)^2}{l^4 z}}+4 \pi ^4 J^2 l^4 \tau +\pi ^4 l^4 q^4 \tau -2 \pi ^2 l^2 q^2 \tau  z^2-\pi ^2 l^4 \tau  z^2-4 \pi  l^2 \tau  z^3-3 \tau  z^4}
		\label{res}
	\end{equation}
	Considering the denominator of the equation as a polynomial function $\mathcal{A}(z)$ we can calculate the poles of $\mathcal{R}_{rc}(z)$.
	\begin{equation}
		\mathcal{A}(z)=l^4 \left(4 \pi ^{3/2} z^2 \sqrt{\frac{4 \pi ^3 J^2 l^2 \left(\pi  l^2+z\right)+\left(\pi  l^2 \left(\pi  q^2+z\right)+z^2\right)^2}{l^4 z}}+\pi ^4 \tau  \left(4 J^2+q^4\right)-\pi ^2 \tau  z^2\right)-2 \pi  l^2 \tau  z^2 \left(\pi  q^2+2 z\right)-3 \tau  z^4
		\label{pole}
	\end{equation}
	To get the winding number, we put $l=0.1, J=1.5$ and $q=0.05$ in equation \ref{pole}.
	\begin{figure}[h]
		\centering
		\includegraphics[height=6cm,width=10cm]{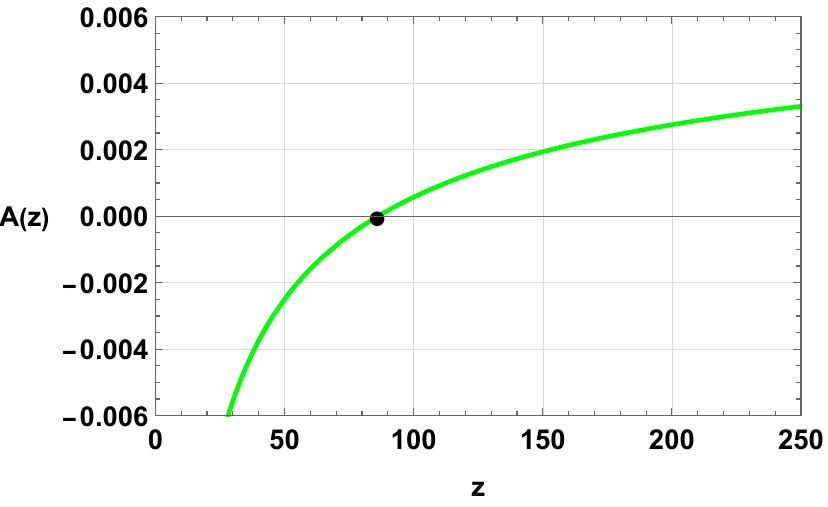}
		\caption{Plot for polynomial function $\mathcal{A}(z)$ for $\tau=0.008$}
		\label{7}
	\end{figure}	
	It is seen from figure \ref{7}, for $\tau=0.008$, the pole is at $z=86.108$. The winding number can be calculated by finding the sign of the residue of the equation \ref{res} around the pole $z=86.108$. We find a positive valued residue in this case. Hence, the winding number or the total topological charge is $1.$ It is found that if the value of $l$ is decreased below $l=0.1$, no of branch and topological charge remains the same.\\
	
	\begin{figure}[h]
		\centering
		\includegraphics[width=0.5\textwidth]{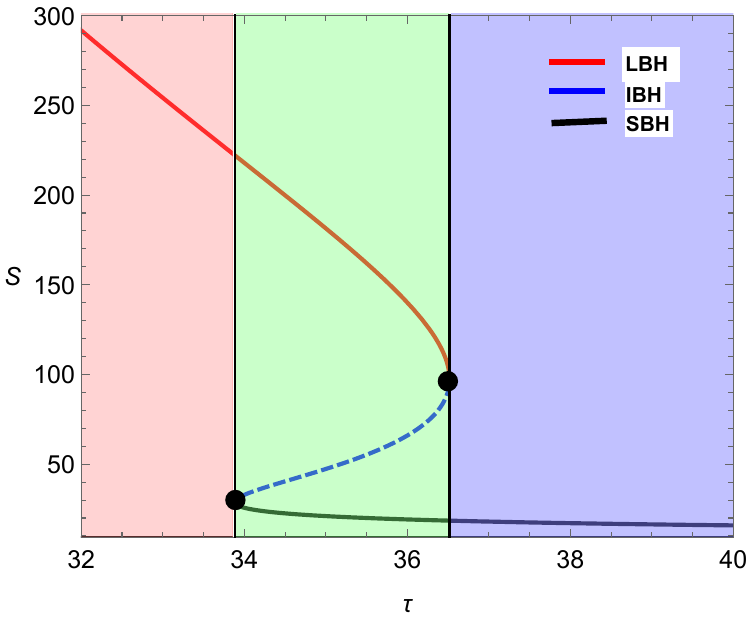}
		\caption{Plots for zero points of $\phi^r$ in the $\tau-S$ plane for rotating charged  black hole with in fixed $(q,J)$ ensemble at $l=10,J=1.5$ and  $q=0.05.$ Two line are drawn at $\tau=36.5182$ and $\tau=33.8834$ which are corresponds to annihilation and generation point respectively.The solid red portion represents a large black hole branch, the blue dashed portion represents an intermediate black hole branch and the solid black portion represents a small black hole branch.}
		\label{8}
	\end{figure}
\begin{figure}[h]	
		\centering
		\begin{subfigure}{0.32\textwidth}
			\includegraphics[width=\linewidth]{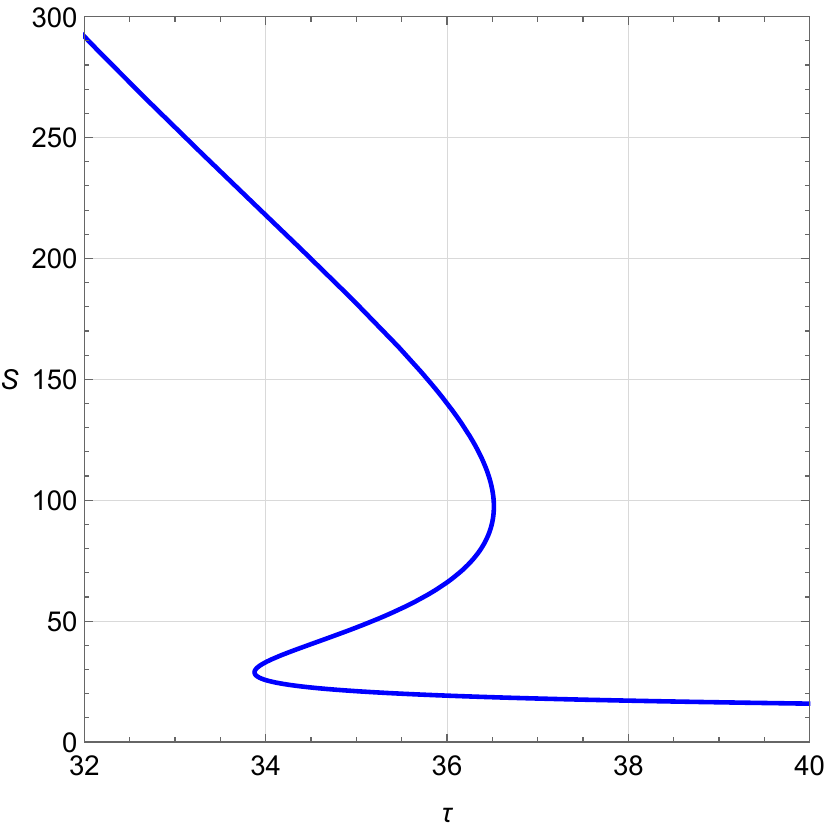}
			\caption{W=1}
			\label{10a}
		\end{subfigure}
		\begin{subfigure}{0.32\textwidth}
			\includegraphics[width=\linewidth]{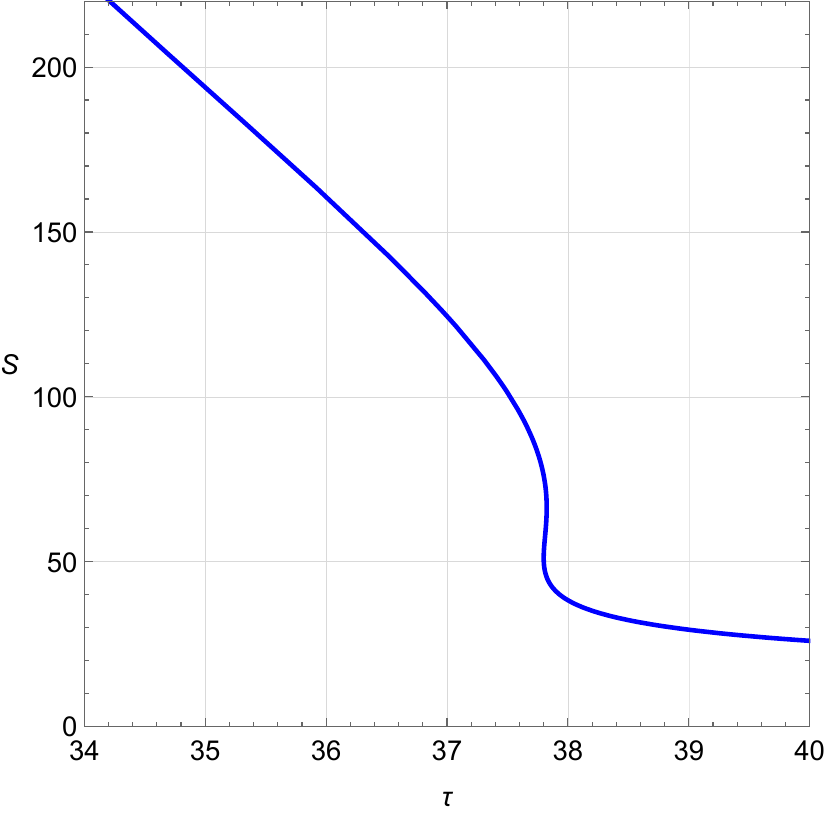}
			\caption{W=1}
			\label{10b}
		\end{subfigure}
		\begin{subfigure}{0.32\textwidth}
			\includegraphics[width=\linewidth]{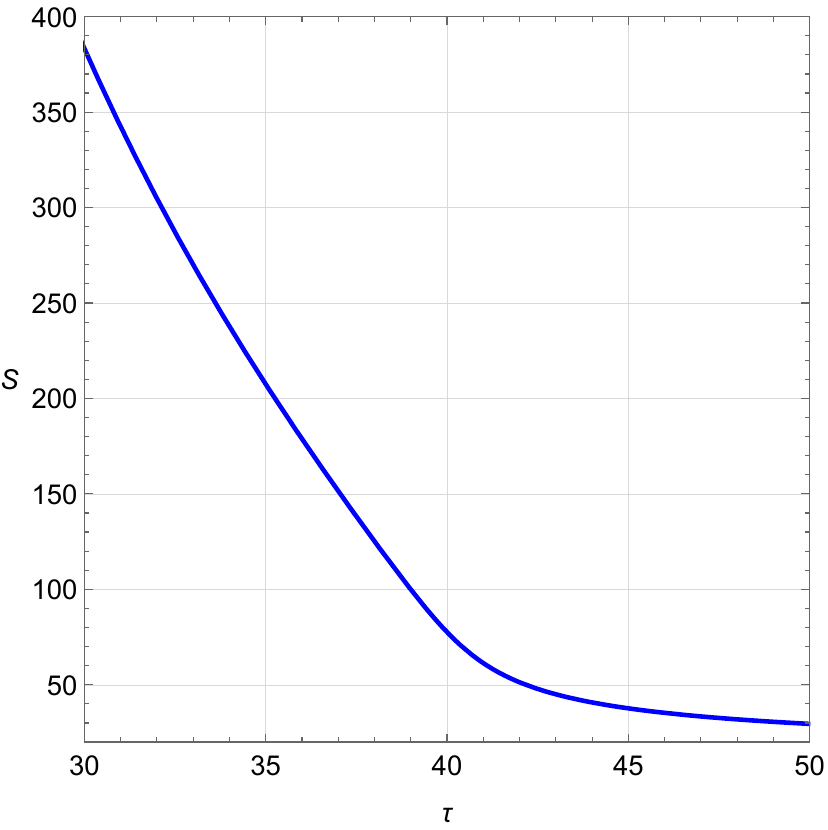}
			\caption{W=1}
			\label{10c}
		\end{subfigure}
		\caption{ $\tau$ vs $S$ plots for rotating charged black hole in fixed $(q,J)$ ensemble when charge $q$ is varied for a fixed length $l=10$ while keeping  $J$ fixed at $J=1.5.$ Figure $\left(a\right)$ shows $\tau$ vs $S$ plot  at $q=0.0001$,$J=1.5$, $l=10$, figure $\left(b\right)$  shows $\tau$ vs $S$ plot at $q=1.3$,$J=1.5$, $l=10$ and figure $\left(c\right)$ shows the same at $q=2$,$J=1.5$, $l=10$. $W$ denotes total topological charge.}
		\label{10}
	\end{figure}
\begin{figure}[h]	
		\centering
		\begin{subfigure}{0.4\textwidth}
			\includegraphics[width=\linewidth]{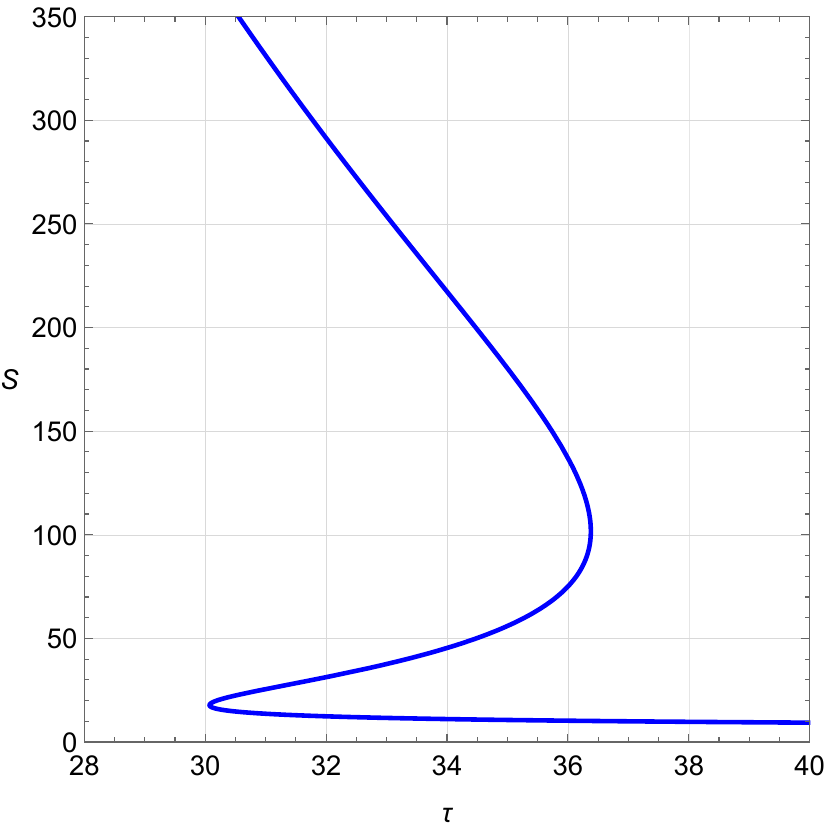}
			\caption{W=1}
			\label{11a}
		\end{subfigure}
		\begin{subfigure}{0.4\textwidth}
			\includegraphics[width=\linewidth]{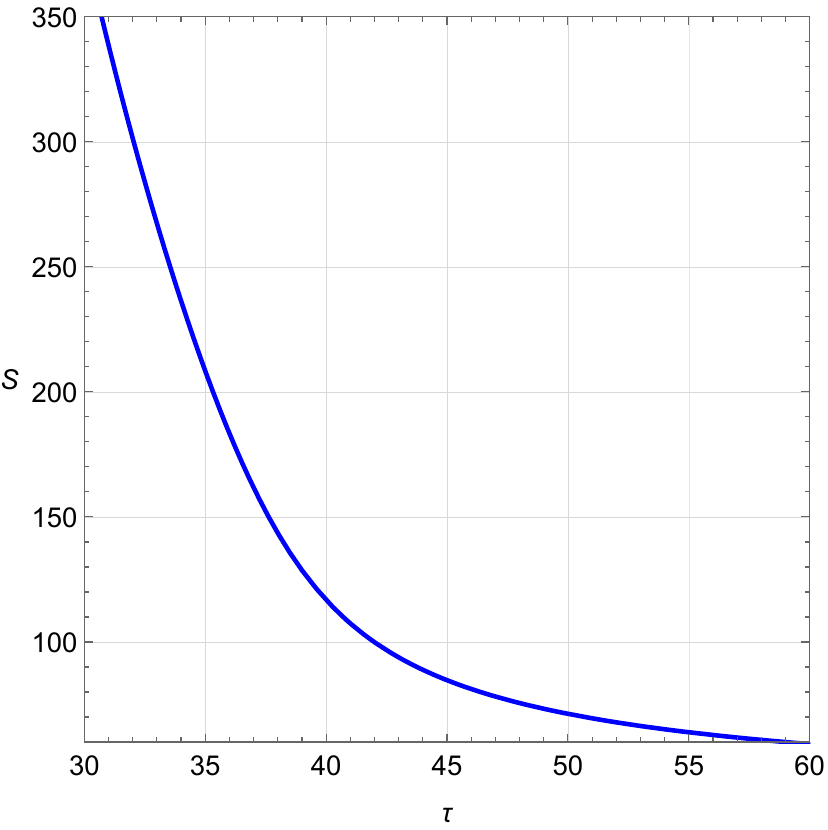}
			\caption{W=1}
			\label{11b}
		\end{subfigure}
		\caption{ $\tau$ vs $S$ plots for rotating charged black hole in fixed $(q,J)$ ensemble when $J$ is varied for a fixed length $l=10$ while keeping the charge fixed at $q=0.05$. Figure $\left(a\right)$  shows $\tau$ vs $S$ plot at $q=0.05$,$J=1$, $l=10$ and figure $\left(b\right)$  shows the same at $q=0.05$,$J=7$, $l=10$. $W$ denotes the corresponding topological charge.}
		\label{11}
	\end{figure}

	In figure.\ref {8}, we have plotted $S$ against $\tau$ when length scale is increased to $l=10$ while keeping $q=0.05$ and $J=1.5$ fixed.Here we observe three black hole branches: a small, an intermediate, and a large black hole branch. We also see a generation point at  $\tau=36.5182, S=97.4031$ and an annihilation point at $\tau=33.8834, S= 28.8408$ which are shown as black dots. To calculate the winding numbers, we put $l=10, J=1.5$ and $q=0.05$ in equation \ref{pole}.
	\begin{figure}[h]
		\centering
		\includegraphics[height=6cm,width=10cm]{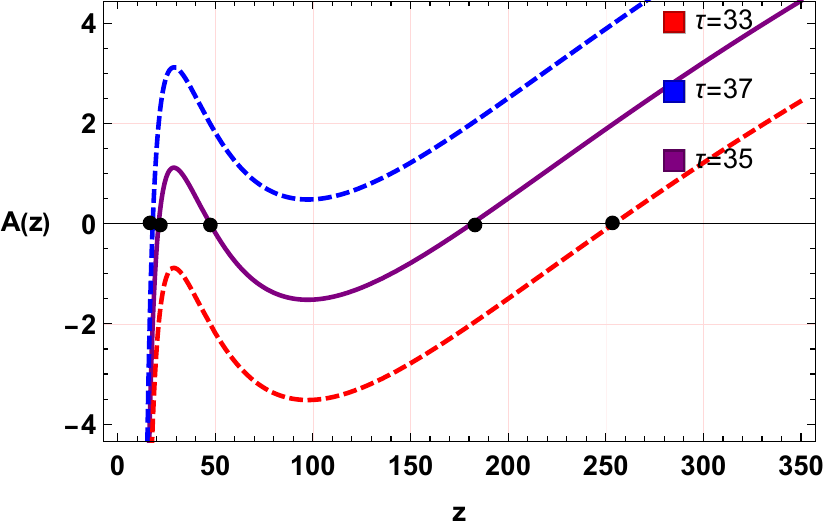}
		\caption{Plot for polynomial function $\mathcal{A}(z)$ for $\tau=33$, $\tau=35$ and  $\tau=37$}
		\label{9}
	\end{figure}
	The plot for the corresponding $\mathcal{A}(z)$ is shown in figure \ref{9}. This clearly shows that for $\tau<33.8834$, there is only one pole, for $33.8834 < \tau < 36.5182 $ there are three poles and for $\tau > 36.5182$ there is one pole. For the large black hole branch, taking $\tau=33$, the pole is at $z=254.214$. Around the pole $z=254.214$, the winding number is found to be $w=+1$. For the intermediate black hole branch, taking $\tau=35$, the poles are at $z_1=21.0537,z_2=47.3137$ and $z_3=181.45$. According to the sign of the residue around these three poles, the winding numbers are found to be $w_1=+1, w_2=-1$ and $w_3=+1$ respectively. Hence the topological charge is:
	$$W=w_1+w_2+w_3=1-1+1=1$$
	Similarly, for a small black hole branch, taking $\tau=33$, the pole is at $z=17.9982$ and according to the sign of residue around the pole, the winding number is found to be $+1$.We have checked that even if the value of $l$ is increased further (i.e. beyond $l=10$), the number of branches and topological charge remains the same.\\
	We repeat the analysis for different values of $J$ and $q$, keeping $l$ constant.We observe no effect on the topological charge. The same is illustrated in figure \ref{10}. In figure \ref{10a}, figure \ref{10b} and figure \ref{10c} we show the effect of change in charge $q$ when $J$ and $l$ are kept fixed at $J=1.5$ and $l=10$. In figure \ref{10a} the charge is changed to a significantly small value $q=0.0001$. Here, we observe three black hole branches and the topological charge is found to be $1.$ In figure \ref{10b} where we set $q=1.3$(figure \ref{10b}), again we find the number of black hole branches equal to three and topological charge to be one. When the value of the charge is changed to  $q=2$ in figure \ref{10c}, the number of branches becomes one but the topological charge remains one.
	For fixed values of $l=10$ and $q=0.05$ the effect of variation in $J$ on the thermodynamic topology is demonstrated in figure \ref{11a} and figure \ref{11b}. For $J=1$, three black hole branches are observed with the sum of the corresponding winding number equal to  $1.$ as shown in figure \ref{11a}.In figure \ref{11b} we set $J=7$ and find a single black hole branch with topological charge equal to $1.$ We have explicitly verified that even for other values of $J$, the topological charge remains the same.\\
	From our analysis, we conclude that the topological charge of the rotating charged black hole in fixed $(q, J)$ ensemble is equal to $+1$ and is unaffected by the variation in the thermodynamic parameters $l,q, J.$\\
	\subsection{Fixed ($\phi$,J) ensemble}
\begin{figure}[h]	
		\centering
		\begin{subfigure}{0.4\textwidth}
			\includegraphics[width=\linewidth]{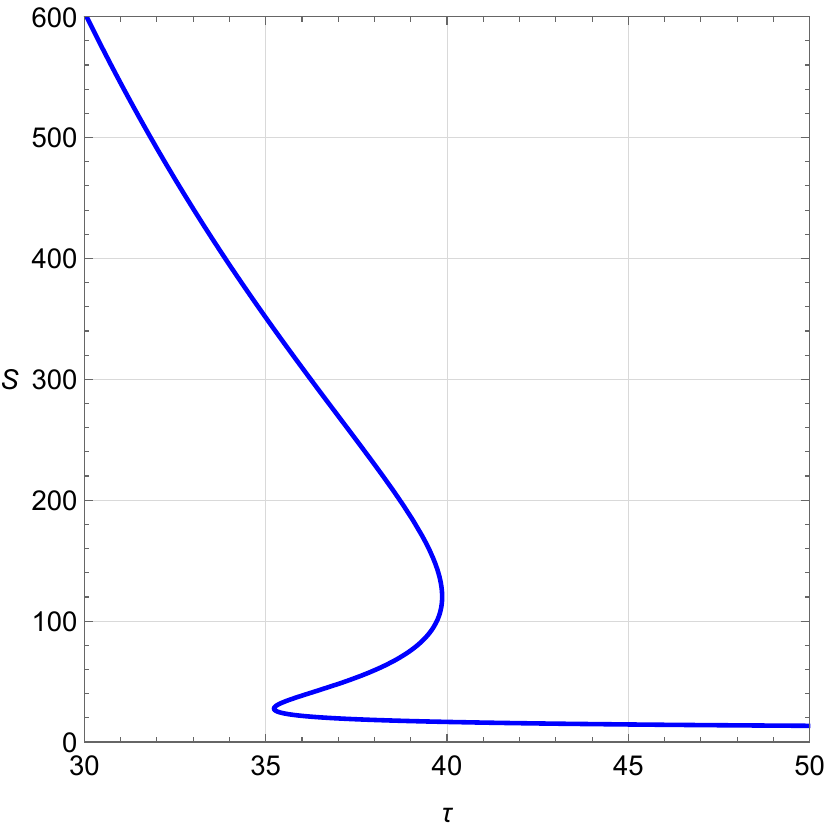}
			\caption{W=1}
			\label{12a}
		\end{subfigure}
		\begin{subfigure}{0.4\textwidth}
			\includegraphics[width=\linewidth]{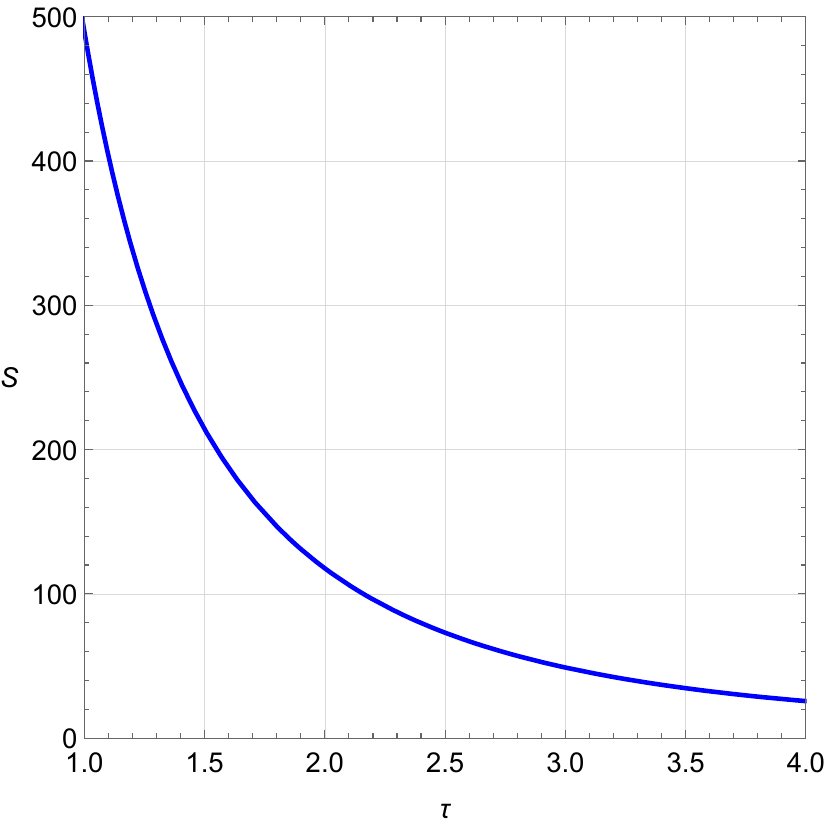}
			\caption{W=1}
			\label{12b}
		\end{subfigure}
		\caption{ $\tau$ vs $S$ plots for a rotating charged black hole in the fixed $(\phi,J)$ ensemble when scalar curvature $R$ is varied  while keeping $J$ and $\phi$ constant at $J=1.5,\phi=0.05.$ Figure $\left(a\right)$ shows $\tau$ vs $S$ plot at $\phi=0.05$, $J=1.5$, $R=-0.1$; figure $\left(b\right)$ shows $\tau$ vs $S$ plot at $\phi=0.05$, $J=1.5$, $R=-4$. $W$ denotes the respective topological charge.}
		\label{12}
	\end{figure}

	In fixed $(\phi, J)$ ensemble, the potential $\phi$ and angular momentum $J$ are kept fixed. The potential $\phi$ is given by :
	\begin{equation}
		\phi=\frac{\partial M}{\partial q}=\frac{\sqrt{\pi } q \left(12 \pi ^2 q^2-R S^2+12 \pi  S\right)}{S \sqrt{\frac{48 \pi ^3 J^2 (12 \pi -R S)+\left(12 \pi ^2 q^2-R S^2+12 \pi  S\right)^2}{S}}}
		\label{phi1}
	\end{equation}
	Solving equation \ref{phi1}, we get an expression for $q$ and find out the new mass($M_{\phi}$) in this ensemble as follows :
	\begin{equation}
		M_{\phi}=M-q \phi
	\end{equation}
	The off-shell free energy is computed using:
	$$\mathcal{F}=M_{\phi}-S/\tau$$
	Following the same procedure as shown in the previous subsection, we calculate the expressions for $\phi^S$ and $\tau$. We plot $\tau$ vs $S$ curve for two different values scalar curvature $R$  as shown in figure \ref{12}. Here, $J$ and $\phi$ are kept  constant with $J=1.5$ and $\phi=0.05.$ In figure \ref{12a}, we set $R=-0.1$ and find three black hole phases phase with topological charge equaling one. In figure \ref{12b} for $R=-4$, we observed a single black hole branch with topological charge still equaling one. We have checked that for the other values of $R$, the topological charge remains equal to one. \\
	
\begin{figure}[h]	
		\centering
		\begin{subfigure}{0.32\textwidth}
			\includegraphics[width=\linewidth]{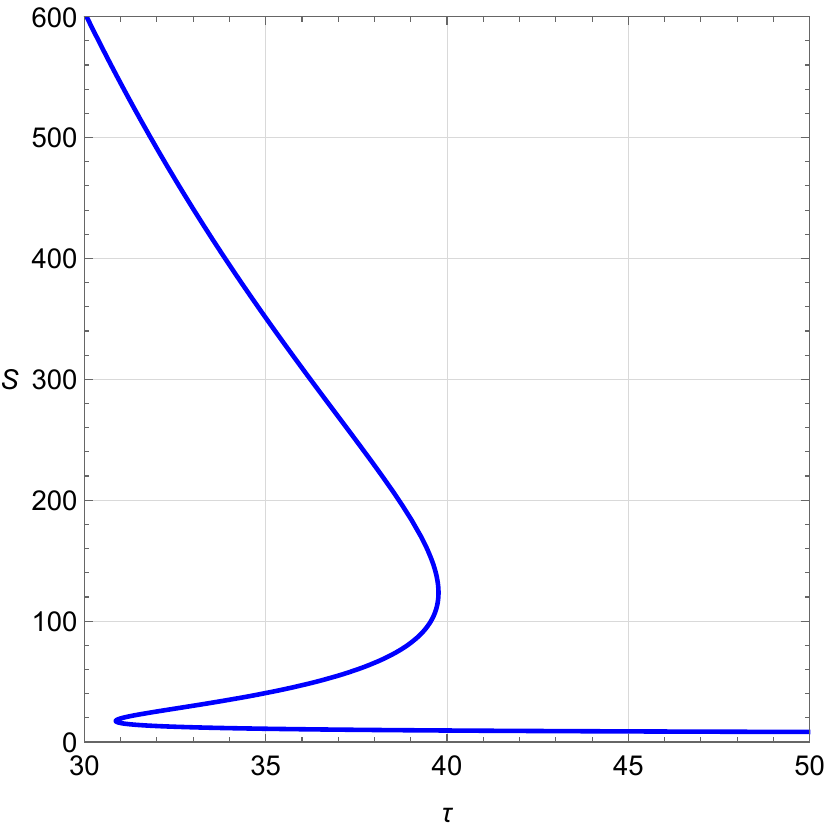}
			\caption{W=1}
			\label{13a}
		\end{subfigure}
		\begin{subfigure}{0.32\textwidth}
			\includegraphics[width=\linewidth]{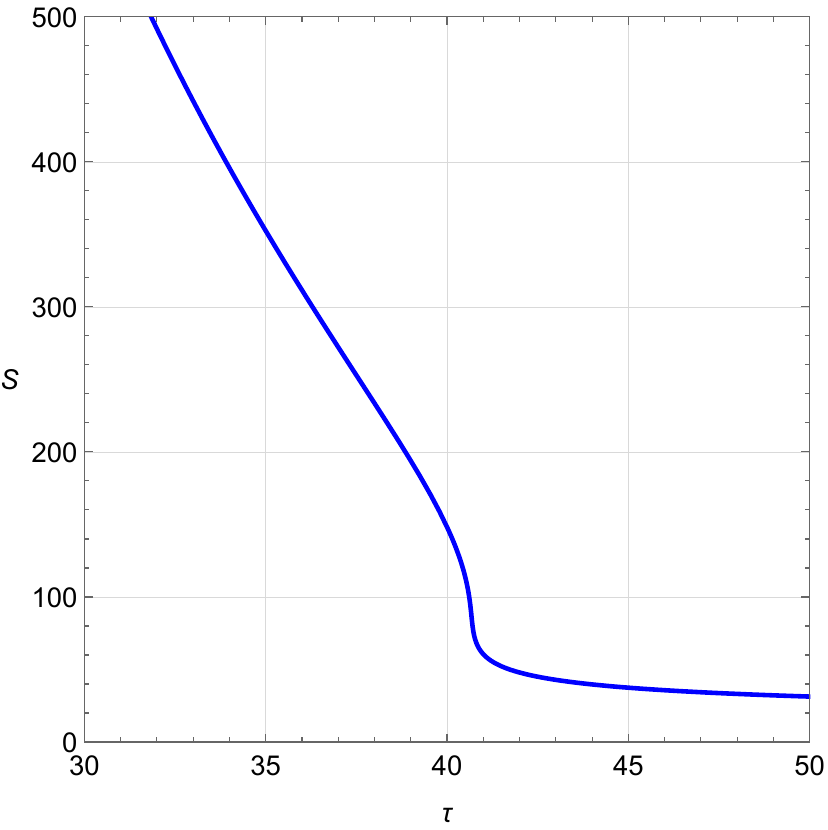}
			\caption{W=1}
			\label{13b}
		\end{subfigure}
		\begin{subfigure}{0.32\textwidth}
			\includegraphics[width=\linewidth]{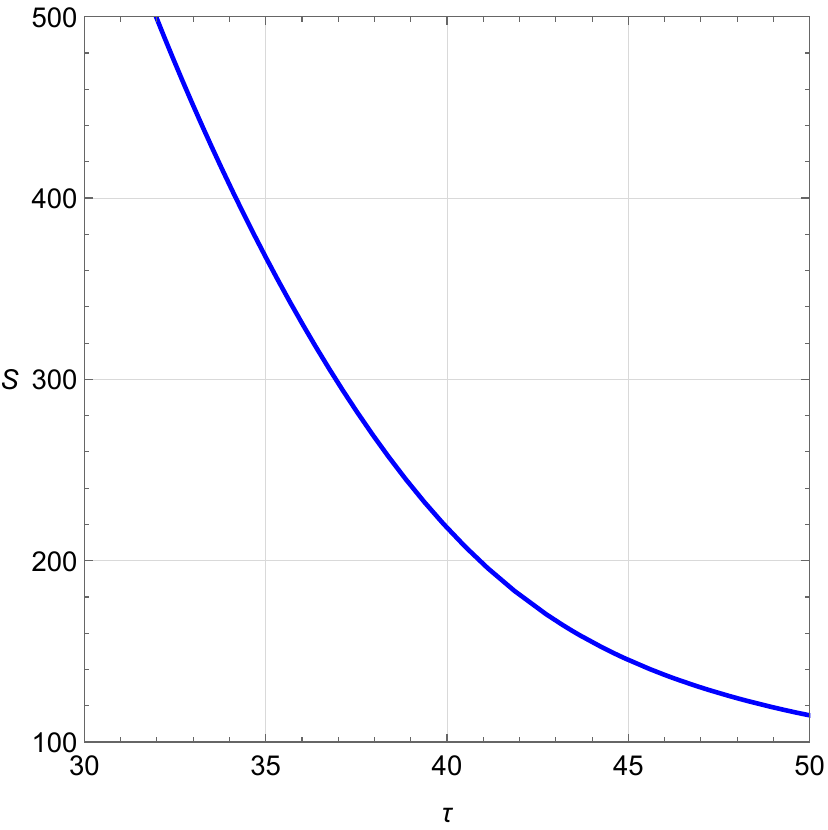}
			\caption{W=1}
			\label{13c}
		\end{subfigure}
		\caption{ $\tau$ vs $S$ plots for rotating charged black hole in fixed $(\phi,J)$ ensemble when $J$ is varied keeping $R=-0.1,\phi=0.05$ fixed. Figure $\left(a\right)$ shows $\tau$ vs $S$ plot  at $\phi=0.05$,$J=1$, $R=-0.1$ and figure $\left(b\right)$  shows $\tau$ vs $S$ plot at $\phi=0.05$,$J=7$, $R=-0.1$ and figure $\left(c\right)$  shows the same at $\phi=0.05$,$J=10$, $R=-0.1$. $W$ denotes the topological charge.}
		\label{13}
	\end{figure}

	Next, we study the influence of changing $J$ on the topological charge with $R$ and $\phi$ kept fixed at $R=-0.1$ and  $\phi=0.05.$ in figure \ref{13}. In figure \ref{13a} we observe three black hole branches leading to a topological charge of $1.$ In figure \ref{13b}, we again encounter three branches and topological charge equal to $1$ for $J=7$. In figure \ref{13c}, with $J=10$ we get one black hole branch with topological charges equaling $1.$ We have explicitly verified that for other values of $J$ the topological charge remains the same. \\
	
\begin{figure}[h]	
		\centering
		\begin{subfigure}{0.4\textwidth}
			\includegraphics[width=\linewidth]{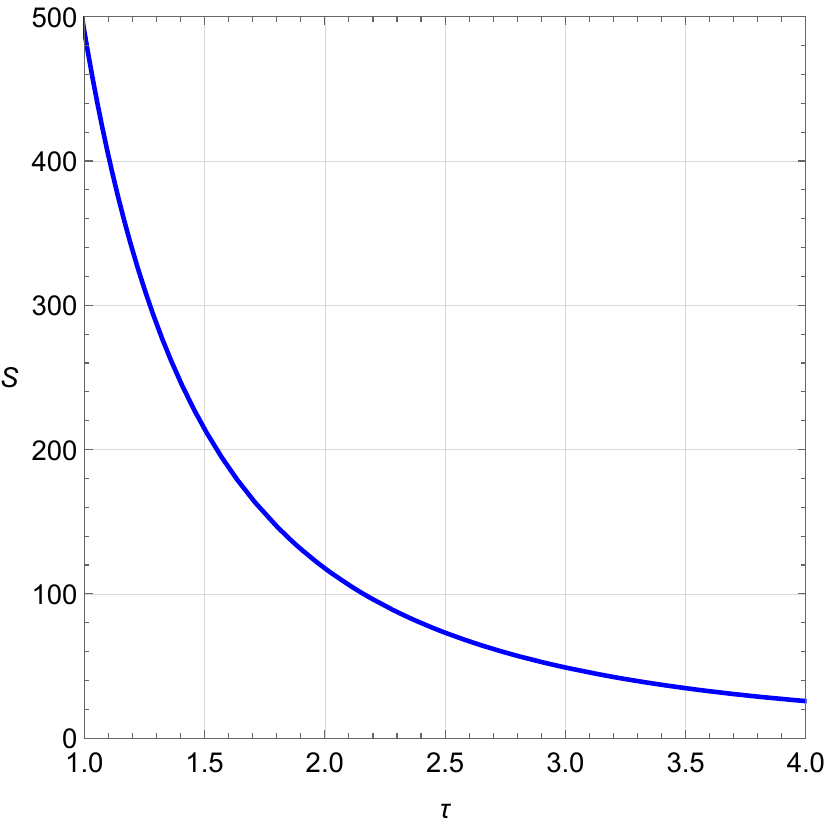}
			\caption{W=1}
			\label{14a}
		\end{subfigure}
		\begin{subfigure}{0.4\textwidth}
			\includegraphics[width=\linewidth]{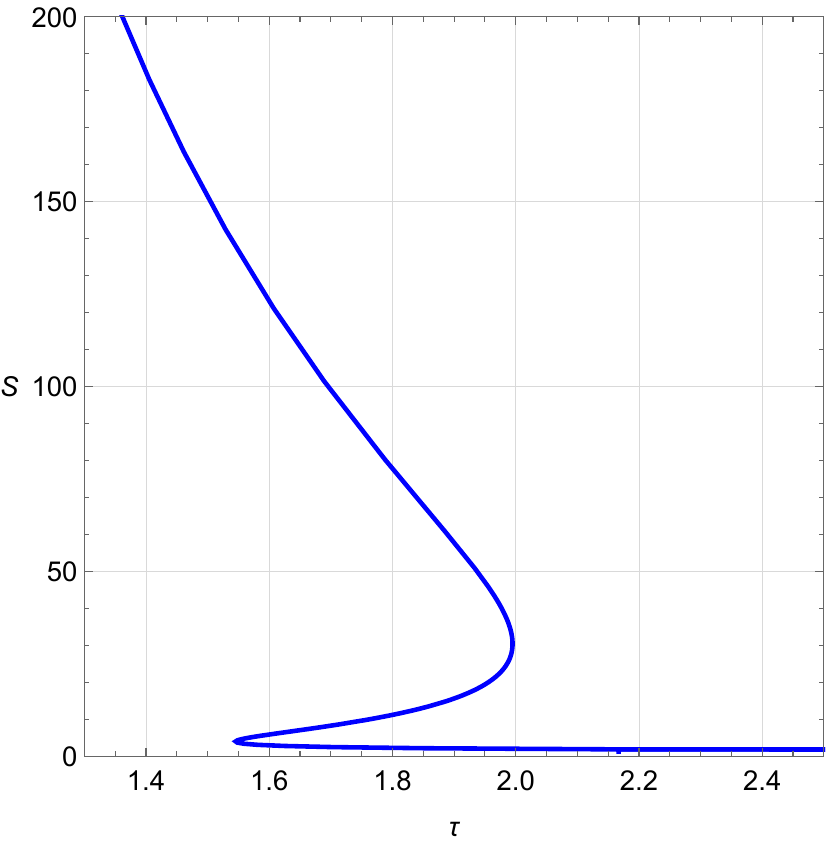}
			\caption{W=1}
			\label{14b}
		\end{subfigure}
		\caption{ $\tau$ vs $S$ plots for rotating charged black hole in fixed $(\phi,J)$ ensemble when $\phi$ is varied keeping $R=-4,J=1.5$ fixed. Figure $\left(a\right)$ shows $\tau$ vs $S$ plot  at $\phi=0.005$,$J=1.5$, $R=-4$ and figure $\left(b\right)$  shows $\tau$ vs $S$ plot at $\phi=3$,$J=1.5$, $R=-4$. $W$ denotes topological charge.}
		\label{14}
	\end{figure}

	Finally, we analyzed the effect  of $\phi$ on topological charge with fixed values of $R$ and $J.$, $\tau$ vs $S$ plots for $J=1.5$ and $R=-0.1$ are shown in figure \ref{14a} and figure \ref{14b} by setting $\phi=0.005$ and $\phi=3$ respectively. While in figure \ref{14a} we find a single black hole branch, in figure \ref{14b}) we see three branches. In both cases, the topological charge is found to be $1$. The same is found to be true for other values of $\phi$ with $J$ and $R$ kept fixed.\\
	Therefore we infer that the topological charge of the rotating charged black hole under consideration in fixed $(\phi, J)$ ensemble is equal to one irrespective of the values of the thermodynamic parameters $\phi, J$ and $R.$

	\subsection{Fixed ($\Omega$, q) ensemble}
	Next, we work in the fixed $(\Omega, q)$ ensemble where the angular frequency $\Omega$ and charge $q$ are kept fixed. We begin with solving the following equation for the angular momentum $J$
	\begin{equation}
		\Omega=\frac{\partial M}{\partial J}=-\frac{2 \pi ^{3/2} J (R S-12 \pi )}{S \sqrt{\frac{48 \pi ^3 J^2 (12 \pi -R S)+\left(12 \pi ^2 q^2-R S^2+12 \pi  S\right)^2}{S}}}
		\label{omega}
	\end{equation}
	From \ref{omega} we obtain the expressions for $J$ as follows:
	\begin{equation}
		J=\frac{\sqrt{S} \Omega  \left(12 \pi ^2 q^2-R S^2+12 \pi  S\right)}{2 \pi ^{3/2} \sqrt{(12 \pi -R S) \left(12 \pi -S \left(R+12 \Omega ^2\right)\right)}}
		\label{j1}
	\end{equation}
	The new mass ($M_{\Omega}$) in this ensemble as :
	\begin{equation}
		M_{\Omega}=M-J \Omega
	\end{equation}
	\begin{equation}
		M_\Omega=\frac{\sqrt{(12 \pi -R S) \left(12 \pi -S \left(R+12 \Omega ^2\right)\right)} \sqrt{\frac{(12 \pi -R S) \left(12 \pi ^2 q^2-R S^2+12 \pi  S\right)^2}{S \left(-R S-12 S \Omega ^2+12 \pi \right)}}-144 \pi ^2 q^2 \sqrt{S} \Omega ^2+12 R S^{5/2} \Omega ^2-144 \pi  S^{3/2} \Omega ^2}{24 \pi ^{3/2} \sqrt{(12 \pi -R S) \left(12 \pi -S \left(R+12 \Omega ^2\right)\right)}}
	\end{equation}
	Accordingly, the off-shell free energy is computed using:
	$$\mathcal{F}=M_{\Omega}-S/\tau$$
	Repeating the procedure alluded to in the previous section, we obtained the expression for $\phi^S$ and $\tau$.\\
	
	\begin{figure}[h]	
		\centering
		\begin{subfigure}{0.4\textwidth}
			\includegraphics[width=\linewidth]{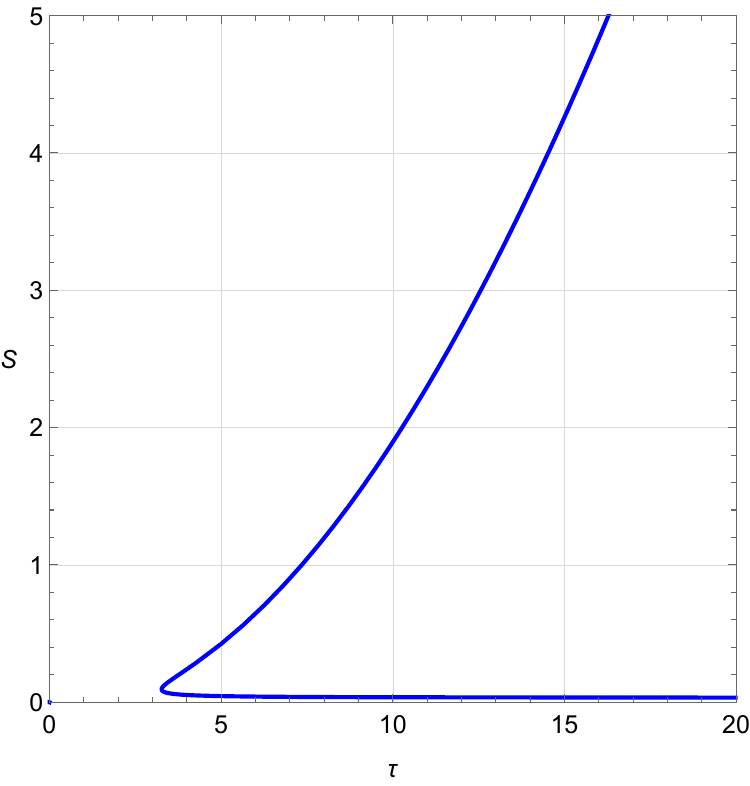}
			\caption{W=0}
			\label{15a}
		\end{subfigure}
		\begin{subfigure}{0.4\textwidth}
			\includegraphics[width=\linewidth]{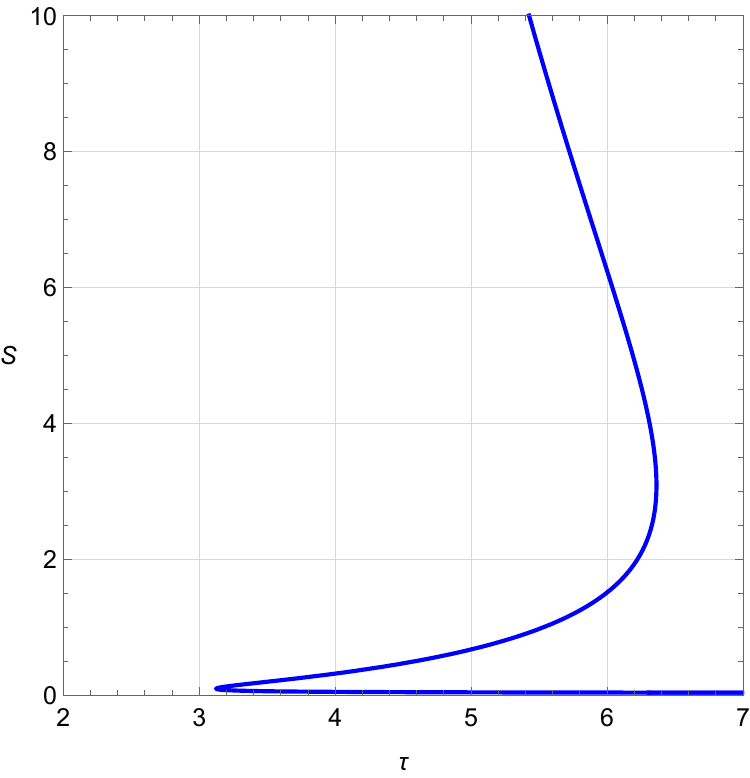}
			\caption{W=1}
			\label{15b}
		\end{subfigure}
		
		\caption{ $\tau$ vs $S$ plots for rotating charged black hole in fixed $(\Omega,q)$ ensemble for different values of $R$. Figure $\left(a\right)$ shows $\tau$ vs $S$ plot  at $\Omega=0.1$,$q=0.1$, $R=-0.01$ and figure $\left(b\right)$  shows the same at $\Omega=0.1$,$q=0.1$, $R=-4$. $W$ denotes the topological charge.
		}
		\label{15}
	\end{figure}
	First, we analyzed the dependence of topological charge on the scalar curvature $R$ keeping $\Omega$ and $q$ fixed.In figure \ref{15a} and \ref{15b}, $\tau$ vs $S$ plots are shown with $\Omega=0.1, q=0.1$ kept fixed and $R=-0.01$ and $R=-4$ respectively. In \ref{15a} two black hole branches and topological charge $0$ are observed. In \ref{15b}, we have three black hole branches totaling a topological charge of $1.$ Therefore, depending on the value of the scalar curvature, the topological charge is found to be either $0$ or $1.$ For other values of $R$, we arrived at the same topological charge($0$ or $1$)\\
	
\begin{figure}[h]	
		\centering
		\begin{subfigure}{0.4\textwidth}
			\includegraphics[width=\linewidth]{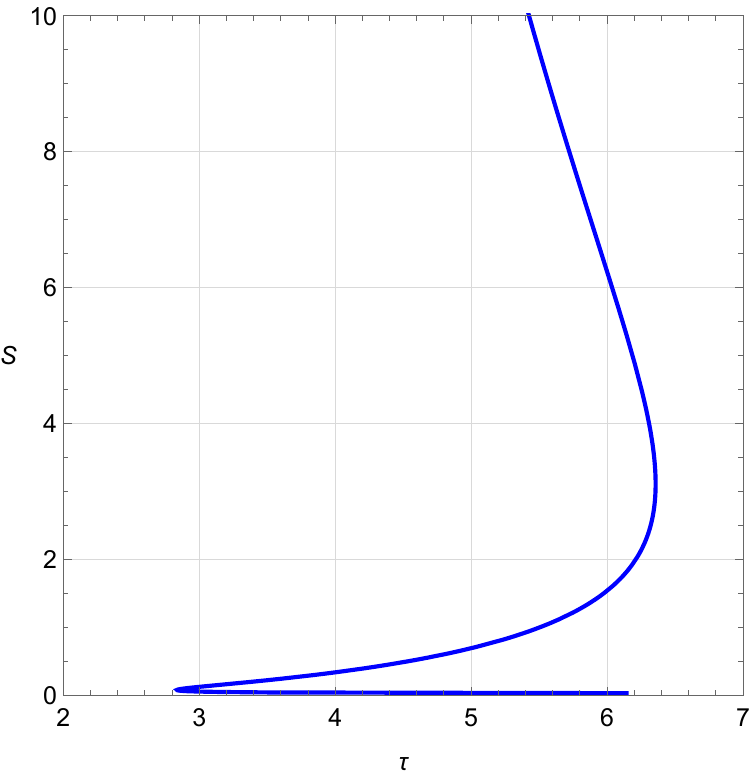}
			\caption{W=1}
			\label{16a}
		\end{subfigure}
		\begin{subfigure}{0.4\textwidth}
			\includegraphics[width=\linewidth]{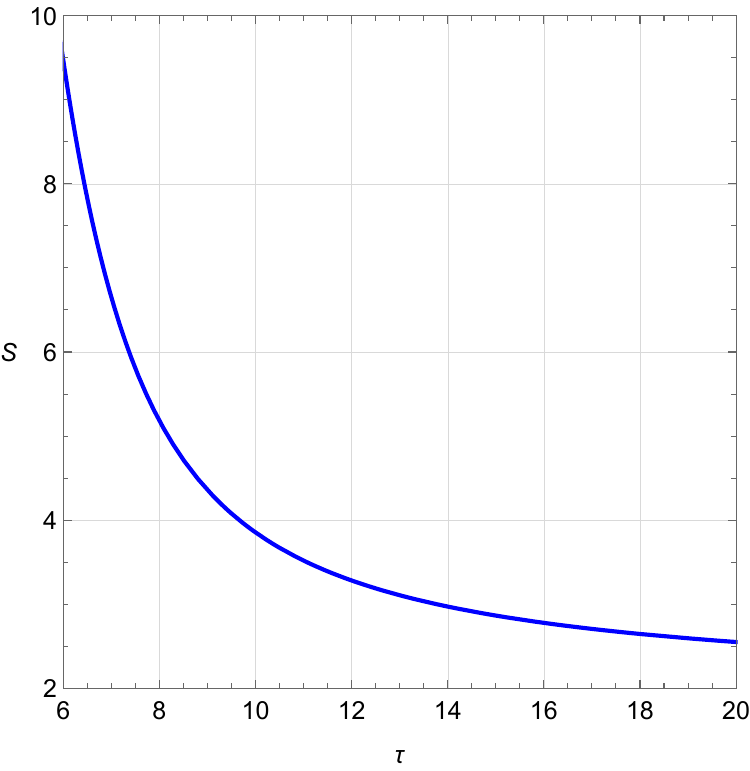}
			\caption{W=1}
			\label{16b}
		\end{subfigure}
		
		\caption{ $\tau$ vs $S$ plots for rotating charged black hole in fixed $(\Omega,q)$ ensemble for different $q$ values when $R=-4$ and $\Omega=0.1$ is kept fixed. Figure $\left(a\right)$ shows $\tau$ vs $S$ plot  at $\Omega=0.1$,$q=0.09$, $R=-4$, figure $\left(b\right)$  shows $\tau$ vs $S$ plot at $\Omega=0.1$,$q=1$, $R=-4$.$W$ denotes topological charge.
		}
		\label{16}
	\end{figure}

	Now we want to understand the impact of change in charge $q$ on the topological charge. In figure \ref{16a} and figure \ref{16b}, we set $\Omega=0.1$,$R=-4$ and $q=0.09$ and $1$ respectively. In figure \ref{16a} we find three black hole branches, in figure \ref{16b} we find a single black hole branch. The topological charge in both cases equals $1$.\\
	
\begin{figure}
		\centering
		\begin{subfigure}{0.4\textwidth}
			\includegraphics[width=\linewidth]{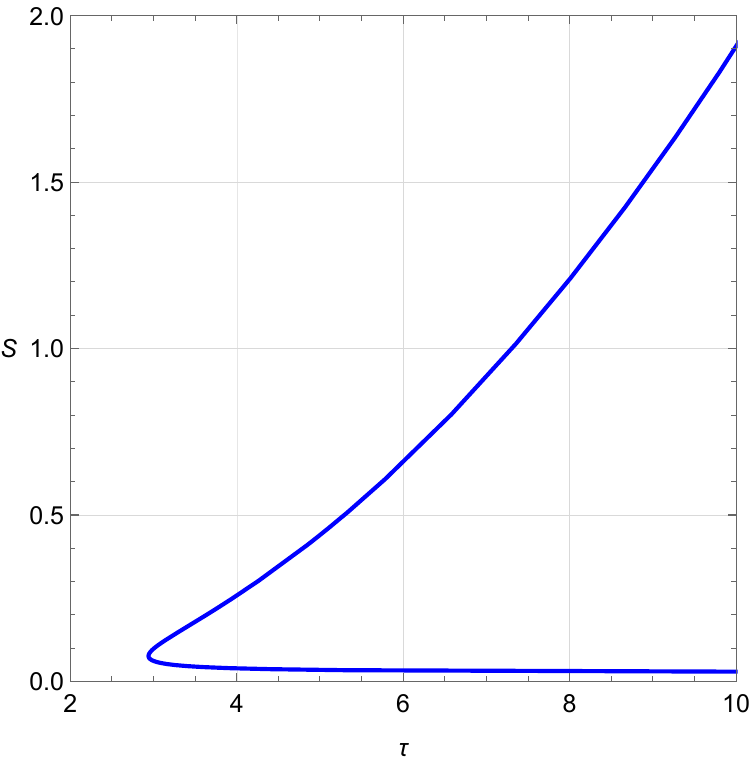}
			\caption{W=0}
			\label{17a}
		\end{subfigure}
		\begin{subfigure}{0.4\textwidth}
			\includegraphics[width=\linewidth]{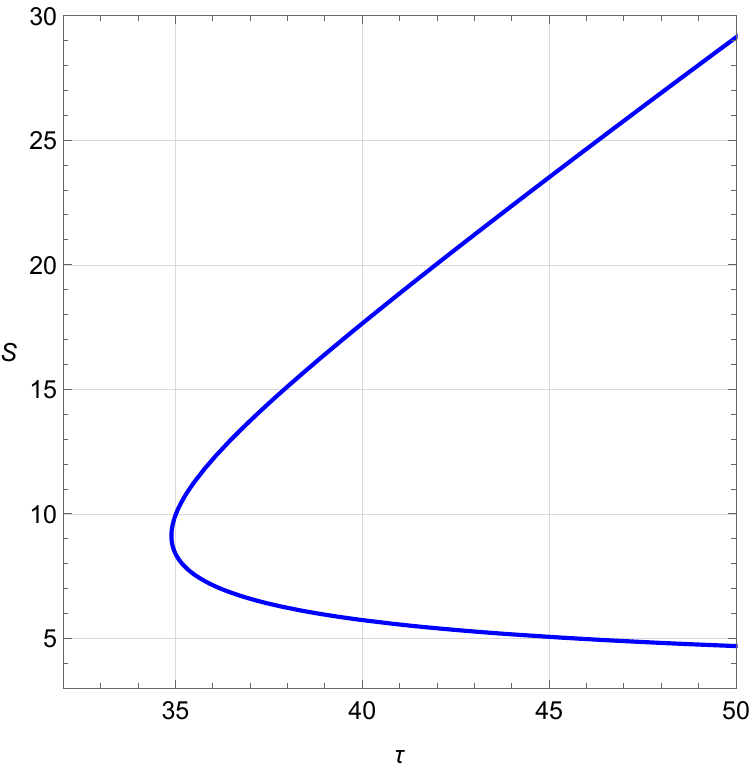}
			\caption{W=0}
			\label{17b}
		\end{subfigure}
		\caption{ $\tau$ vs $S$ plots for rotating charged black hole in fixed $(\Omega,q)$ ensemble for different $q$ value at $R=-0.01$ and $\Omega=0.1$. Figure $\left(a\right)$ shows $\tau$ vs $S$ plot  at $q=0.09$, figure $\left(b\right)$ shows $\tau$ vs $S$ plot at $q=1$.The topological charge for all the cases is $1.$ $W$ denotes the topological charge.
		}
		\label{17}
	\end{figure}

	In figure \ref{17a} and \ref{17b} we repeat the same analysis with $\Omega=0.1$ $R=-0.01$ and $q=0.09$ and $q=1$ respectively. The number of branches and the topological charge in both cases are found to be identical(two black hole branches and topological charge $0$). Hence it is found that the topological charge does not change with the charge $q$ although the number of black hole branches may vary with a variation in $q.$\\
	
\begin{figure}[h]	
		\centering
		\begin{subfigure}{0.4\textwidth}
			\includegraphics[width=\linewidth]{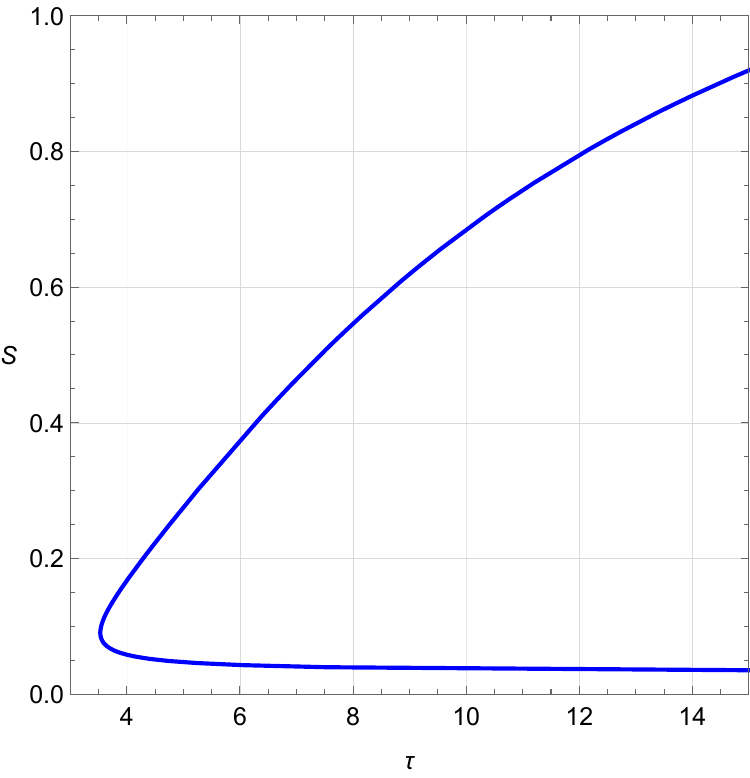}
			\caption{W=0}
			\label{18a}
		\end{subfigure}
		\begin{subfigure}{0.4\textwidth}
			\includegraphics[width=\linewidth]{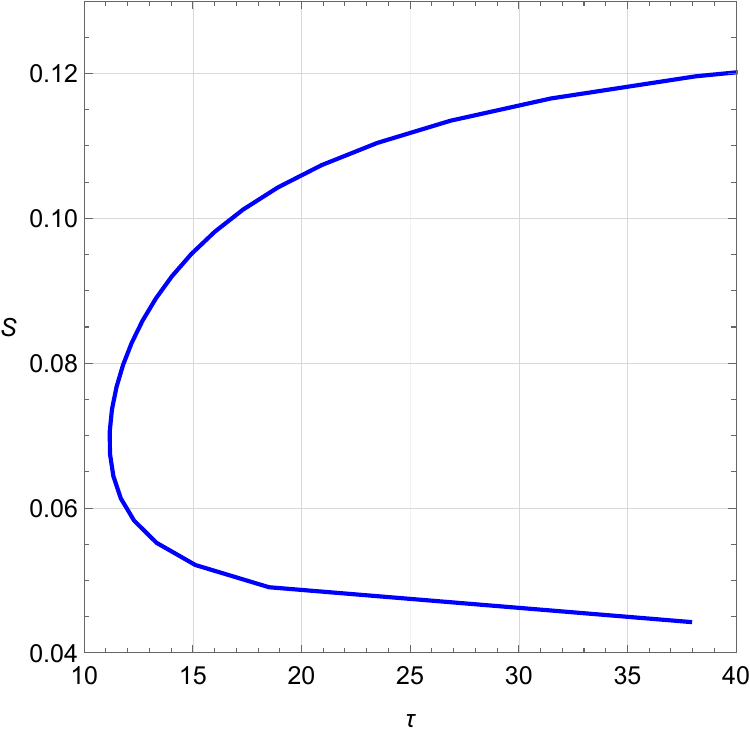}
			\caption{W=0}
			\label{18b}
		\end{subfigure}
		\caption{ $\tau$ vs $S$ plots for rotating charged black hole in fixed $(\Omega,q)$ ensemble for different $\Omega$ value at $R=-0.01$ and $q=0.1$. Figure $\left(a\right)$ shows $\tau$ vs $S$ plot  at $\Omega=1$, figure $\left(b\right)$ shows $\tau$ vs $S$ plot at $\Omega=3$.The topological charge for all the cases is $0.$
		}
		\label{18}
	\end{figure}

	Finally, we examine the role of $\Omega$ in determining the topological charge. For that, we keep $R$ fixed $R=-0.01$ and $q=0.1$ in figures \ref{18a} and figure \ref{18b}. In Figure \ref{18a} and \ref{18b} we set $\Omega=1$ and $\Omega=3$ respectively. As seen from the figure the topological charge remains unaffected by the variation in $\Omega$ and is equal to $0.$ In conclusion our results indicate that the topological charge of the rotating charged black hole in fixed $(\Omega,q)$ is either $0$ or $1$ depending on the values of $R.$ The other thermodynamic parameters $\Omega$ and $q$ however, do not have any impact on the topological charge.\\ 
	
	\subsection{Fixed ($\Omega$, $\Phi$) ensemble}
	The last ensemble in which we conduct our analysis is the fixed $(\Omega,\phi)$ ensemble. In this ensemble $\Omega$ and $\phi$ are kept fixed. First, We substitute J from equation \ref{j1} in the expression for mass in equation \ref{smr} as follows:
	\begin{equation}
		M=\sqrt{\frac{(R S-12 \pi ) \left(12 \pi ^2 q^2-R S^2+12 \pi  S\right)^2}{576 \pi ^3 S \left(S \left(R+12 \Omega ^2\right)-12 \pi \right)}}
		\label{mj}	
	\end{equation}
	Now equation \ref{mj} becomes independent of variable $J$. From equation \ref{mj}, we compute $\phi$ as:
	\begin{equation}
		\phi=\frac{dM_J}{dq}=\frac{\sqrt{\pi } q \sqrt{\frac{(R S-12 \pi ) \left(12 \pi ^2 q^2-R S^2+12 \pi  S\right)^2}{S \left(S \left(R+12 \Omega ^2\right)-12 \pi \right)}}}{12 \pi ^2 q^2-R S^2+12 \pi  S}
		\label{phi}
	\end{equation}
	Accordingly, $q$ is given by equation \ref{phi} as:
	$$q=\frac{\sqrt{S} \phi  \sqrt{R S+12 S \Omega ^2-12 \pi }}{\sqrt{\pi } \sqrt{R S-12 \pi }}$$
	The new expression for angular momentum($J$) is given by:
	$$J=\frac{S^{3/2} \Omega  \left(R^2 S^2-12 \pi  R S \phi ^2-24 \pi  R S-144 \pi  S \Omega ^2 \phi ^2+144 \pi ^2 \phi ^2+144 \pi ^2\right)}{2 \pi ^{3/2} (12 \pi -R S) \sqrt{(12 \pi -R S) \left(-R S-12 S \Omega ^2+12 \pi \right)}}$$
	Finally, the modified mass in fixed ($\Omega$,$\Phi$) ensemble is written as :
	\begin{equation}
		\tilde{M}=M-q \phi-J \Omega
	\end{equation}
	or
	\begin{multline}
		\tilde{M}=
		\frac{\sqrt{\frac{S \left(R^2 S^2-12 \pi  S \left(R \left(\phi ^2+2\right)+12 \Omega ^2 \phi ^2\right)+144 \pi ^2 \left(\phi ^2+1\right)\right)^2}{(12 \pi -R S) \left(12 \pi -S \left(R+12 \Omega ^2\right)\right)}}}{24 \pi ^{3/2}} 
		- \frac{S^{3/2} \Omega ^2 \left(\frac{12 \pi  \phi ^2 \left(S \left(R+12 \Omega ^2\right)-12 \pi \right)}{R S-12 \pi }-R S+12 \pi \right)}{2 \pi ^{3/2} \sqrt{(12 \pi -R S) \left(12 \pi -S \left(R+12 \Omega ^2\right)\right)}} \\
		- \frac{\sqrt{S} \phi ^2 \sqrt{S \left(R+12 \Omega ^2\right)-12 \pi }}{\sqrt{\pi } \sqrt{R S-12 \pi }}
		\label{mOP}
	\end{multline}
	Using equation \ref{mOP}, $\mathcal{F}$, $\phi^S$ and $\tau$ are constructed  following standard procedure. \\
	
\begin{figure}[h]	
		\centering
		\begin{subfigure}{0.4\textwidth}
			\includegraphics[width=\linewidth]{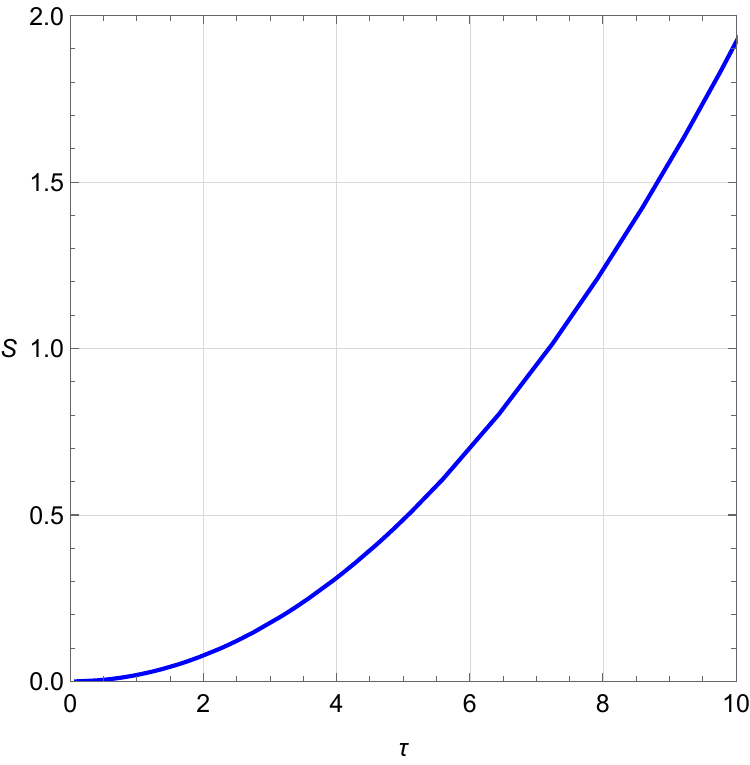}
			\caption{W=-1}
			\label{19a}
		\end{subfigure}
		\begin{subfigure}{0.4\textwidth}
			\includegraphics[width=\linewidth]{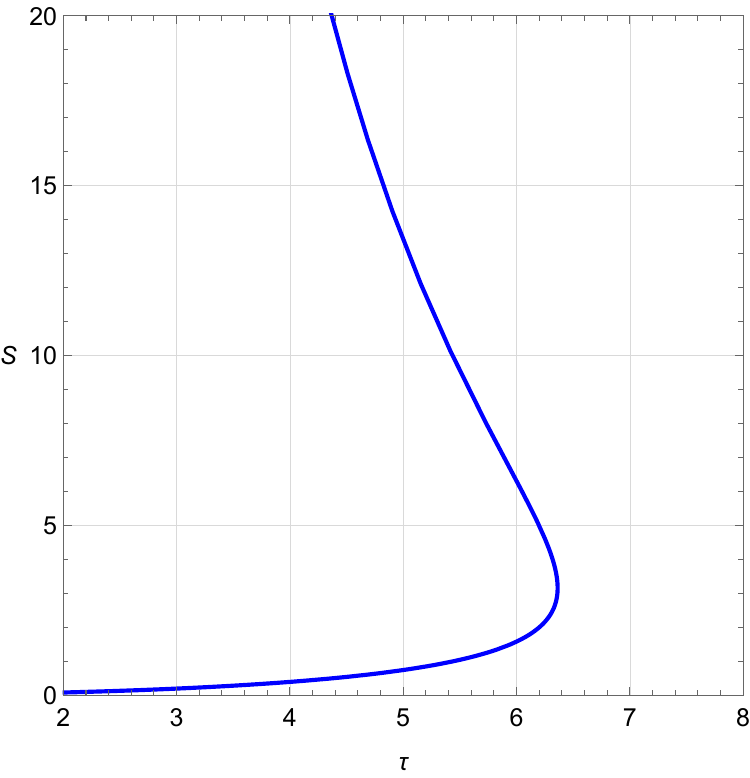}
			\caption{W=0}
			\label{19b}
		\end{subfigure}
		
		\caption{ $\tau$ vs $S$ plots for rotating charged black hole in fixed $(\Omega,\phi)$ ensemble for different values of $R$. Figure $\left(a\right)$ shows $\tau$ vs $S$ plot at $\Omega=0.1$,$\phi=0.1$, $R=-0.01$, figure $\left(b\right)$ shows $\tau$ vs $S$ plot at $\Omega=0.1$,$\phi=0.1$, $R=-4$. $W$ denotes the topological charge.
		}
		\label{19}
	\end{figure}

	We plot $\tau$ vs $S$ curve for different values of $R$ as shown in figure \ref{19} keeping $\Omega=0.1,\phi=0.1$ constant. In figure \ref{19a}, with $\Omega0.1, \phi=0.1$ and $R=-0.01$, one black hole branch with topological charge $W=-1$ is observed. With the same values of $\Omega$ and $\phi$ but for different values of $R$ at $R=-4$, in figure \ref{19b} two black hole branches and topological charge $0$ are found.\\
	
\begin{figure}[h]	
		\centering
		\begin{subfigure}{0.4\textwidth}
			\includegraphics[width=\linewidth]{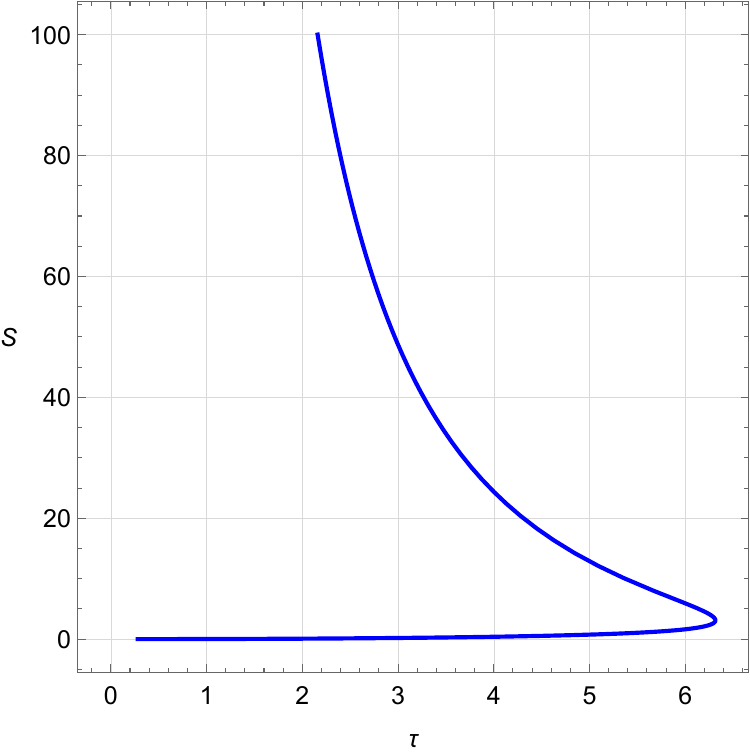}
			\caption{W=0}
			\label{20a}
		\end{subfigure}
		\begin{subfigure}{0.4\textwidth}
			\includegraphics[width=\linewidth]{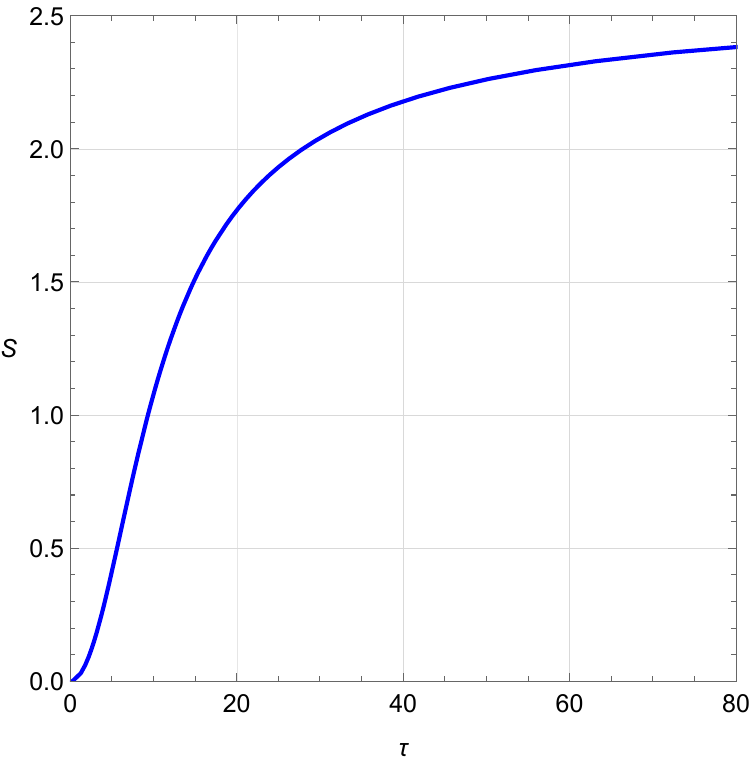}
			\caption{W=-1}
			\label{20b}
		\end{subfigure}
		
		\caption{ $\tau$ vs $S$ plots for rotating charged black hole in fixed $(\Omega,\phi)$ ensemble for different $\Omega$ value at $R=-4$ scale when $\phi=0.1$ is kept fixed. Figure $\left(a\right)$ shows $\tau$ vs $S$ plot  at $\Omega=0.001$,$\phi=0.1$, $R=-4$, figure $\left(b\right)$  shows $\tau$ vs $S$ plot at $\Omega=1$,$\phi=0.1$, $R=-4$.$W$ denotes total Topological charge.
		}
		\label{20}
	\end{figure}

	Next in figures \ref{20a} and figure \ref{20b}, we fix $\phi=0.1$,$R=-4$ and vary $\Omega.$ In figure \ref{20a}  setting $\Omega=0.001$ we get two black hole branches and topological charge $W=0.$ In figure \ref{20b} with $\Omega=1$ a single blackl hole branch with topological charge $W=-1$ is observed.\\ 
	
\begin{figure}[h]	
		\centering
		\begin{subfigure}{0.4\textwidth}
			\includegraphics[width=\linewidth]{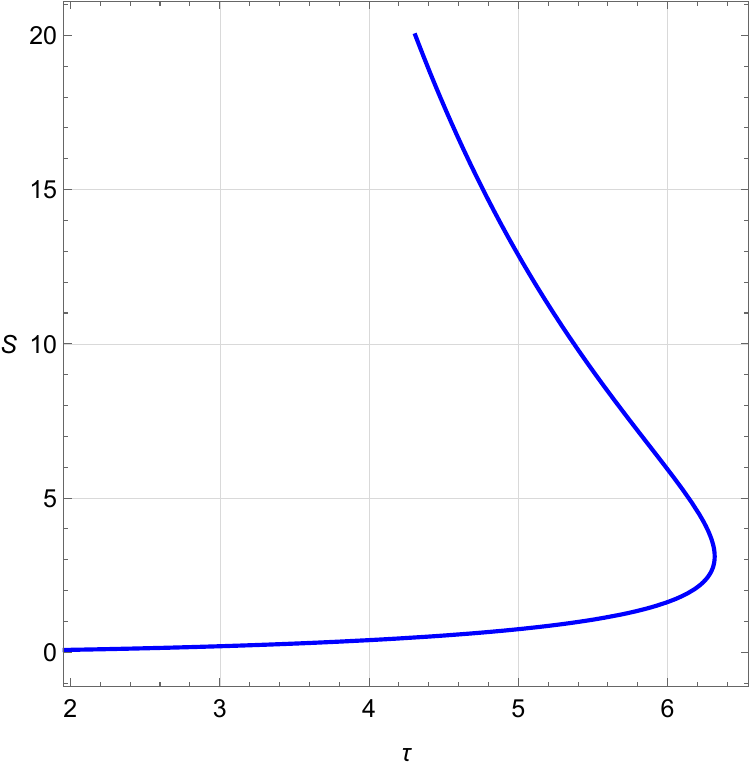}
			\caption{W=0}
			\label{21a}
		\end{subfigure}
		\begin{subfigure}{0.4\textwidth}
			\includegraphics[width=\linewidth]{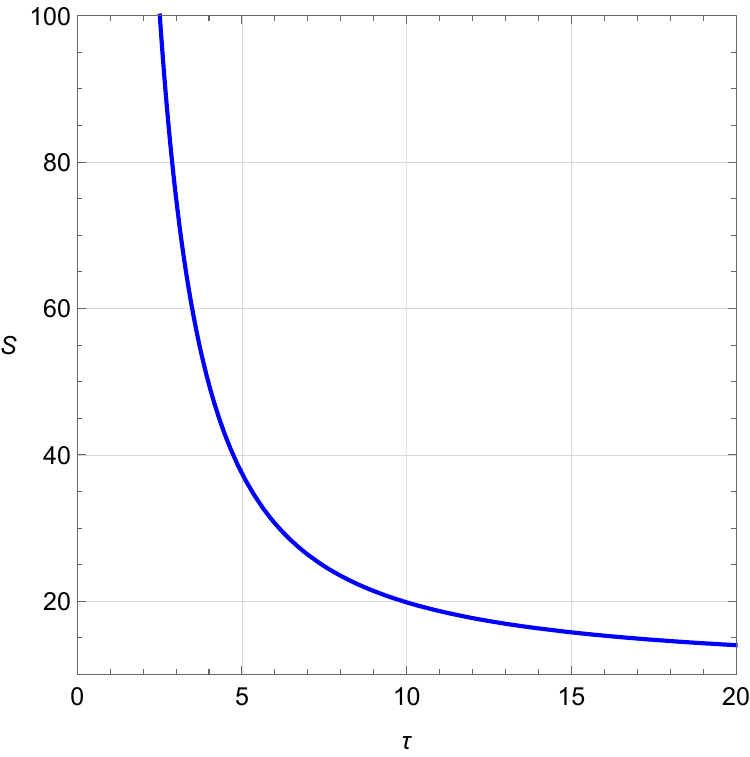}
			\caption{W=1}
			\label{21b}
		\end{subfigure}
		\caption{ $\tau$ vs $S$ plots for rotating charged black hole in fixed $(\Omega,\phi)$ ensemble for different $\phi$ value when $R=-4$ and $\Omega=0.1$ is kept fixed. Figure $\left(a\right)$ shows $\tau$ vs $S$ plot  at $\phi=0.001$,$\Omega=0.1$, $R=-4$, figure $\left(b\right)$  shows $\tau$ vs $S$ plot at $\phi=1$,$\Omega=0.1$, $R=-4$. $W$ denotes the Topological charge.
		}
		\label{21}
	\end{figure}

	Finally, we probe the thermodynamic topology with reference to a variation in the potential $\phi$ in figure \ref{21a} and \ref{21b} we set $\Omega=0.1$, $R=-4$ and $\phi=0.001$ and $\phi=2$ respectively. While in the first case, we find two black hole branches and winding number $W=0$, in the latter case a single black hole branch with topological charge $1$ is found.\\
	
\begin{figure}[h]	
		\centering
		\begin{subfigure}{0.4\textwidth}
			\includegraphics[width=\linewidth]{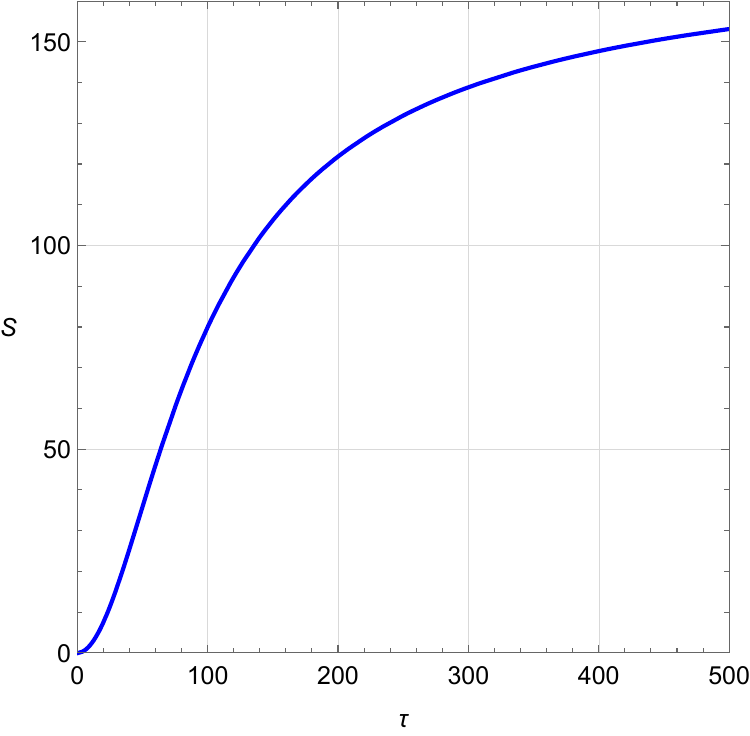}
			\caption{}
			\label{22a}
		\end{subfigure}
		\begin{subfigure}{0.4\textwidth}
			\includegraphics[width=\linewidth]{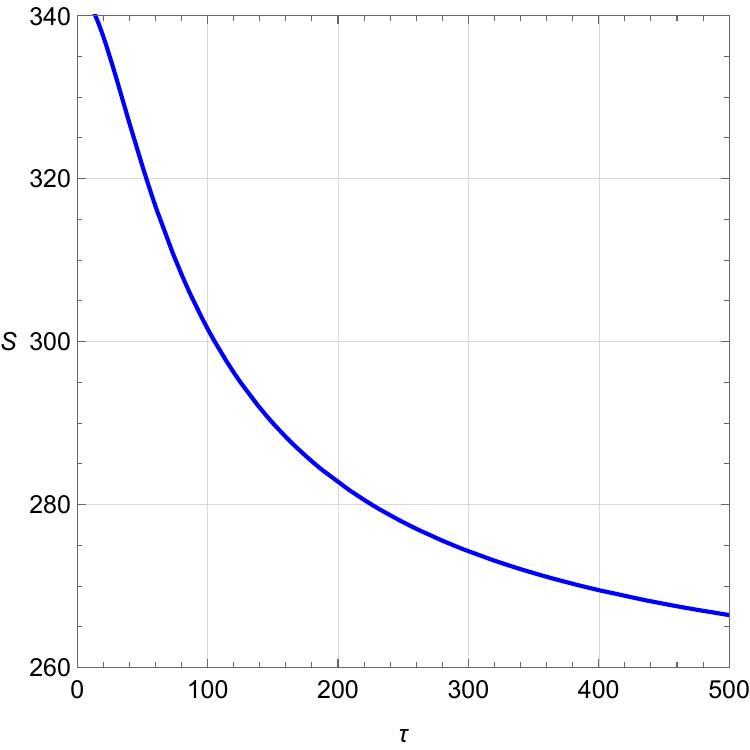}
			\caption{}
			\label{22b}
		\end{subfigure}
		\caption{ $\tau$ vs $S$ plots for rotating charged black hole in fixed $(\Omega,\phi)$ ensemble for different $\phi$ value when $R=-0.01$ and $\Omega=0.1$ is kept fixed. Figure $\left(a\right)$ shows $\tau$ vs $S$ plot  at $\phi=0.001$,$\Omega=0.1$, $R=-0.01$, figure $\left(b\right)$  shows $\tau$ vs $S$ plot at $\phi=2$,$\Omega=0.1$, $R=-0.01$. $W$ denotes the Topological charge.
		}
		\label{22}
	\end{figure}

	We continue our study at a different value of $R$ equal to $-0.01$ at figure \ref{22a} and figure \ref{22b}. In figure \ref{22a}, $\Omega=0.1$ and $\phi=0.001$. In figure \ref{22b}, $\Omega$ is again fixed at $0.1$ but $\phi$ is changed $\phi=2$  for $\phi=0.001$ a single black hole branch with topological charge $W=-1$ is seen. For $\phi=2$ we again encounter a single black hole branch, but this time with a topological charge of $+1.$\\
	
	So to summarize, the topological charge for the rotating charged black hole in fixed $(\Omega,\phi)$ ensemble is $-1$,$0$ or $+1$ depending on all the thermodynamic parameters $R,\Omega$ and $\phi.$ 
	\vspace{5cm}
	
	\section{Conclusion}
	In this study, we have analyzed the thermodynamic topology of a static black hole, a charged, static black hole, and a charged rotating black hole within the framework of $f(R)$ gravity. We have considered two distinct ensembles for charged static black holes: the fixed charged$(q)$ ensemble and the fixed potential$(\phi)$ ensemble. For charged, rotating black holes, we have explored four different ensembles: fixed $(q, J)$, fixed $(\phi, J)$, fixed $(q,\Omega)$, and fixed $(\phi,\Omega)$ ensembles.Considering these black holes as topological defects at thermodynamic spaces, We have computed the associated winding numbers or the topological charges to study the local and global topologies of these black holes.\\
	
	It has been observed that for the static black hole, the topological charge remains constant at $-1$, irrespective of the three models we have considered and the respective thermodynamic parameters of the model.\\
	
	In the case of the static charged black hole in a fixed charge ensemble, the topological charge is computed to be zero and it does not change with variations in the charge $q$ and the curvature radius $l$. However, in the fixed potential ensemble, for the charged static black hole, the topological charge is found to be $-1$. In this scenario as well, the topological charge remains unaffected by variations in the potential $\phi$ and the curvature radius $l$.\\
	
	In the case of the rotating charged black hole, we have considered four ensembles and the results obtained in those ensembles can be summarized as follows : 
	For fixed $(q, J)$ ensemble, the topological charge is found to be $1$ and it does not vary with charge $q$, angular momentum $J$ and $l$. However, for different length scales,  the number of branches in $\tau$ vs $S$ plot change with variation of $q, J$ and $l$. Although it does not result in a change of the topological charge. \\
	
	In case of fixed $(\phi, J)$ ensemble, the topological charge is found to be $1$ and it does not depend upon the values of potential $\phi$, $J$ and scalar curvature $R.$ Again the number of branches in $\tau$ vs $S$ plot varies with changes in $\phi, J$ and $R$ values keeping the topological charge unchanged.\\
	
	In the case of fixed $(q,\Omega)$ ensemble, the topological charge is $1$ or $0$ depending on the value of scalar curvature $R$. Although the number of branches of $\tau$ vs $S$ plot changes with variation of $q$ and $\Omega$ values for a fixed $R$, the topological charge remains unaltered.\\
	
	Finally, in fixed $(\Omega, \phi)$ ensemble, the topological charge is $-1, 0$ or $1$ depending on the values of $R, \Omega$ and $\phi$. The results for the charged, rotating black hole case are summarized in the following table.\\
	
	\clearpage
	\begin{table}[h]
		\resizebox{0.65\textwidth}{!}{
			\begin{tabular}{|c|c|c|c|c|c|}
				\hline
				\multirow{9}{*}{\textbf{\begin{tabular}[c]{@{}c@{}}Fixed $(q,J)$\\ ensemble\end{tabular}}}                          & \multicolumn{1}{l|}{\textbf{q}}      & \multicolumn{1}{l|}{\textbf{J}} & \multicolumn{1}{l|}{\textbf{l}} & \multicolumn{1}{l|}{\textbf{\begin{tabular}[c]{@{}l@{}}Number of\\ branches in\\ $\tau$ vs $S$ plot\end{tabular}}}  & \multicolumn{1}{l|}{\textbf{Topological charge}} \\ \cline{2-6} 
				& 0.05                                 & 1.5                             & 0.1                             & 1                                                                                                                   & 1                                                \\ \cline{2-6} 
				& 0.05                                 & 7                               & 0.1                             & 1                                                                                                                   & 1                                                \\ \cline{2-6} 
				& 0.001                                & 1.5                             & 0.1                             & 1                                                                                                                   & 1                                                \\ \cline{2-6} 
				& 2                                    & 1.5                             & 0.1                             & 1                                                                                                                   & 1                                                \\ \cline{2-6} 
				& 0.05                                 & 1.5                             & 10                              & 3                                                                                                                   & 1                                                \\ \cline{2-6} 
				& 0.05                                 & 7                               & 10                              & 1                                                                                                                   & 1                                                \\ \cline{2-6} 
				& 0.001                                & 1.5                             & 10                              & 3                                                                                                                   & 1                                                \\ \cline{2-6} 
				& 2                                    & 1.5                             & 10                              & 1                                                                                                                   & 1                                                \\ \hline
				\multicolumn{1}{|l|}{\multirow{9}{*}{\textbf{\begin{tabular}[c]{@{}l@{}}Fixed $(\phi,J)$\\ ensemble\end{tabular}}}} & \multicolumn{1}{l|}{\textbf{$\phi$}} & \multicolumn{1}{l|}{\textbf{J}} & \multicolumn{1}{l|}{\textbf{R}} & \multicolumn{1}{l|}{\textbf{\begin{tabular}[c]{@{}l@{}}Number of\\  branches in\\ $\tau$ vs $S$ plot\end{tabular}}} & \multicolumn{1}{l|}{\textbf{Topological charge}} \\ \cline{2-6} 
				\multicolumn{1}{|l|}{}                                                                                              & 0.05                                 & 1.5                             & -0.1                            & 3                                                                                                                   & 1                                                \\ \cline{2-6} 
				\multicolumn{1}{|l|}{}                                                                                              & 0.05                                 & 10                              & -0.1                            & 1                                                                                                                   & 1                                                \\ \cline{2-6} 
				\multicolumn{1}{|l|}{}                                                                                              & 0.005                                & 1.5                             & -0.1                            & 3                                                                                                                   & 1                                                \\ \cline{2-6} 
				\multicolumn{1}{|l|}{}                                                                                              & 3                                    & 1.5                             & -0.1                            & 3                                                                                                                   & 1                                                \\ \cline{2-6} 
				\multicolumn{1}{|l|}{}                                                                                              & 0.05                                 & 1.5                             & -4                              & 1                                                                                                                   & 1                                                \\ \cline{2-6} 
				\multicolumn{1}{|l|}{}                                                                                              & 0.05                                 & 3                               & -4                              & 1                                                                                                                   & 1                                                \\ \cline{2-6} 
				\multicolumn{1}{|l|}{}                                                                                              & 0.005                                & 1.5                             & -4                              & 1                                                                                                                   & 1                                                \\ \cline{2-6} 
				\multicolumn{1}{|l|}{}                                                                                              & 3                                    & 1.5                             & -4                              & 3                                                                                                                   & 1                                                \\ \hline
				\multirow{11}{*}{\textbf{Fixed $(q,\Omega)$}}                                                                       & \textbf{q}                           & \textbf{$\Omega$}  & \textbf{R}                      & \textbf{\begin{tabular}[c]{@{}c@{}}Number of\\ branch in\\ $\tau$ vs $S$ plot\end{tabular}}                         & \textbf{Topological charge}                      \\ \cline{2-6} 
				& 0.1                                  & 0.1                             & -0.01                           & 2                                                                                                                   & 0                                                \\ \cline{2-6} 
				& 0.1                                  & 0.01                            & -0.01                           & 2                                                                                                                   & 0                                                \\ \cline{2-6} 
				& 0.1                                  & 3                               & -0.01                           & 2                                                                                                                   & 0                                                \\ \cline{2-6} 
				& 0.09                                 & 0.1                             & -0.01                           & 2                                                                                                                   & 0                                                \\ \cline{2-6} 
				& 1                                    & 0.1                             & -0.01                           & 2                                                                                                                   & 0                                                \\ \cline{2-6} 
				& 0.1                                  & 0.1                             & -4                              & 3                                                                                                                   & 1                                                \\ \cline{2-6} 
				& 0.1                                  & 0.001                           & -4                              & 3                                                                                                                   & 1                                                \\ \cline{2-6} 
				& 0.1                                  & 1                               & -4                              & 3                                                                                                                   & 1                                                \\ \cline{2-6} 
				& 0.09                                 & 0.1                             & -4                              & 3                                                                                                                   & 1                                                \\ \cline{2-6} 
				& 1                                    & 0.1                             & -4                              & 1                                                                                                                   & 1                                                \\ \hline
				\multirow{11}{*}{\textbf{Fixed $(\Omega,\phi)$}}                                                                    & \textbf{$\phi$}                      & \textbf{$\Omega$}               & \textbf{R}                      & \textbf{\begin{tabular}[c]{@{}c@{}}Number of\\ branches in\\ $\tau$ vs $S$ plot\end{tabular}}                       & \textbf{Topological charge}                      \\ \cline{2-6} 
				& 0.1                                  & 0.1                             & -0.01                           & 1                                                                                                                   & -1                                               \\ \cline{2-6} 
				& 0.1                                  & 0.001                           & -0.01                           & 1                                                                                                                   & -1                                               \\ \cline{2-6} 
				& 0.1                                  & 1                               & -0.01                           & 1                                                                                                                   & -1                                               \\ \cline{2-6} 
				& 0.001                                & 0.1                             & -0.01                           & 1                                                                                                                   & -1                                               \\ \cline{2-6} 
				& 2                                    & 0.1                             & -0.01                           & 1                                                                                                                   & 1                                                \\ \cline{2-6} 
				& 0.1                                  & 0.1                             & -4                              & 2                                                                                                                   & 0                                                \\ \cline{2-6} 
				& 0.1                                  & 0.001                           & -4                              & 2                                                                                                                   & 0                                                \\ \cline{2-6} 
				& 0.1                                  & 1                               & -4                              & 1                                                                                                                   & -1                                               \\ \cline{2-6} 
				& 0.001                                & 0.1                             & -4                              & 2                                                                                                                   & 0                                                \\ \cline{2-6} 
				& 1                                    & 0.1                             & -4                              & 1                                                                                                                   & 1                                                \\ \hline
			\end{tabular}
		}
	\end{table}
	
	\clearpage
	Therefore, we conclude that the thermodynamic topologies of the charged static black hole and charged rotating black hole are influenced by the choice of ensemble. In addition, the thermodynamic topology of the charged rotating black hole also depends on the thermodynamic parameters.\\
	
	It will be interesting to study the thermodynamic topology of various black hole systems in other modified theories of gravity. We plan to do so in our future work.\\
	

\begin{thebibliography}{99}
		\bibitem{1} S.W Hawking,\textit{Gravitational Radiation from Colliding Black Holes},
		PhysRevLett.\textbf{26}.1344,1971.
		\bibitem{2} Jacob D. Bekenstein. \textit{Black holes and entropy}, Phys. Rev. D, \textbf{7}:2333–
		2346,  1973
		\bibitem{3} James M. Bardeen, B. Carter, and S. W. Hawking. \textit{The Four laws of
			black hole mechanics}, Commun. Math. Phys,\textit{31}, 1973.
		\bibitem{4} S.W.Hawking. \textit{Black hole explosions?},Nature.\textbf{248}.5443,1974
		\bibitem{5} Jacob D. Bekenstein.\textit{ Generalized second law of thermodynamics in
			black-hole physics.} Phys. Rev. D, \textbf{9}:3292–3300, Jun 1974.
		\bibitem{6}  S.W. Hawking, \textit{Particle Creation by Black Holes}, Commun. Math. Phys., 43:199–220, 1975. [Erratum: Commun.Math.Phys. 46, 206 (1976)]
		\bibitem{7} Robert M. Wald. \textit{Entropy and black-hole thermodynamics.} Phys. Rev.
		\textbf{D}, 20:1271–1282,1979.
		\bibitem{8} Jacob D. Bekenstein. Black-hole thermodynamics. Physics Today \textbf{33}.24-31 1980.
		\bibitem{9} W. Israel. \textit{Third law of black-hole dynamics: A formulation and proof.}
		Phys. Rev. Lett., \textbf{57}:397–399,1986.
		\bibitem{10} David Kastor, Sourya Ray, and Jennie Traschen. \textit{Enthalpy and
			the mechanics of AdS black holes}, Classical and Quantum Gravity \textbf{
			26}(19):195011, 2009.
		\bibitem{11} Sharmila Gunasekaran, David Kubizˇn´ak, and Robert B. Mann. \textit{Extended phase space thermodynamics for charged and rotating black
			holes and born-infeld vacuum polarization.} Journal of High Energy
		Physics, \textbf{2012}(11),  2012.
		\bibitem{12} Brian P.\textit{ Where is the PdV in the first law of black hole thermodynamics?} Open Questions in Cosmology. InTech, 2012.
		\bibitem{13} Deyou Chen, Gan qingyu, and Jun Tao. \textit{The modified first laws of
			thermodynamics of anti-de sitter and de sitter space–times.}, Nuclear
		Physics B, \textbf{918}:115–128, 2017.
		\bibitem{14} David Kubizˇn´ak and Robert B. Mann. \textit{P-v criticality of charged AdS
			black holes.} Journal of High Energy Physics, \textbf{2012}(7),2012.
		\bibitem{15} Natacha Altamirano, David Kubizˇn ´ak, and Robert B. Mann. \textit{Reentrant
			phase transitions in rotating anti–de sitter black holes.} Physical Review
		D, \textbf{88}(10),2013.
		\bibitem{16} Natacha Altamirano, David Kubizˇn´ak, Robert B Mann, and Zeinab
		Sherkatghanad. \textit{Kerr-ads analogue of triple point and solid/liquid/gas
			phase transition.} Classical and Quantum Gravity, \textbf{31}(4):042001,
		2014.
		\bibitem{17} Shao-Wen Wei and Yu-Xiao Liu. \textit{Triple points and phase diagrams in
			the extended phase space of charged gauss-bonnet black holes in ads
			space.}, Phys. Rev. D, \textbf{90}:044057,2014.
		\bibitem{18} Antonia M. Frassino, David Kubiznak, Robert B. Mann, and Fil Simovic,
		,\textit{Multiple reentrant phase transitions and triple points in lovelock thermodynamics.}, Journal of High Energy Physics, \textbf{2014}(9),2014.
		\bibitem{19} Rong-Gen Cai, Li-Ming Cao, Li Li, and Run-Qiu Yang. \textit{P-v criticality
			in the extended phase space of gauss-bonnet black holes in AdS space.}
		Journal of High Energy Physics, \textbf{2013}(9), 2013
		\bibitem{20} Hao Xu, Wei Xu, and Liu Zhao. \textit{Extended phase space thermodynamics
			for third-order lovelock black holes in diverse dimensions.}, The European
		Physical Journal C, \textbf{74}(9),2014.
		\bibitem{21} Brian P Dolan, Anna Kostouki, David Kubizˇn´ak, and Robert B Mann.
		\textit{Isolated critical point from lovelock gravity.} Classical and Quantum
		Gravity, \textbf{31}(24):242001, 2014
		\bibitem{22} Robie A. Hennigar, W. G. Brenna, and Robert B. Mann. \textit{P-v criticality
			in quasitopological gravity.}, Journal of High Energy Physics, \textbf{2015}(7), 2015.
		\bibitem{23} Robie Hennigar and Robert Mann. \textit{Reentrant phase transitions and van
			der waals behaviour for hairy black holes.} Entropy\textbf{17}(12):8056–8072, 2015.
		\bibitem{24} Robie A. Hennigar, Robert B. Mann, and Erickson Tjoa. \textit{Superfluid
			black holes.} Physical Review Letters \textbf{118}(2),2017.
		\bibitem{25} De-Cheng Zou,Ruihong Yue, and Ming Zhang. \textit{Reentrant phase transitions of higher-dimensional ads black holes in drgt massive gravity.}
		The European Physical Journal C, \textbf{77}(4),2017.
		\bibitem{new} Gogoi, Naba Jyoti and Phukon, Prabwal. \textit{Thermodynamic geometry of 5D $R$-charged black holes in extended thermodynamic space.} Phys. Rev. D, \textbf{103}(126008),2021. doi:10.1103/PhysRevD.103.126008.
		\bibitem{rp1}
		J.~Sadeghi, M.~Shokri, S.~Gashti Noori and M.~R.~Alipour,
		\textit{RPS thermodynamics of Taub\textendash{}NUT AdS black holes in the presence of central charge and the weak gravity conjecture}.
		Gen. Rel. Grav. \textbf{54} (2022)
		\bibitem{rp2}
		Y.~Ladghami, B.~Asfour, A.~Bouali, A.~Errahmani and T.~Ouali,
		\textit{4D-EGB black holes in RPS thermodynamics},
		Phys. Dark Univ. \textbf{41} (2023),
		\bibitem{rp3}
		X.~Kong, Z.~Zhang and L.~Zhao,
		\textit{Restricted phase space thermodynamics of charged AdS black holes in conformal gravity},
		Chin. Phys. C \textbf{47} (2023)
		\bibitem{rp4}
		M.~R.~Alipour, J.~Sadeghi and M.~Shokri,
		\textit{WGC and WCCC of black holes with quintessence and cloud strings in RPS space},
		Nucl. Phys. B \textbf{990} (2023)
		\bibitem{rp5}
		T.~Wang and L.~Zhao,
		\textit{Black hole thermodynamics is extensive with variable Newton constant},
		Phys. Lett. B \textbf{827} (2022)
		\bibitem{rp6}
		G.~Zeyuan and L.~Zhao,
		\textit{Restricted phase space thermodynamics for AdS black holes via holography},
		Class. Quant. Grav. \textbf{39} (2022)
		\bibitem{rp7}
		Z.~Gao, X.~Kong and L.~Zhao,
		\textit{Thermodynamics of Kerr-AdS black holes in the restricted phase space},
		Eur. Phys. J. C \textbf{82} (2022)
		\bibitem{rp8}
		S.~Dutta and G.~S.~Punia,
		\textbf{String theory corrections to holographic black hole chemistry},
		Phys. Rev. D \textbf{106}, no.2, 026003 (2022)
		\bibitem{rp9}
		T.~F.~Gong, J.~Jiang and M.~Zhang,
		\textit{Holographic thermodynamics of rotating black holes},
		JHEP \textbf{06}, 105 (2023)
		\bibitem{rp10}
		W.~Cong, D.~Kubiznak, R.~B.~Mann and M.~R.~Visser,
		\textit{Holographic CFT phase transitions and criticality for charged AdS black holes},
		JHEP \textbf{08}, 174 (2022)
		\bibitem{rp11}
		M.~R.~Visser,
		\text{Holographic thermodynamics requires a chemical potential for color},
		Phys. Rev. D \textbf{105}, no.10, 106014 (2022)
		\bibitem{28} Shao-Wen Wei and Yu-Xiao Liu. \textit{Topology of black hole thermodynamics.} Phys. Rev. D \textbf{105}:104003, 2022.
		\bibitem{29} Shao-Wen Wei, Yu-Xiao Liu, and Robert B. Mann.\textit{ Black hole solutions as topological thermodynamic defects.} Phys. Rev. Lett.\textbf{129}:191101,2022.
		\bibitem{30} C.~Fairoos,
		\textit{Topological Interpretation Black Hole Phase Transition in Gauss-Bonnet Gravity},
		[arXiv:2311.04050 [gr-qc]].
		\bibitem{31} M.~Rizwan and K.~Jusufi,
		\textit{Topological classes of thermodynamics of black holes in perfect fluid dark matter background}
		Eur. Phys. J. C \textbf{83}, no.10, 944 (2023)
		doi:10.1140/epjc/s10052-023-12126-1
		[arXiv:2310.15182 [gr-qc]]
		\bibitem{32} A.~Mehmood and M.~U.~Shahzad,
		\textit{Thermodynamic Topological Classifications of Well-Known Black Holes},
		[arXiv:2310.09907 [hep-th]]
		\bibitem{33} C.~W.~Tong, B.~H.~Wang and J.~R.~Sun,
		\textit{Topology of black hole thermodynamics via R\'enyi statistics},
		[arXiv:2310.09602 [gr-qc]].
		\bibitem{34} M.~U.~Shahzad, A.~Mehmood, S.~Sharif and A.~\"Ovg\"un,
		\textit{Criticality and topological classes of neutral Gauss\textendash{}Bonnet AdS black holes in 5D},
		Annals Phys. \textbf{458}, 169486 (2023)
		doi:10.1016/j.aop.2023.169486.
		\bibitem{35} Z.~Q.~Chen and S.~W.~Wei,
		\textit{Thermodynamics, Ruppeiner geometry, and topology of Born-Infeld black hole in asymptotic flat spacetime},
		Nucl. Phys. B \textbf{996}, 116369 (2023)
		doi:10.1016/j.nuclphysb.2023.116369.
		\bibitem{36} C.~Liu, R.~Li, K.~Zhang and J.~Wang,
		\textit{Generalized free energy and dynamical state transition of the dyonic AdS black hole in the grand canonical ensemble},
		[arXiv:2309.13931 [gr-qc]].
		\bibitem{37} F.~Barzi, H.~El Moumni and K.~Masmar,
		\textit{R\'enyi Topology of Charged-flat Black Hole: Hawking-Page and Van-der-Waals Phase Transitions},
		[arXiv:2309.14069 [hep-th]].
		\bibitem{38} Y.~Z.~Du, H.~F.~Li, Y.~B.~Ma and Q.~Gu,
		\textit{Topology and phase transition for EPYM AdS black hole in thermal potential},
		[arXiv:2309.00224 [hep-th]].
		\bibitem{39} F.~Demami, H.~El Moumni, K.~Masmar and S.~Mazzou,
		\textit{Thermodynamics and phase transition structure of charged black holes in f(R) gravity background from R\'enyi statistics perception},
		Nucl. Phys. B \textbf{994}, 116316 (2023)
		doi:10.1016/j.nuclphysb.2023.116316
		\bibitem{40} R.~Li, C.~Liu, K.~Zhang and J.~Wang,
		\textit{Topology of the landscape and dominant kinetic path for the thermodynamic phase transition of the charged Gauss-Bonnet-AdS black holes},
		Phys. Rev. D \textbf{108}, no.4, 044003 (2023)
		doi:10.1103/PhysRevD.108.044003
		[arXiv:2302.06201 [gr-qc]].
		\bibitem{41} D.~Wu,
		\textit{Topological classes of thermodynamics of the four-dimensional static accelerating black holes},
		Phys. Rev. D \textbf{108}, no.8, 084041 (2023)
		doi:10.1103/PhysRevD.108.084041
		[arXiv:2307.02030 [hep-th]]
		\bibitem{42} Y.~S.~Wang, Z.~M.~Xu and B.~Wu,
		\textit{Thermodynamic phase transition and winding number for the third-order Lovelock black hole},
		[arXiv:2307.01569 [gr-qc]].
		\bibitem{43} D.~Chen, Y.~He and J.~Tao,
		\textit{Topological classes of higher-dimensional black holes in massive gravity},
		Eur. Phys. J. C \textbf{83}, no.9, 872 (2023)
		doi:10.1140/epjc/s10052-023-11983-0
		[arXiv:2306.13286 [gr-qc]].
		\bibitem{44} J.~Sadeghi, S.~Noori Gashti, M.~R.~Alipour and M.~A.~S.~Afshar,
		\textit{Bardeen black hole thermodynamics from topological perspective},
		Annals Phys. \textbf{455}, 169391 (2023)
		doi:10.1016/j.aop.2023.169391
		[arXiv:2306.05692 [hep-th]]. 
		\bibitem{45} D.~Wu,
		\textit{Consistent thermodynamics and topological classes for the four-dimensional Lorentzian charged Taub-NUT spacetimes},
		Eur. Phys. J. C \textbf{83}, no.7, 589 (2023)
		doi:10.1140/epjc/s10052-023-11782-7
		[arXiv:2306.02324 [gr-qc]].
		\bibitem{46} T.~N.~Hung and C.~H.~Nam,
		\textit{Topology in thermodynamics of regular black strings with Kaluza\textendash{}Klein reduction},
		Eur. Phys. J. C \textbf{83}, no.7, 582 (2023)
		doi:10.1140/epjc/s10052-023-11768-5
		[arXiv:2305.15910 [gr-qc]].
		\bibitem{47} M.~Y.~Zhang, H.~Chen, H.~Hassanabadi, Z.~W.~Long and H.~Yang,
		\textit{Topology of nonlinearly charged black hole chemistry via massive gravity},
		Eur. Phys. J. C \textbf{83}, no.8, 773 (2023)
		doi:10.1140/epjc/s10052-023-11933-w
		[arXiv:2305.15674 [gr-qc]].
		\bibitem{48} N.~J.~Gogoi and P.~Phukon,
		\textit{Topology of thermodynamics in R-charged black holes},
		Phys. Rev. D \textbf{107}, no.10, 106009 (2023)
		doi:10.1103/PhysRevD.107.106009
		\bibitem{49} Z.~M.~Xu, Y.~S.~Wang, B.~Wu and W.~L.~Yang,
		\textit{Thermodynamic phase transition of the AdS black holes from the perspective of the complex analysis},
		[arXiv:2305.05916 [gr-qc]].
		\bibitem{50} M.~R.~Alipour, M.~A.~S.~Afshar, S.~Noori Gashti and J.~Sadeghi,
		\textit{Topological classification and black hole thermodynamics},
		Phys. Dark Univ. \textbf{42}, 101361 (2023)
		doi:10.1016/j.dark.2023.101361
		[arXiv:2305.05595 [gr-qc]].
		\bibitem{51} P.~K.~Yerra, C.~Bhamidipati and S.~Mukherji,
		\textit{Topology of critical points in boundary matrix duals},
		[arXiv:2304.14988 [hep-th]].
		\bibitem{52} N.~J.~Gogoi and P.~Phukon,
		\textit{Thermodynamic topology of 4D dyonic AdS black holes in different ensembles},
		Phys. Rev. D \textbf{108}, no.6, 066016 (2023)
		doi:10.1103/PhysRevD.108.066016
		[arXiv:2304.05695 [hep-th]].
		\bibitem{53} C.~Fairoos and T.~Sharqui,
		\textit{Topological nature of black hole solutions in dRGT massive gravity},
		Int. J. Mod. Phys. A \textbf{38}, no.25, 2350133 (2023)
		doi:10.1142/S0217751X23501336
		[arXiv:2304.02889 [gr-qc]]. 
		\bibitem{54} R.~Li and J.~Wang,
		\textit{Generalized free energy landscapes of charged Gauss-Bonnet-AdS black holes in diverse dimensions},
		Phys. Rev. D \textbf{108}, no.4, 044057 (2023)
		doi:10.1103/PhysRevD.108.044057
		[arXiv:2304.03425 [gr-qc]].
		\bibitem{55} M.~Zhang and J.~Jiang,
		\textit{Bulk-boundary thermodynamic equivalence: a topology viewpoint},
		JHEP \textbf{06}, 115 (2023)
		doi:10.1007/JHEP06(2023)115
		[arXiv:2303.17515 [hep-th]].
		\bibitem{56} Y.~Du and X.~Zhang,
		\textit{Topological classes of black holes in de-Sitter spacetime},
		Eur. Phys. J. C \textbf{83}, no.10, 927 (2023)
		doi:10.1140/epjc/s10052-023-12114-5
		[arXiv:2303.13105 [gr-qc]].
		\bibitem{57} S.~W.~Wei, Y.~P.~Zhang, Y.~X.~Liu and R.~B.~Mann,
		\textit{Static spheres around spherically symmetric black hole spacetime},
		Phys. Rev. Res. \textbf{5}, no.4, 043050 (2023)
		doi:10.1103/PhysRevResearch.5.043050
		[arXiv:2303.06814 [gr-qc]].
		\bibitem{58} Y.~Du and X.~Zhang,
		\textit{Topological classes of BTZ black holes},
		[arXiv:2302.11189 [gr-qc]].
		\bibitem{59} Q.~Yu, Q.~Xu and J.~Tao,
		\textit{Thermodynamics and microstructures of Euler\textendash{}Heisenberg black hole in a cavity},
		Commun. Theor. Phys. \textbf{75}, no.9, 095402 (2023)
		doi:10.1088/1572-9494/ace4b3
		[arXiv:2302.09821 [gr-qc]].
		\bibitem{60} N.~Chatzifotis, P.~Dorlis, N.~E.~Mavromatos and E.~Papantonopoulos,
		\textit{Thermal stability of hairy black holes},
		Phys. Rev. D \textbf{107}, no.8, 084053 (2023)
		doi:10.1103/PhysRevD.107.084053
		[arXiv:2302.03980 [gr-qc]].
		\bibitem{61}D.~Wu,
		\textit{Classifying topology of consistent thermodynamics of the four-dimensional neutral Lorentzian NUT-charged spacetimes},
		Eur. Phys. J. C \textbf{83}, no.5, 365 (2023)
		doi:10.1140/epjc/s10052-023-11561-4
		[arXiv:2302.01100 [gr-qc]].
		\bibitem{62} D.~Wu and S.~Q.~Wu,
		\textit{Topological classes of thermodynamics of rotating AdS black holes},
		Phys. Rev. D \textbf{107}, no.8, 084002 (2023)
		doi:10.1103/PhysRevD.107.084002
		[arXiv:2301.03002 [hep-th]].
		\bibitem{63} D.~Wu,
		\textit{Topological classes of rotating black holes},
		Phys. Rev. D \textbf{107}, no.2, 024024 (2023)
		doi:10.1103/PhysRevD.107.024024
		[arXiv:2211.15151 [gr-qc]].
		\bibitem{64} C.~Fang, J.~Jiang and M.~Zhang,
		\textit{Revisiting thermodynamic topologies of black holes},
		JHEP \textbf{01}, 102 (2023)
		doi:10.1007/JHEP01(2023)102
		[arXiv:2211.15534 [gr-qc]].
		\bibitem{65} Z.~Y.~Fan,
		\textit{Topological interpretation for phase transitions of black holes}
		Phys. Rev. D \textbf{107}, no.4, 044026 (2023)
		doi:10.1103/PhysRevD.107.044026
		[arXiv:2211.12957 [gr-qc]].
		\bibitem{66} C.~Liu and J.~Wang,
		\textit{Topological natures of the Gauss-Bonnet black hole in AdS space},
		Phys. Rev. D \textbf{107}, no.6, 064023 (2023)
		doi:10.1103/PhysRevD.107.064023
		[arXiv:2211.05524 [gr-qc]].
		\bibitem{67} N.~C.~Bai, L.~Li and J.~Tao,
		\textit{Topology of black hole thermodynamics in Lovelock gravity}
		Phys. Rev. D \textbf{107}, no.6, 064015 (2023)
		doi:10.1103/PhysRevD.107.064015
		[arXiv:2208.10177 [gr-qc]].
		\bibitem{68} P.~K.~Yerra, C.~Bhamidipati and S.~Mukherji,
		\textit{Topology of critical points and Hawking-Page transition},
		Phys. Rev. D \textbf{106}, no.6, 064059 (2022)
		doi:10.1103/PhysRevD.106.064059
		[arXiv:2208.06388 [hep-th]].
		\bibitem{69}
		J.~Sadeghi, M.~A.~S.~Afshar, S.~Noori Gashti and M.~R.~Alipour,
		\textit{Thermodynamic topology of black holes from bulk-boundary, extended, and restricted phase space perspectives},
		Annals Phys. \textbf{460} (2024), 169569
		doi:10.1016/j.aop.2023.169569
		[arXiv:2312.04325 [hep-th]]
		\bibitem{70}
		N.~J.~Gogoi and P.~Phukon,
		\textit{Thermodynamic topology of 4D Euler-Heisenberg-AdS black hole in different ensembles},
		[arXiv:2312.13577 [hep-th]]
		\bibitem{71}
		B.~Hazarika and P.~Phukon,
		\textit{Thermodynamic Topology of $D=4,5$ Horava Lifshitz Black Hole in Two Ensembles},
		[arXiv:2312.06324 [hep-th]].
		\bibitem{72}
		M.~Y.~Zhang, H.~Chen, H.~Hassanabadi, Z.~W.~Long and H.~Yang,
		\textit{Thermodynamic topology of Kerr-Sen black holes via R\'enyi statistics},
		[arXiv:2312.12814 [gr-qc]].
		\bibitem{73}
		D.~Chen, Y.~He, J.~Tao and W.~Yang,
		\textit{Topology of Ho\v{r}ava-Lifshitz black holes in different ensembles},
		[arXiv:2311.11606 [hep-th]].
		\bibitem{g1} Y. S. Duan, \textit{The structure of the topological current},SLAC-PUB-3301,1984.
		\bibitem{g2} Yi-Shi Duan and Mo-Lin Ge. \textit{SU(2) Gauge Theory and Electrodynamics with N Magnetic Monopoles},Sci.Sin \textbf{9},1072,1979
		
		\bibitem{f1} Nojiri, S.D. Odintsov,\textit{ Unified cosmic history in modified gravity: from F(R) theory to Lorentz non-invariant models}. Phys. Rep.
		\textbf{505}, 59–144 (2011). https://doi.org/10.1016/j.physrep.2011.04.
		001. arXiv:1011.0544 [gr-qc]
		
		\bibitem{f2} T.P. Sotiriou, V. Faraoni, \textit{f(R) theories of gravity}. Rev. Mod.
		Phys. \textbf{82}, 451–497 (2010). https://doi.org/10.1103/RevModPhys.
		82.451. arXiv:0805.1726 [gr-qc]
		\bibitem{f3} A. De Felice, S. Tsujikawa, \textit{f(R) theories}. Living Rev. Relativ.\textbf{13},3 (2010). https://doi.org/10.12942/lrr-2010-3. arXiv:1002.4928
		[gr-qc]
		\bibitem{f4} S. Chakraborty, S. SenGupta, \textit{Spherically symmetric brane spacetime with bulk f(r) gravity}. Eur. Phys. J. C \textbf{75}(1), 11 (2015). https://
		doi.org/10.1140/epjc/s10052-014-3234-3
		\bibitem{f5} S. Nojiri, S.D. Odintsov, V.K. Oikonomou, \textit{Modified gravity theories on a nutshell: inflation, bounce and late-time evolution}. Phys.
		Rep. \textbf{692}, 1–104 (2017). https://doi.org/10.1016/j.physrep.2017.
		06.001. arXiv:1705.11098 [gr-qc]
		\bibitem{new10}
		S.~D.~Odintsov, V.~K.~Oikonomou, I.~Giannakoudi, F.~P.~Fronimos and
		E.~C.~Lymperiadou,\textit{Recent Advances on Inflation},
		[arXiv:2307.16308 [gr-qc]].
		
		\bibitem{f6} S. Nojiri, S.D. Odintsov, \textit{Newton law corrections and instabilities in f(R) gravity with the effective cosmological constant epoch}.
		Phys. Lett. B \textbf{652}, 343–348 (2007). https://doi.org/10.1016/j.
		physletb.2007.07.039. arXiv:0706.1378 [hep-th]
		\bibitem{new1}
		S.~Nojiri and S.~D.~Odintsov,
		\textit{Modified gravity with negative and positive powers of the curvature: Unification of the inflation and of the cosmic acceleration},
		Phys. Rev. D \textbf{68} (2003), 123512
		doi:10.1103/PhysRevD.68.123512
		[arXiv:hep-th/0307288 [hep-th]].
		\bibitem{new2}
		S.~Nojiri and S.~D.~Odintsov,
		\textit{Modified f(R) gravity consistent with realistic cosmology: From matter dominated epoch to dark energy universe},
		Phys. Rev. D \textbf{74} (2006), 086005
		doi:10.1103/PhysRevD.74.086005
		[arXiv:hep-th/0608008 [hep-th]].
		
		\bibitem{new3}
		G.~Cognola, E.~Elizalde, S.~Nojiri, S.~D.~Odintsov, L.~Sebastiani and S.~Zerbini,
		\textit{A Class of viable modified f(R) gravities describing inflation and the onset of accelerated expansion},
		Phys. Rev. D \textbf{77} (2008), 046009
		doi:10.1103/PhysRevD.77.046009
		[arXiv:0712.4017 [hep-th]].
		\bibitem{new4}
		S.~Capozziello, S.~Nojiri, S.~D.~Odintsov and A.~Troisi,
		\textit{Cosmological viability of f(R)-gravity as an ideal fluid and its compatibility with a matter dominated phase},
		Phys. Lett. B \textbf{639} (2006), 135-143
		doi:10.1016/j.physletb.2006.06.034
		[arXiv:astro-ph/0604431 [astro-ph]]
		\bibitem{new5}
		J.~de Haro, S.~Nojiri, S.~D.~Odintsov, V.~K.~Oikonomou and S.~Pan,
		\textit{Finite-time cosmological singularities and the possible fate of the Universe},
		Phys. Rept. \textbf{1034} (2023), 1-114
		doi:10.1016/j.physrep.2023.09.003
		[arXiv:2309.07465 [gr-qc]].
		
		\bibitem{new6}
		E.~Elizalde, G.~G.~L.~Nashed, S.~Nojiri and S.~D.~Odintsov,
		\textit{Spherically symmetric black holes with electric and magnetic charge in extended gravity: physical properties, causal structure, and stability analysis in Einstein\textquoteright{}s and Jordan\textquoteright{}s frames},
		Eur. Phys. J. C \textbf{80} (2020) no.2, 109
		doi:10.1140/epjc/s10052-020-7686-3
		[arXiv:2001.11357 [gr-qc]].
		\bibitem{new7}
		S.~Nojiri and S.~D.~Odintsov,
		\textit{Regular multihorizon black holes in modified gravity with nonlinear electrodynamics},
		Phys. Rev. D \textbf{96} (2017) no.10, 104008
		doi:10.1103/PhysRevD.96.104008
		[arXiv:1708.05226 [hep-th]].
		\bibitem{new8}
		A.~Addazi, S.~Nojiri and S.~Odintsov,
		\textit{Evaporation and antievaporation instability of a Schwarzschild\textendash{}de Sitter braneworld: The case of five-dimensional $F(R)$ gravity},
		Phys. Rev. D \textbf{95} (2017) no.12, 124020
		doi:10.1103/PhysRevD.95.124020
		[arXiv:1705.03265 [gr-qc]].
		\bibitem{new9}
		S.~Nojiri and S.~D.~Odintsov,
		\textit{Anti-Evaporation of Schwarzschild-de Sitter Black Holes in $F(R)$ gravity},
		Class. Quant. Grav. \textbf{30} (2013), 125003
		doi:10.1088/0264-9381/30/12/125003
		[arXiv:1301.2775 [hep-th]].
		\bibitem{new11} 
		S.~Nojiri and S.~D.~Odintsov,
		\textit{Instabilities and anti-evaporation of Reissner\textendash{}Nordstr\"om black holes in modified $F(R)$ gravity},
		Phys. Lett. B \textbf{735}, 376-382 (2014)
		doi:10.1016/j.physletb.2014.06.070
		[arXiv:1405.2439 [gr-qc]].
		\bibitem{new12} S.~Nojiri and S.~D.~Odintsov,
		\textit{Regular multihorizon black holes in modified gravity with nonlinear electrodynamics},
		Phys. Rev. D \textbf{96}, no.10, 104008 (2017)
		doi:10.1103/PhysRevD.96.104008
		[arXiv:1708.05226 [hep-th]].
		\bibitem{f7} S. Nojiri, S.D. Odintsov, \textit{Unifying inflation with LambdaCDM
			epoch in modified f(R) gravity consistent with Solar System
			tests}. Phys. Lett. B \textbf{657}, 238–245 (2007). https://doi.org/10.1016/
		j.physletb.2007.10.027. arXiv:0707.1941 [hep-th]
		\bibitem{f8} S. Nojiri, S.D. Odintsov, \textit{Modified f(R) gravity unifying
			R**m inflation with Lambda CDM epoch}. Phys. Rev. D \textbf{77},
		026007 (2008). https://doi.org/10.1103/PhysRevD.77.026007.
		arXiv:0710.1738 [hep-th]
		\bibitem{f9a} S. Capozziello, A. Stabile, A. Troisi, \textit{The Newtonian Limit of f(R) gravity}. Phys. Rev. D \textbf{76}, 104019 (2007). https://doi.org/10.1103/
		PhysRevD.76.104019. arXiv:0708.0723 [gr-qc]
		\bibitem{f9b} T. Multamaki, I. Vilja, Phys. Rev. D \textbf{74}, 064022 (2006). https://
		doi.org/10.1103/PhysRevD.74.064022. arXiv:astro-ph/0606373
		[astro-ph]
		\bibitem{f10} L. Sebastiani, S. Zerbini, Eur. Phys. J. C \textbf{71}, 1591 (2011). https://doi.org/10.1140/epjc/s10052-011-1591-8. arXiv:1012.5230 [gr-qc]
		\bibitem{f11a} Z. Amirabi, M. Halilsoy, S. Habib Mazharimousavi, Eur.
		Phys. J. C \textbf{76}(6), 338 (2016). https://doi.org/10.1140/epjc/
		s10052-016-4164-z. arXiv:1509.06967 [gr-qc]
		\bibitem{f11b} G.G.L. Nashed, Int. J. Mod. Phys. D \textbf{27}(7), 1850074 (2018).
		https://doi.org/10.1142/S0218271818500748
		\bibitem{f12} G.G.L. Nashed, Eur. Phys. J. Plus \textbf{133}(1), 18 (2018). https://doi.org/10.1140/epjp/i2018-11849-7
		\bibitem{f13} G.G.L. Nashed, Adv. High Energy Phys. \textbf{2018}, 7323574 (2018).
		https://doi.org/10.1155/2018/7323574
		\bibitem{f14} A. de la Cruz-Dombriz, A. Dobado, A.L. Maroto, Phys. Rev. D \textbf{80},
		124011 (2009) [Erratum: Phys. Rev. D 83, 029903 (2011)]. https://
		doi.org/10.1103/PhysRevD.80.124011. arXiv:0907.3872 [gr-qc]
		\bibitem{f15} T. Moon, Y.S. Myung, E.J. Son, Gen. Relativ. Gravit. \textbf{43},
		3079–3098 (2011). https://doi.org/10.1007/s10714-011-1225-3.
		arXiv:1101.1153 [gr-qc]
		\bibitem{f16} A. de la Cruz-Dombriz, D. Saez-Gomez, Entropy \textbf{14}, 1717–1770
		(2012). https://doi.org/10.3390/e14091717. arXiv:1207.2663
		[gr-qc]
		\bibitem{f17} G.G.L. Nashed, E.N. Saridakis, Phys. Rev. D \textbf{102}(12),
		124072 (2020). https://doi.org/10.1103/PhysRevD.102.124072.
		arXiv:2010.10422 [gr-qc]
		\bibitem{f18} G.G.L. Nashed, S. Nojiri, Fortsch. Phys. \textbf{71}(2–3),
		2200091 (2023). https://doi.org/10.1002/prop.202200091.
		arXiv:2206.04836 [gr-qc]
		\bibitem{f19} T. Shirafuji, G.G.L. Nashed, Y. Kobayashi, Prog. Theor.
		Phys. \textbf{96}, 933–948 (1996). https://doi.org/10.1143/PTP.96.933.
		arXiv:gr-qc/9609060 [gr-qc]
		\bibitem{f20} G.G.L. Nashed, Astrophys. Space Sci. \textbf{330}, 173 (2010). https://
		doi.org/10.1007/s10509-010-0375-1. arXiv:1503.01379 [gr-qc]
		G.G.L. Nashed, S. Nojiri, Phys. Rev. D 104(12), 124054
		(2021). https://doi.org/10.1103/PhysRevD.104.124054.
		arXiv:2103.02382 [gr-qc]
		\bibitem{f21} G.G.L. Nashed, Phys. Lett. B \textbf{815}, 136133 (2021). https://doi.org/10.1016/j.physletb.2021.136133. arXiv:2102.11722 [gr-qc]
		\bibitem{f22} G.G.L. Nashed, S. Nojiri, Phys. Rev. D \textbf{102}, 124022
		(2020). https://doi.org/10.1103/PhysRevD.102.124022.
		arXiv:2012.05711 [gr-qc]
		\bibitem{f23} G.G.L. Nashed, S. Nojiri, Phys. Lett. B \textbf{820}, 136475
		(2021). https://doi.org/10.1016/j.physletb.2021.136475.
		arXiv:2010.04701 [hep-th]
		\bibitem{f24} G.G.L. Nashed, S. Capozziello, Phys. Rev. D \textbf{99}(10),
		104018 (2019). https://doi.org/10.1103/PhysRevD.99.104018.
		arXiv:1902.06783 [gr-qc]
		\bibitem{f25} T. Multamaki, I. Vilja, Phys. Rev. D \textbf{76}, 064021 (2007). https://
		doi.org/10.1103/PhysRevD.76.064021. arXiv:astro-ph/0612775
		[astro-ph]
		\bibitem{f26} S.H. Mazharimousavi, M. Halilsoy, T. Tahamtan, Eur.
		Phys. J. C \textbf{72}, 1958 (2012). https://doi.org/10.1140/epjc/
		s10052-012-1958-5. arXiv:1109.3655 [gr-qc]
		\bibitem{f27} S.H. Mazharimousavi, M. Halilsoy, Phys. Rev. D \textbf{84},
		064032 (2011). https://doi.org/10.1103/PhysRevD.84.064032.
		arXiv:1105.3659 [gr-qc]
		\bibitem{f28} S. Habib Mazharimousavi, M. Halilsoy, T. Tahamtan, Eur.
		Phys. J. C \textbf{72}, 1851 (2012). https://doi.org/10.1140/epjc/
		s10052-011-1851-7. arXiv:1110.5085 [gr-qc]
		\bibitem{f29} L. Hollenstein, F.S.N. Lobo, Phys. Rev. D \textbf{78}, 124007 (2008).
		https://doi.org/10.1103/PhysRevD.78.124007. arXiv:0807.2325
		[gr-qc]
		\bibitem{f30} M.E. Rodrigues, E.L.B. Junior, G.T. Marques, V.T. Zanchin, Phys.
		Rev. D \textbf{94}(2), 024062 (2016). https://doi.org/10.1103/PhysRevD.
		94.024062. arXiv:1511.00569 [gr-qc]
		\bibitem{f31} R.A. Hurtado, R. Arenas, Phys. Rev. D \textbf{102}(10), 104019
		(2020). https://doi.org/10.1103/PhysRevD.102.104019.
		arXiv:2002.06059 [gr-qc]
		\bibitem{f32} S. Capozziello, M. De laurentis, A. Stabile, Class. Quantum Gravity \textbf{27}, 165008 (2010). https://doi.org/10.1088/0264-9381/27/16/
		165008. arXiv:0912.5286 [gr-qc]
		\bibitem{f33} S.H. Hendi, Phys. Lett. B 690, 220–223 (2010). https://doi.org/
		10.1016/j.physletb.2010.05.035. arXiv:0907.2520 [gr-qc]
		\bibitem{f34} S.H. Hendi, B. Eslam Panah, S.M. Mousavi, Gen. Rel-
		ativ. Gravit. \textbf{44}, 835–853 (2012). https://doi.org/10.1007/
		s10714-011-1307-2. arXiv:1102.0089 [hep-th]
		\bibitem{f35} T.R.P. Carames, E.R. Bezerra de Mello, Eur. Phys. J. C \textbf{64}, 113–
		121 (2009). https://doi.org/10.1140/epjc/s10052-009-1115-y.
		arXiv:0901.0814 [gr-qc]
		\bibitem{f36} B. Eslam Panah, and M. E. Rodrigues, \textit{Topological phantom AdS black holes in F(R) gravity}, Eur. Phys. J. C \textbf{83} (2023) 237.
		\bibitem{f37}B. Eslam Panah, \textit{Two-dimensional Lifshitz-like AdS black holes in F(R) gravity},J. Math. Phys. \textbf{63}, 112502 (2022).
		
		\bibitem{f38}B. Eslam Panah \textit{Analytic electrically charged black holes in F(R)-ModMax theory},B. Eslam Panah, Progress of Theoretical and Experimental Physics, ptae\textbf{012}, 2024.
		
		\bibitem{f39} S.P.Sarmah and U.D.Goswami,\textit{Propagation and Fluxes of Ultra High Energy Cosmic Rays in $f(R)$ Gravity Theory},2023,arXiv:2303.16678[hep-th]
		\bibitem{f40} S.P.Sarmah and U.D.Goswami,\textit{Anisotropies of Diffusive Ultra-high Energy Cosmic Rays in $f(R)$ Gravity Theory},2023,arXiv:2309.14361[hep-th]
		
		\bibitem{fr1} Soroushfar, Saheb , Saffari, Reza Kamvar, Negin. \textit{Thermodynamic geometry of black holes in f(R) gravity}, European Physical Journal \textbf{76},2016 
		\bibitem{fr2} Moon, Taeyoon and Myung, Yun Soo and Son, Edwin J.\textit{$f(R)$ black holes},Springer Science and Business Media LLC,\textbf{43},2011
		\bibitem{rc1} Larraaga,\textit{A rotating charged black hole solution in f(R) gravity},\textbf{78},2012
		\bibitem{ec1}
		S.~Mahapatra, P.~Phukon and T.~Sarkar,\textit{On Black Hole Entropy Corrections in the Grand Canonical Ensemble},Phys. Rev. D \textbf{84}, 044041 (2011)
		\bibitem{ec2}
		N.~J.~Gogoi, G.~K.~Mahanta and P.~Phukon,
		\textit{Geodesics in geometrothermodynamics (GTD) type II geometry of 4D asymptotically anti-de-Sitter black holes},
		Eur. Phys. J. Plus \textbf{138}
		\bibitem{ec3}
		C.~S.~Peca and J.~P.~S.~Lemos,
		\textit{Thermodynamics of Reissner-Nordstrom anti-de Sitter black holes in the grand canonical ensemble}
		Phys. Rev. D \textbf{59}, 124007 (1999)
		\bibitem{ec4}
		T.~V.~Fernandes and J.~P.~S.~Lemos,
		\textit{Grand canonical ensemble of a d-dimensional Reissner-Nordstr\"om black hole in a cavity}
		Phys. Rev. D \textbf{108}, no.8, 084053 (2023)
		\bibitem{ec5}
		Z.~Wang, H.~Ren, J.~Chen and Y.~Wang,
		\textit{Thermodynamics and phase transition of Bardeen black hole via R\'enyi statistics in grand canonical ensemble and canonical ensemble},
		Eur. Phys. J. C \textbf{83}, no.6, 527 (2023)
		\bibitem{ec6}
		C.~Liu, R.~Li, K.~Zhang and J.~Wang,
		\textit{Generalized free energy and dynamical state transition of the dyonic AdS black hole in the grand canonical ensemble},
		JHEP \textbf{11}, 068 (2023)
		\bibitem{m0} Multam\"aki, T. and Vilja, I.
		\textit{Spherically symmetric solutions of modified field equations in $f(R)$ theories of gravity},Phys. Rev. D,textbf{74},American Physical Society,
		\bibitem{m1}
		R. Saffari, S. Rahvar, Phys. Rev. D \textbf{77}, 104028 (2008).
		arXiv:0708.1482 [astro-ph]
		\bibitem{m2} . S. Soroushfar, R. Saffari, J. Kunz, C. Lämmerzahl, Phys. Rev. D
		\textbf{92}(4), 044010 (2015). arXiv:1504.07854 [gr-qc]
		\bibitem{m4} 5. R. Tharanath, J. Suresh, N. Varghese, V.C. Kuriakose, Gen. Relativ.Gravit. \textbf{46}, 1743 (2014). arXiv:1404.6789 [gr-qc]
	\end{thebibliography}
\end{document}